\documentclass[12pt,letterpaper,twoside,singlespace]{mythesis} % DRAFT
\input epsf                   %%%%%%%% added by Roma
\usepackage{graphicx}
\usepackage{subfigure}
\title{\bf Search for Gamma Ray Emission from Galactic Plane with Milagro.}
\thesisauthor{Roman Fleysher}
\degree{Doctor of Philosophy}
\department{Physics}
\thesisadviser{Peter Nemethy}     %%%%%%%%%%%%%%%%%%%%%%%%%%%%%%%%%%%%%%
\submitmonth{January}             % GSAS allows January May or September
\submityear{2003}                 %%%%%%%%%%%%%%%%%%%%%%%%%%%%%%%%%%%%%%

\begin{document}
    \pagenumbering{roman}
    \maketitle
    \copyrightpage        % optional                  counted, no number
                          % blank page(optional)      counted, no number
                          % frontispiece(optional)    counted, no number
                          % dedication(optional)      numbered
    
\begin{preface}{Dedication}

\begin{center}

{\em I dedicate this work to my grandfather Aron Frenkel and
     my parents \linebreak Lidiya Frenkel and Miron Fleysher.

In memory of my grandmother Sonya Kolesnikova.}
\end{center}

\end{preface}

    \begin{preface}{Acknowledgements}

There are many people who have played important roles in what I have
accomplished today. Perhaps the most relevant is my NYU thesis advisor
Peter Nemethy who introduced me to the subject of Galactic Gamma Rays and
to the Milagro experiment. I will never forget the first month in Los
Alamos when Peter had introduced me to the people of the Milagro
collaboration and asked some of them to walk me through the Milagro
hardware. He watched that the new (to me) acronyms such as PE, VME, TOT,
ADC, TDC, DAQ are translated and explained. This allowed me quickly
understand the general picture of the project. Peter is also the first
member of the collaboration who treated me as a colleague rather than as a
student. Latter, during preparation of this manuscript, regular meetings
with Peter were essential to the progress of the data analysis. I have
learned a great deal from these interactions. I thank him for all of this.

I would like to thank Todd Haines who, effectively, was my advisor while I
was at Los Alamos. Todd is famous in the collaboration for his
phrase: "This will not work!". I had always considered this phrase as an
attention call signal. It drew my attention to the critic that followed
from his lips. I feel fortunate that I was exposed to it. This
automatically broadened my view of a problem. And, as always, a decision
can be made only when several choices are considered. He is a person of
great physical erudition and professional concern for students. He is a
big chapter in my life which I hope will stay open for many years to come.

Another member of the Milagro collaboration and also a professor at NYU is
Allen Mincer. I have met Allen even before becoming a graduate student at
physics department at NYU; he was a Director of Graduate Studies at that
time. In my interactions with Allen I quickly understood that he is a
person of great knowledge and high moral. I value his desire to get to the
roots of any problem, be it physics related or not. One rarely meets
people like him. I thank both Allen and the heavens for this.

I would also like to thank several senior members of the Milagro
collaboration: Gaurang Yodh, Cy Hoffman, Don Coyne, David Berley Jordan
Goodman and Gus Sinnis from whom I learned many things about cosmic ray
physics and life in general. I would also like to mention the other
members of the collaboration. I must recognize that, involuntary, my
interaction with a member was often limited by the Parkinson's law:

\begin{center}
  interaction $\sim (a-d)$
\end{center}

where $a$ is the age and $d$ is the average distance to the collaboration
member (both expressed in appropriate units). The collaboration spans the
entire United States from the East coast to the West one, but time is on
our side!!

I appreciate the many post docs and graduate students who made my time
spent at Los Alamos more enjoyable. These are Isabel Leonor, Diane Evans,
Rob Atkins, Joe McCullough, Morgan Wascko, Kelin Wang, Richard Miller,
Stefan Westerhoff, Andy Smith, Julie McEnery and Frank Samuelson.

I would also like to thank someone whom I seldom acknowledge but often
rely on, my life-long collaborator, opponent and proponent, my
twin-brother Zorik. I am also indebted to my parents and my grandparents
for my very existence and for the development and support of my interest
in physics. Thank you.

I was very honored by the presence of professors Peter Nemethy, Todd
Haines, Allen Mincer, Patrick Huggins and Alberto Sirlin at the defense
of my dissertation.

\begin{quote}
Roman Fleysher \\*
New York University \\*
September 18, 2002
\end{quote}

\end{preface}
         % acknoledgement            numbered
%    \include{preface}    % preface                   numbered
    \begin{preface}{Abstract}

The majority of galactic gamma rays are produced by interaction of cosmic
rays with matter. This results in a diffuse radiation concentrated in the
galactic plane where the flux of cosmic rays and the density of material
(mostly atomic, molecular and ionized hydrogen) is high. The interactions
producing gamma rays include, among others, the decay of $\pi^{0}$'s
produced in spallation reactions. Gamma emission from the plane has indeed
been detected in the energy range up to 30 GeV by space-based detectors.
Above 1 GeV, the observed intensity is notably higher than expected in
simple models, possibly implying an enhancement at the TeV region as well.
Observations at TeV energies, for which the flux is too low for satellite
detection, can be done with ground based telescopes. Milagro is a large
aperture water Cherenkov detector for extensive air showers, collecting
data from a solid angle of more than two steradians in the overhead sky at
energies near 1 TeV. A 2000-2001 data set from Milagro has been used to
search for the emission of diffuse gamma rays from the galactic disk. An
excess has been observed from the region of the Milagro inner Galaxy
defined by $l \in (20^{\circ}, 100^{\circ})$ and $|b| < 5^{\circ}$ with
the significance $2.3 \cdot 10^{-4}$. The emission from the region of the
Milagro outer Galaxy defined by $l \in (140^{\circ}, 220^{\circ})$ and
$|b| < 5^{\circ}$ is not inconsistent with being that of background only.
Under the assumption that EGRET measurements in 10-30 GeV range can be
extended to TeV region with a simple power law energy spectrum, the
integral gamma ray flux with energies above 1 TeV for the region of inner
Galaxy is measured to be $F(>1 TeV) = (9.5 \pm 2.0) \cdot 10^{-10} \;
cm^{-1}\; sr^{-1} \; s^{-1}$ with spectral index $\alpha_{\gamma} = 2.59
\pm 0.07$. The 99.9\% upper limit for the diffuse emission in the region
of outer Galaxy is set at $F(>1 TeV) < 4.5 \cdot 10^{-10} \; cm^{-1}\;
sr^{-1} \; s^{-1}$ using a differential spectral index of $\alpha_{\gamma}
= 2.49$. The upper limit for the outer Galaxy is consistent with the
extrapolation of EGRET measurements between 1 and 30 GeV. Extrapolation of
the EGRET measurements between 1 and 30 GeV for the region of inner Galaxy
using constant power law spectral index is incompatible with the Milagro
data. This indicates softening of the spectrum at energy between 10 GeV
and 1 TeV. These observations may be used to constrain some models of
Galactic gamma ray emission.

\end{preface}
    % abstract                  numbered

         % !!!! all numbered pages should be in the table of contents !!!!
    \tableofcontents          % table of contents         numbered
    \listoffigures            % table of figures if present       numbered
    \listoftables             % table of tables if present        numbered
    \listofappendices        % table of appendices if present   numbered

%%%%%%%%%%%%%%%%%%%%%%%%%%%%%%%%%%%%%%%%%%%%%%%%%%%%%%%%%%%%%%%%%%%%%
%               Start thesis body with this command
%%%%%%%%%%%%%%%%%%%%%%%%%%%%%%%%%%%%%%%%%%%%%%%%%%%%%%%%%%%%%%%%%%%%%

\begin{thesisbody}

% MAIN TEXT GOES HERE....
% INTORUCTION might be chapter 0. No number though...

\chapter{Introduction}

%\\\\\\\\\\\\\\\\\\\\\\\\\\\\\\\\\\\\\\\\\\\
%\section{Introduction}

Most cosmic rays are accelerated by unknown objects in our Gala\-xy and
are trapped (for about 100 million years) by Galactic magnetic fields. The
interaction of high energy cosmic rays with the interstellar material
produces $\gamma$-rays by a combination of electron bremsstrahlung,
inverse Compton and nucleon-nucleon processes. The nucleon-nucleon
interactions give rise to $\pi^{0}$'s which decay to gamma rays and are
expected to dominate the flux at energies above several GeV. In this
manner, the regions of enhanced density (clouds of mostly atomic and
molecular hydrogen) act as passive targets, converting some fraction of
impinging cosmic rays into gamma rays.  This should appear as a diffuse
glow concentrated in the narrow band along the Galactic equator.  Indeed,
such an emission was detected by the space-borne detectors SAS 2
\cite{sas2}, COS B \cite{cos-b} and EGRET \cite{egret} at energies up to
30 GeV. Figure~\ref{fig:introduction:egret} presents the EGRET all sky
survey plotted in Galactic coordinates. The Galactic plane is clearly
visible.

% ////////////////////////////////////////////////////////////
% ======  Pictures.....===========
\begin{center}
\begin{figure}
\includegraphics[width=5in]{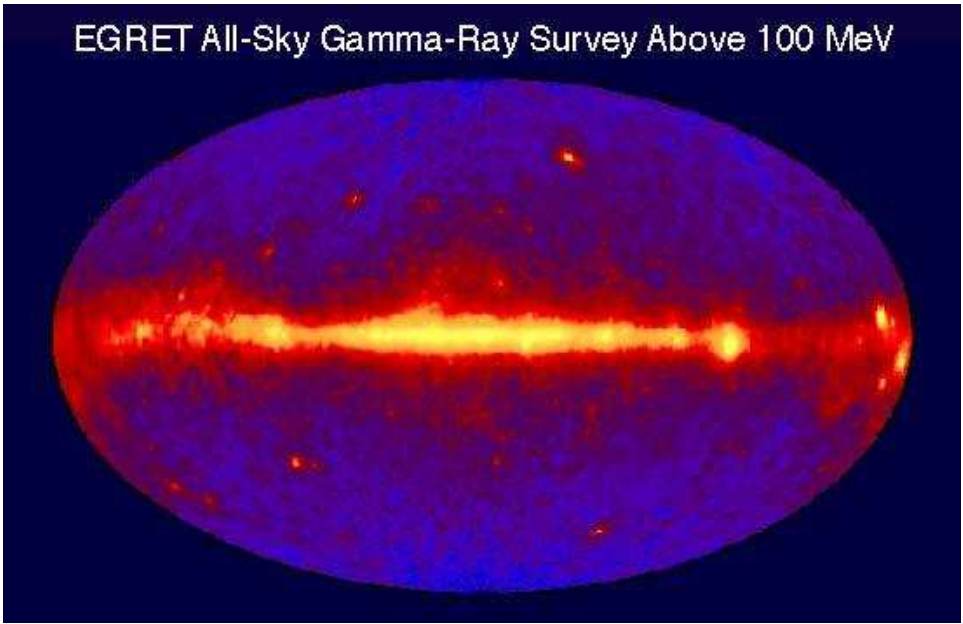}
\caption{All sky gamma ray survey presented in Galactic
coordinates. The Galactic center is in the middle of the
map~\cite{egretweb}.}
\label{fig:introduction:egret}
\end{figure}
\end{center}

However, observations with present satellite based instruments at higher
energies are not possible due to the rapidly decreasing flux of
$\gamma$-rays, requiring bigger effective area of the detectors.
Therefore, the use of ground-based arrays is need\-ed to observe the
diffuse Galactic radiation.  Inasmuch as the Galactic cosmic ray spectrum
extends beyond $10^{15}$ eV, the diffuse Galactic emission should extend
well beyond the energy threshold of Milagro ($\sim 400$ GeV). A number of
authors have estimated the expected diffuse very high energy gamma-ray
flux from the Galactic plane (see for example~\cite{galactic_flux}): they
generally predict a flux within $\pm 5^{\circ}$ of the Galactic equator in
latitude that is $\sim 10^{-4}-10^{-5}$ of the cosmic ray flux for the
regions of the outer gala\-xy.\footnote{The outer Galaxy is defined as the
region with galactic longitude $l$, $40^{\circ}<l<320^{\circ}$.} The shape
of the gamma-ray spectrum is predicted by the same authors to have power
law form $\frac{dN}{dE}\sim E^{-\alpha}$, with spectral index
$\alpha=2.7$. However, at TeV energies the contribution from source cosmic
rays, considered by \cite{galactic_flux2}, may increase the expected
diffuse $\gamma$-ray flux by almost an order of magnitude compared to
$\pi^{0}$-decay model predictions. It is also possible that the spectrum
of cosmic rays in the interstellar medium is substantially harder compared
with the local one measured directly in the solar neighborhood
\cite{galactic_flux3} which will lead to higher diffuse $\gamma$-ray flux
as well.

At present, gamma rays from the galactic plane have not been detected
above EGRET energies (only upper limits were set). The only measurements
that approach the required sensitivity are above 180 TeV, performed by the
CASA-MIA experiment~\cite{casa-mia}. The best measurement in the 1 TeV
region, which is two orders of magnitude less sensitive, is due to
Whipple~\cite{whipple}. The present state of theoretical predictions and
experimental measurements is summarized in
figure~\ref{fig:introduction:ong} \cite{ong}.  Milagro, a detector
designed to cover the energy gap in the few TeV region between other
existing instruments, should be able to detect the diffuse very high
energy Galactic emission and possibly its spatial distribution and provide
an enhanced understanding of Galactic cosmic rays. The sky coverage of
Milagro is illustrated in figure~\ref{fig:coverage}. Because Milagro is
located in the northern hemisphere at a latitude of $36^{\circ}$, the
Galactic center is not in its field of view. However, a considerable
portion of the outer disk is visible to Milagro. For a year's exposure,
Milagro is sensitive to a gamma ray flux of about that theoretically
predicted.

% ////////////////////////////////////////////////////////////
% ======  Pictures.....===========
\begin{center}
\begin{figure}
\includegraphics[width=5in]{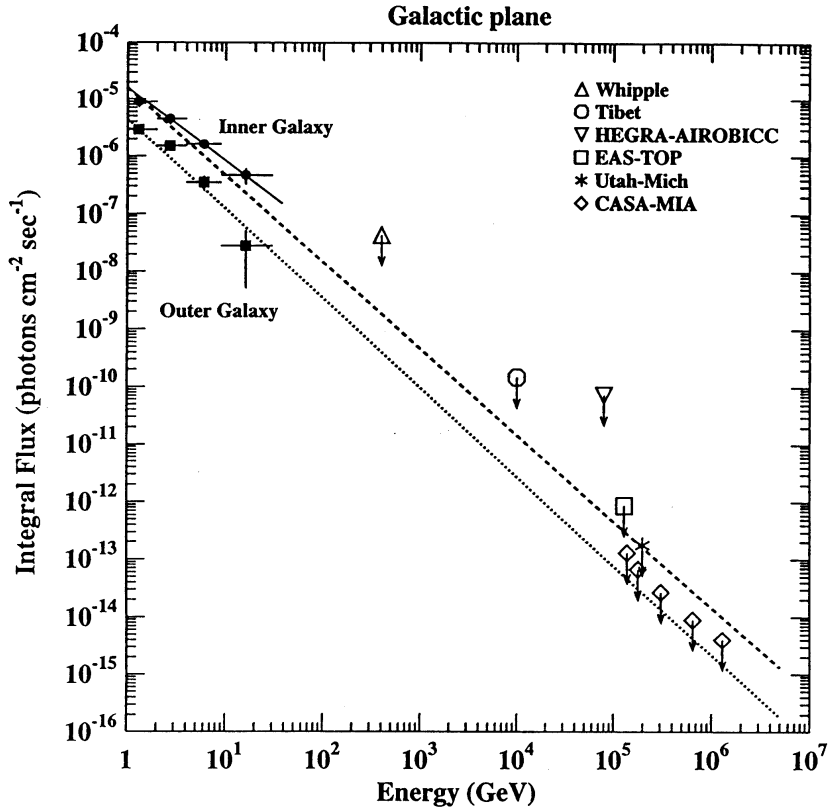}
\caption{Theoretical predictions and observations of the diffuse 
          $\gamma$-ray flux from the Galactic plane as a function of
          energy. EGRET measurements for the inner and outer regions are
          shown by filled points with error bars. The solid line
          represents the fit to the EGRET data for the inner Galaxy,
          indicating that  the observed spectrum is significantly harder 
          than expected (dashed line). The dotted line shows predictions
          for the outer Galaxy region. Identification of other instruments
          and their upper limits are shown as open symbols~\cite{ong}.}
\label{fig:introduction:ong}
\end{figure}
\end{center}

% ////////////////////////////////////////////////////////////
% ======  Pictures.....===========
\begin{center}
\begin{figure}
\includegraphics[width=5in]{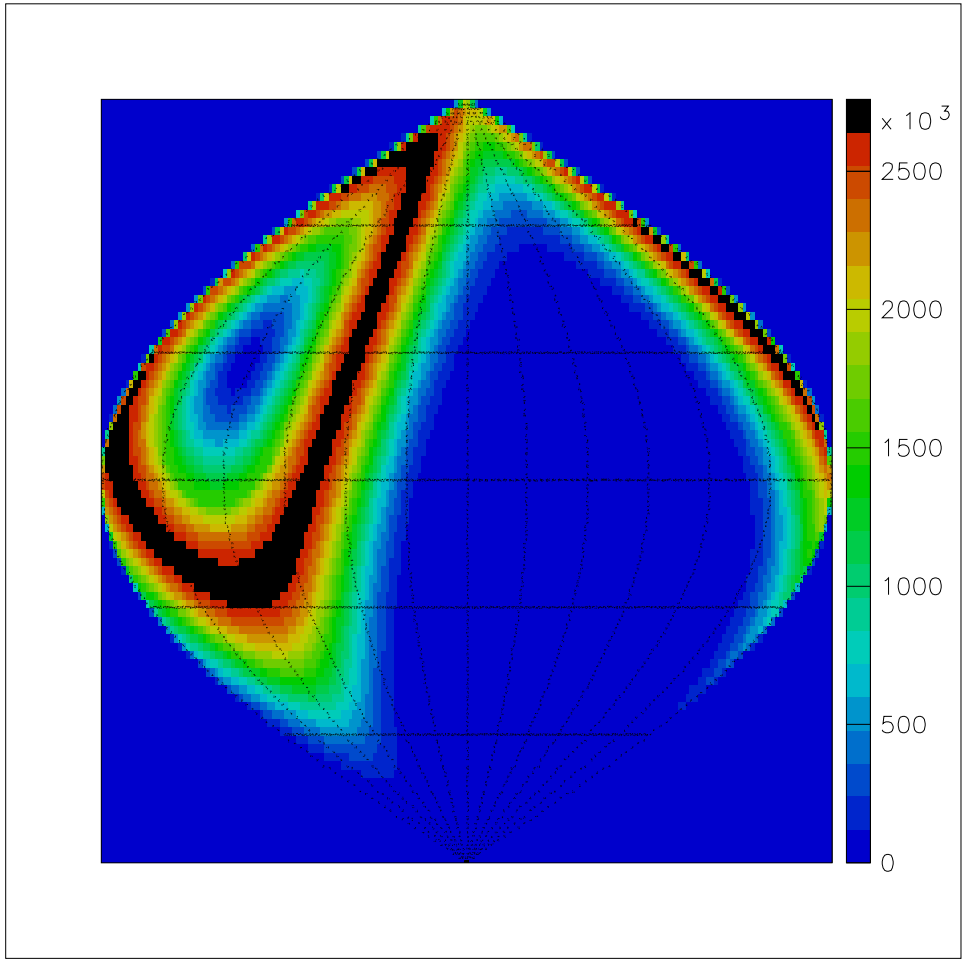}
 \caption{The density of events from Milagro (in arbitrary units) plotted
          in a galactic coordinate system, sine equal area projection. 
          Grid lines are plotted every $30^{\circ}$ in longitude and
          latitude. The Galactic center, not visible by Milagro, is in the
          center of the map. Galactic longitude increases to the left.}
 \label{fig:coverage}

\end{figure}
\end{center}

\chapter{Diffuse Galactic Gamma Ray Emission.}

% ////////////////////////////////////////////////////////////
% ====== Pictures.....===========
\begin{figure}
\centering
  \subfigure[The heavy solid line is the flux without contribution from 
             positron annihilation, heavy dashed line takes it into
             account.]{ 
    \label{fig:diffuse:aharonian_1}
    \includegraphics[width=2.7in]{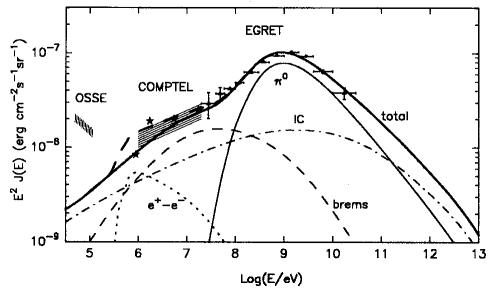}
  }
  \subfigure[The heavy solid line is the flux from the main component
             (figure \ref{fig:diffuse:aharonian_1}), heavy dashed line
             takes also into account a possible population of electrons
             accelerated to energies 250 TeV. Measurements from X-rays to
             very high energy gamma rays and upper limits are also shown
             on the plot.]{
    \label{fig:diffuse:aharonian_2}
    \includegraphics[width=2.5in]{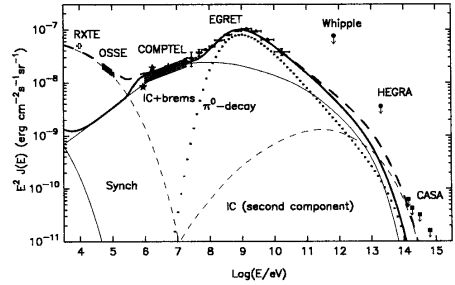}
  }

\caption{The fluxes of diffuse radiation produced by both electronic and 
         nucleonic components of cosmic rays in the inner Galaxy
         ($315^{\circ} < l < 45^{\circ}$) calculated for a hard power law
         spectrum of electrons (with index 2.15) and protons (with index
         2.1 and gradual turnover to 2.75 above few GeV). Contributions
         from different emission mechanisms are shown and 
         labeled. (Adopted from \cite{galactic_flux3})}
\label{fig:diffuse:aharonian}
\end{figure}

% ////////////////////////////////////////////////////////////
% ======  Pictures.....===========
\begin{figure}
\centering
\includegraphics[width=4.5in]{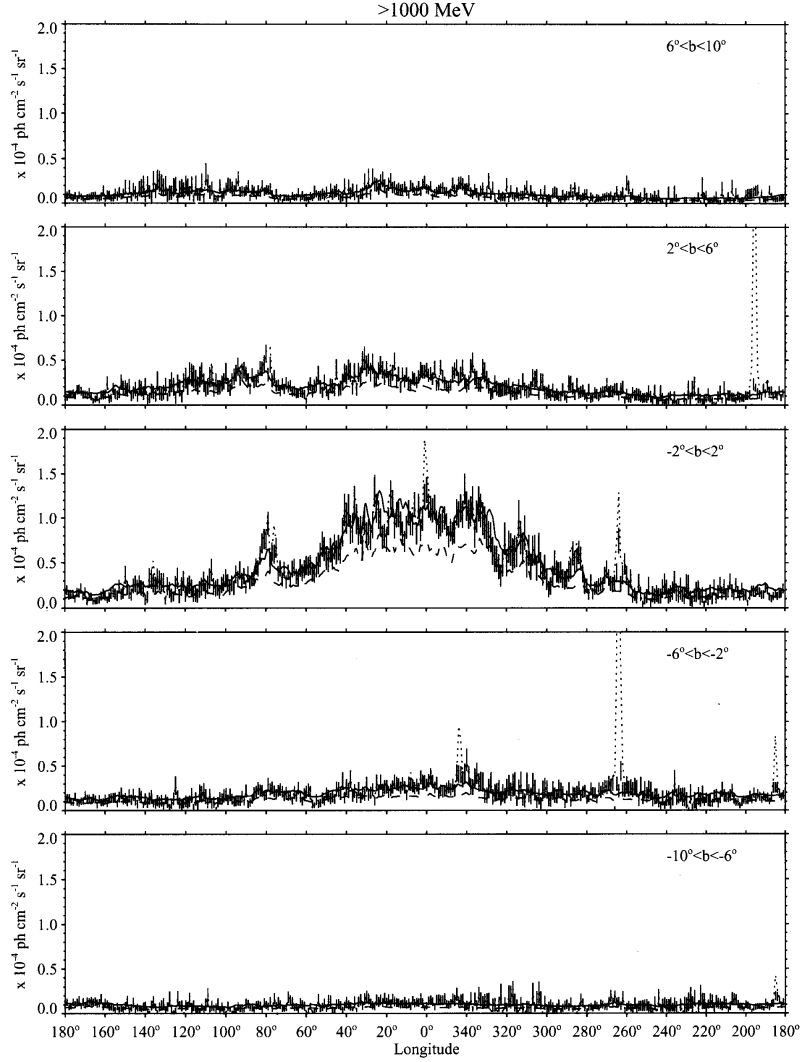}
\caption{Longitude dependence of diffuse gamma ray emission averaged
         over $4^{\circ}$ wide latitude intervals in the energy range
         between 1 and 30 GeV measured by EGRET. The emission including
         point sources is shown as dotted line. The solid line is the best
         fit model calculated for the same energy range, convolved with
         the EGRET resolution function and averaged over the same latitude
         interval. The isotropic diffuse component added to the model for
         this energy range is $0.12 \cdot 10^{-5} 
         \; photons\; cm^{-2}\; s^{-1}\; sr^{-1}$. The model without the
         60 \% increase is shown as a dashed line. (Adopted from
         \cite{egret}.)} 
\label{fig:diffuse:egret_longitude} 
\end{figure}

% ////////////////////////////////////////////////////////////
% ======  Pictures.....===========
\begin{figure}
\centering
\includegraphics[width=5.0in]{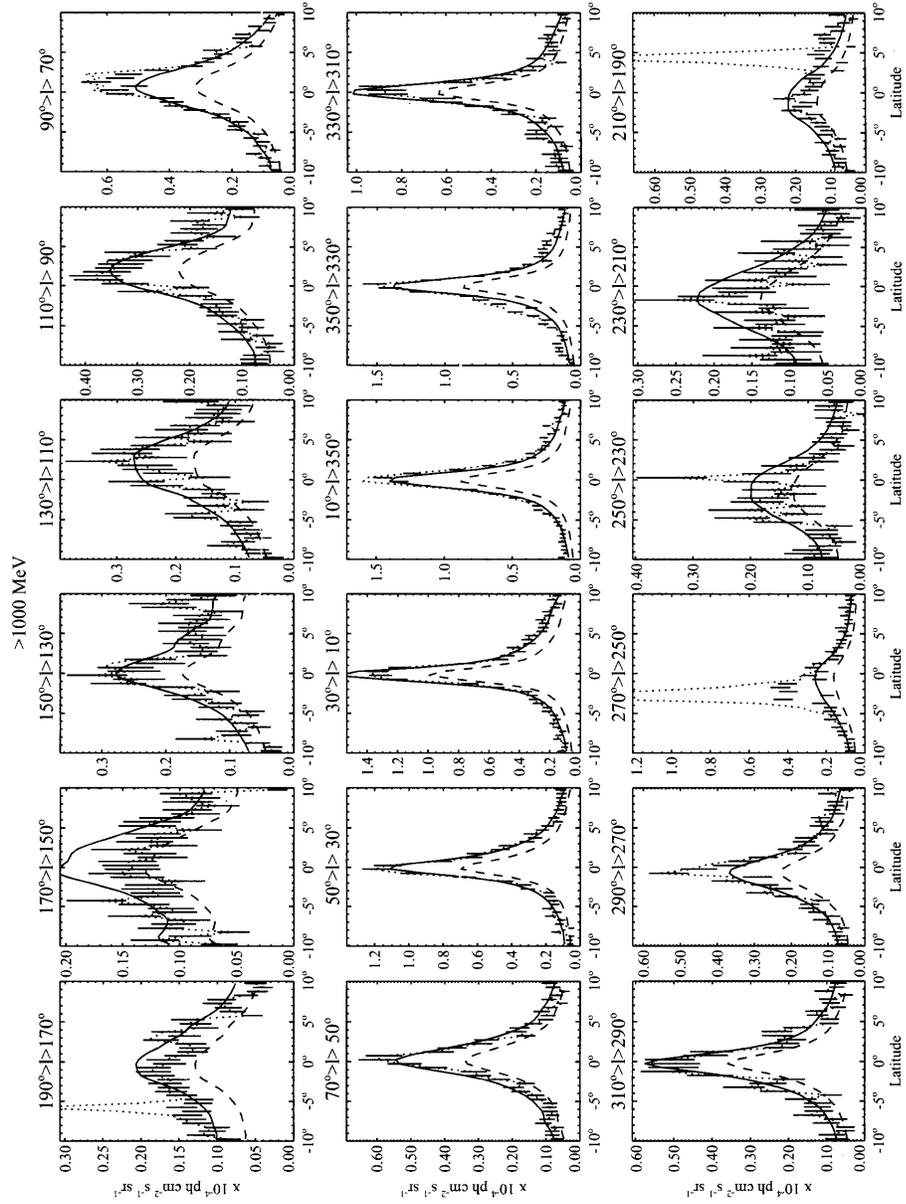}
\caption{Latitude dependence of diffuse gamma ray emission averaged over
         $20^{\circ}$ wide longitude intervals in the energy range between
         1 and 30 GeV measured by EGRET. See caption of figure
         \ref{fig:diffuse:egret_longitude} for details.}
\label{fig:diffuse:egret_latitude}
\end{figure}

% ////////////////////////////////////////////////////////////
% ======  Pictures.....===========
\begin{figure}
\centering
\includegraphics[width=3.5in]{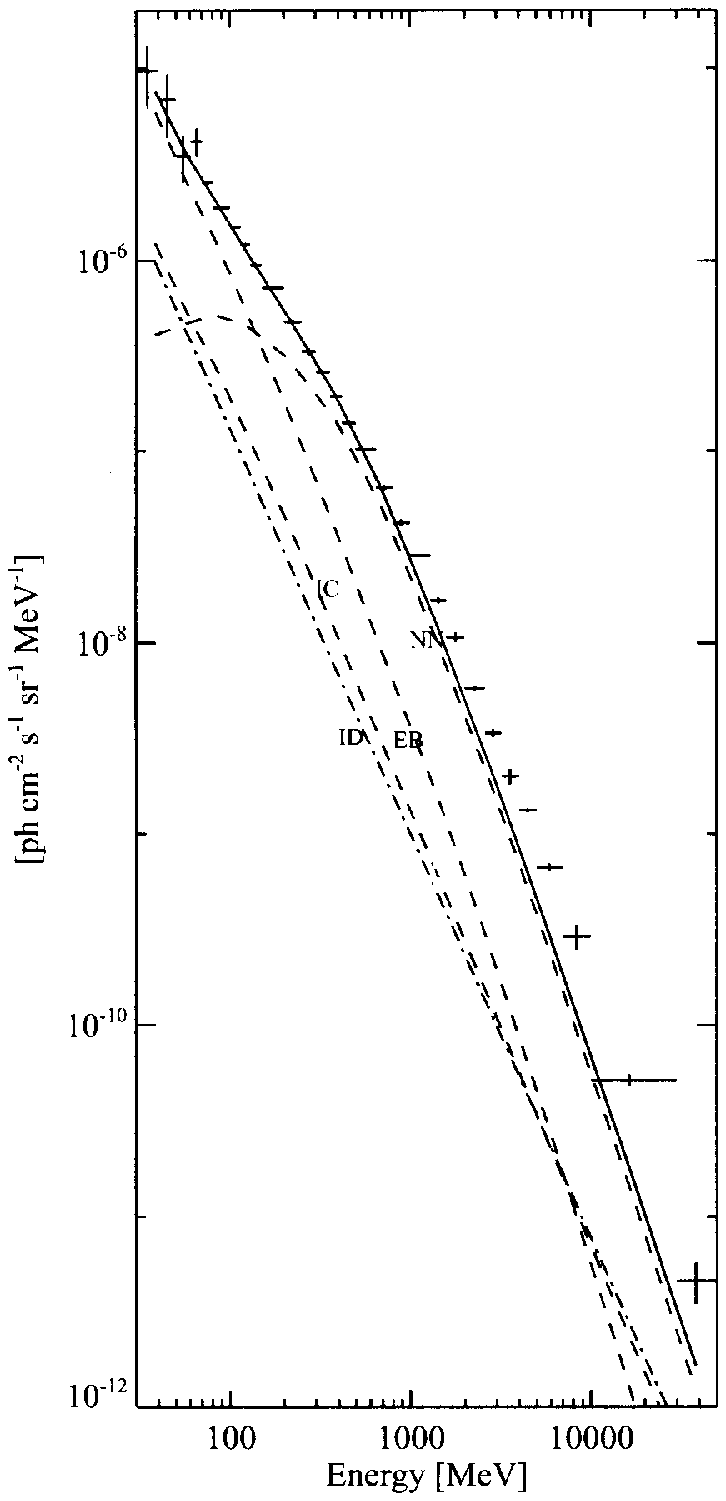}
\caption{Average diffuse gamma ray spectrum of inner Galaxy region,
         $300^{\circ} < l < 60^{\circ}$, $|b|<10^{\circ}$ measured by 
         EGRET. The contribution from point sources had been removed. The
         model calculations plus the isotropic diffuse emission
         (dash-dotted line ID) is shown as solid line. The individual
         components of the calculation are also shown as dashed
         lines: nucleon-nucleaon (NN), electron bremsstrahlung (EB),
         inverse Compton scattering of electrons(IC) contributions. At
         energies above 1 GeV the data points are significantly higher
         than the model calculations.(Adopted from \cite{egret}.)}
\label{fig:diffuse:egret_inner_spectrum}
\end{figure}

High energy gamma rays produced by interactions of cosmic rays with the
interstellar material should provide tracers of cosmic rays because their
trajectories are undeflected by interstellar magnetic fields and because
they traverse the Galaxy without significant attenuation. (Neutrinos would
provide even better tracers because of their weak interactions.)
Therefore, diffuse gamma ray emission of the Galactic disk carries unique
information about the production sites, the fluxes and the spatial
distribution of Galactic cosmic rays. Indeed, the separation of different
emission mechanisms in a broad energy region from a few KeV to few hundred
TeV in different parts of the Galaxy would provide an important insight
into the problem of the origin of cosmic rays and of their propagation in
the interstellar medium. This is illustrated on figure
\ref{fig:diffuse:aharonian} \cite{galactic_flux3}. Diffuse gamma radiation
in the center of the Galaxy or in its halo has been proposed as a probe of
annihilating dark matter particles as well \cite{neutralino_urban}.

The observations of the diffuse gamma ray radiation conducted in 1990's by
the EGRET detector aboard Compton Gamma Ray Observatory \cite{egret}
resulted in good quality data over three decades in energy of gamma rays.
The results support a Galactic origin of cosmic rays and strong
correlation between the high energy gamma ray flux and the column density
of the galactic hydrogen. The latter demonstrated the existence of a truly
diffuse radiation based on the earlier data from SAS-2 and COS B
\cite{bloemen}.

The extraction of the diffuse component of the gamma ray emission from the
EGRET data, that is the radiation produced by cosmic ray electrons,
protons and nuclei interacting with the ambient gas and photon fields, is
obscured by the emission from resolved and an uncertain number of
unresolved point sources as well as by the diffuse isotropic emission of
presumably extragalactic origin. An example of longitudinal intensity
distribution of the observed emission, including the isotropic emission
and excluding the point source contribution is shown on figure
\ref{fig:diffuse:egret_longitude}. Figure \ref{fig:diffuse:egret_latitude}
shows the latitude distribution of the intensity for the same energy
range. The dotted lines represent the observed emission including the
point sources, the solid lines are the calculated intensities using the
model described in \cite{egret,model_bertsch}. In this model, the EGRET
data together with radio data was used to develop a three-dimensional
picture of both gas and cosmic ray densities in the Galaxy. The diffuse
gamma ray emission for energy $E_{\gamma}$ from the galactic longitude $l$
and latitude $b$, $|b|<10^{\circ}$ is given by \cite{egret,model_bertsch}:

\[
  j(E_{\gamma}, l,b) = \frac{1}{4\pi} \int 
                       \left[ 
                          c_{e}(\rho, l, b) q_{em}(E_{\gamma}) +
                          c_{n}(\rho, l, b) q_{nm}(E_{\gamma})
                       \right] \times 
\]
\[
                       \times \left[
                          n_{HI}(\rho, l, b) + n_{HII}(\rho, l, b) +
                          n_{H_{2}}(\rho, l, b)
                       \right] d \rho +
\]

\[ 
                     + \frac{1}{4\pi} \sum_{i} \int
                        c_{e}(\rho, l, b) q_{p}^{(i)}(E_{\gamma}, \rho)
                        u_{p}^{(i)}(\rho, l, b) d \rho
\]
\[
                      (photons \; cm^{-2}\; sr^{-1}\; GeV^{-1})
\]

where $\rho$ is the distance of the line of sight in the direction of $l$
and $b$ measured from the Sun. The first integral represents the gamma ray
production due to cosmic ray interactions with matter where
$q_{em}(E_{\gamma})$ and $q_{nm}(E_{\gamma})$ are the electron
bremsstrahlung and nucleon-nucleon production functions per target
hydrogen atom based on the cosmic ray spectrum measured in the vicinity of
the Sun. The functions $c_{e}(\rho, l, b)$ and $c_{n}(\rho, l, b)$ are the
ratios of the cosmic ray electron and proton densities at the given
location to their densities in the vicinity of the Sun. The electron and
proton spectra are assumed to have the same shape as measured locally near
the Sun and therefore the $c$'s are independent of energy. The ratios are
also assumed to be be equal and independent of $b$: $c_{e}(\rho, l, b) =
c_{n}(\rho, l, b) = c(\rho, l)$, $c(\rho =0, l) =1$. The quantities
$n_{HI}(\rho, l, b)$ and $n_{H_{2}}(\rho, l, b)$ are the atomic and
molecular hydrogen densities expressed as atoms per unit volume derived
from the 21 cm hyperfine transition emission line (HI) and 2.6 mm
kinematic $J=1\rightarrow 0$ transition line of CO surveys. The CO
intensities are scaled by $2X$ where $X=N(H_{2})/W_{CO}$ is the
proportionality constant between the column density of molecular hydrogen
and the integrated intensity of the CO line. The distribution of ionized
hydrogen $n_{HII}$ is taken from a model \cite{egret} and is shown to have
small contribution to the diffuse gamma ray emission compared to that of
HI and $H_{2}$. The second integral describes the contribution from
inverse Compton interactions between cosmic ray electrons and interstellar
photons where $q_{p}^{(i)}(E_{\gamma}, \rho)$ is the inverse Compton
production function based on the local electron spectrum, and the
summation is over discrete wavelength bands ($i$) of cosmic blackbody
radiation, infrared, optical and ultraviolet that arise from within our
Galaxy with corresponding photon energy density distributions
$u_{p}^{(i)}(\rho, l, b)$.

This model accurately matches the observed emission as seen by EGRET for
all Galactic longitudes over the energy range from 30 MeV to about 30 GeV.
The model underestimates the observations at energies above 1 GeV in the
region of inner Galaxy $300^{\circ} < l < 60^{\circ}$, $|b| < 10^{\circ}$.
(The model calculations without the 60~\% increase in this region is shown
as a dashed line on figures \ref{fig:diffuse:egret_longitude},
\ref{fig:diffuse:egret_latitude}. This is also illustrated on figure
\ref{fig:diffuse:egret_inner_spectrum}.) Some authors \cite{pohl_esposito}
have suggested that because electron propagation is limited by radiative
losses the local measurements of the electron spectrum may not be
representative for the entire Galaxy. In this case, a harder cosmic ray
electron spectrum could be used to explain the observed excess. It is also
possible \cite{galactic_flux3} that the proton spectrum in the inner
Galaxy is harder then observed in the Solar neighborhood, at least in the
region below few GeV. Another possibility is contribution from unresolved
sources. Assuming these are supernova remnants, it was shown
\cite{galactic_flux2} that their spatially averaged contribution to the
diffuse gamma ray flux at 1 TeV should exceed the above model predictions
\cite{egret} extended to 1 TeV by almost an order of magnitude. It is
therefore of interest to search for the diffuse emission from the Galactic
plane at energies around 1 TeV.

\chapter{Extensive Air Showers.\label{chapter:eas}}

% ////////////////////////////////////////////////////////////
\section{Longitudinal Development of Extensive Air \\ Showers.}

The desire to detect low fluxes of gamma rays at energy around 1 TeV using
ground based telescopes impels consideration of propagation of the photons
to the ground level, where they can be registered.

A high energy primary gamma-ray entering the atmosphere interacts with it,
initiating the production of secondary particles which in turn create
tertiary particles and so on. Such electromagnetic cascades are propagated
predominantly by photons and electrons. The basic high energy processes
that make up the cascade are pair production and bremsstrahlung occurring
in the field of nuclei of air which produce successive generations of
electrons, positrons and photons. Charged particles are removed from the
shower development by ionization losses, photons --- by Compton
scattering. The number of particles increases until their energies
decrease to the critical energy $E_{c} \approx 80 MeV$, when ionization
and scattering become the main energy loss mechanisms. This stage of the
development is called {\em shower maximum}.

Inasmuch as for ultrarelativistic particles the radiation length $X_{0}$
for the bremsstrahlung process (the amount of the material traversed over
which particle's energy is decreased by a factor of $e$,
$E=E_{0}e^{-x/X_{0}}$) is approximately equal to the gamma ray interaction
length for electron-positron pair production \cite{Longair}, it provides a
convenient scale, and in the case of air it is about $X_{0} \approx 37
g/cm^{2}$.

The cascade development provides an interesting example of a stochastic
process which is prohibitively difficult for analytic calculations.
Although several approximations have been considered \cite{Rossi_Greisen},
numerical methods are usually needed to obtain results of practical use.

Recent simulations by DiSciascio et al \cite{Sciascio} show that the
average number of photons and electrons with the energy greater than
$E_{th}$ in the shower initiated by a photon with energy $E_{0}$ is well
described by a modified Greisen formula:

\begin{equation}
 N(E_{0},E_{th},t) = A(E_{th})\frac{0.31}{\sqrt{y}}
                           e^{t_{1}(1-1.5 \ln s_{1})}
\label{equation:eas_longitudal}
\end{equation}

where $t_{1}$ is the modified depth $t=x/X_{0}$ from the top of the
atmosphere measured along the trajectory of the primary particle and
expressed in the units of radiation length:

 \[ t_{1} = t + a(E_{th}) \]

The parameter $s_{1}$ represents the age of the shower and increases as it
develops starting at $s_{1}=0$, $s_{1}=1$ at the maximum, $s_{1}>1$ in the
declining stage of the cascade:

\[ s_{1} = \frac{3t_{1}}{t_{1}+2y}\]
\[ y = \ln \frac{E_{0}}{E_{c}}\]

The parameterization is valid in the depth range $4<t<24$ for primary
photon energies $0.1<E_{0}<10^{3}$ TeV. The coefficients $A(E_{th})$ and
$a(E_{th})$ are given in the table \ref{table:eas:longitude}. The
dependence is illustrated on figure~\ref{fig:eas:longitude}. The Milagro
detector is situated at the vertical depth of about 20 radiation lengths,
thus within the range of the simulations for photons with zenith angles
between 0 and 32 degrees.

\begin{table}[htbp]
\begin{center} 
\begin{tabular}{|c|c|c|c|c|} \hline
  $E_{th}$, & \multicolumn{2}{c|}{electrons} & \multicolumn{2}{c|}{photons} \\ \cline{2-5}
  MeV       &     A   &    a    &  A   &   a    \\  \hline
 1          &    0.92 &   0.00  & 4.80 & -0.88  \\
 5          &    0.75 &   0.19  & 2.98 & -0.69  \\
 10         &    0.63 &   0.35  & 2.13 & -0.57  \\
 20         &    0.50 &   0.53  & 1.45 & -0.36  \\ \hline
\end{tabular}
\end{center}
\caption{ Coefficients $A$ and $a$ for modified depth calculation
          \cite{Sciascio} in longitudinal development case, equation
          \ref{equation:eas_longitudal}.}
\label{table:eas:longitude}
\end{table}

% ////////////////////////////////////////////////////////////
\begin{figure}
\centering
  \subfigure[Longitudinal development of electron component.]{
    \label{fig:eas:longitude_electrons}
    \includegraphics[width=2.7in]{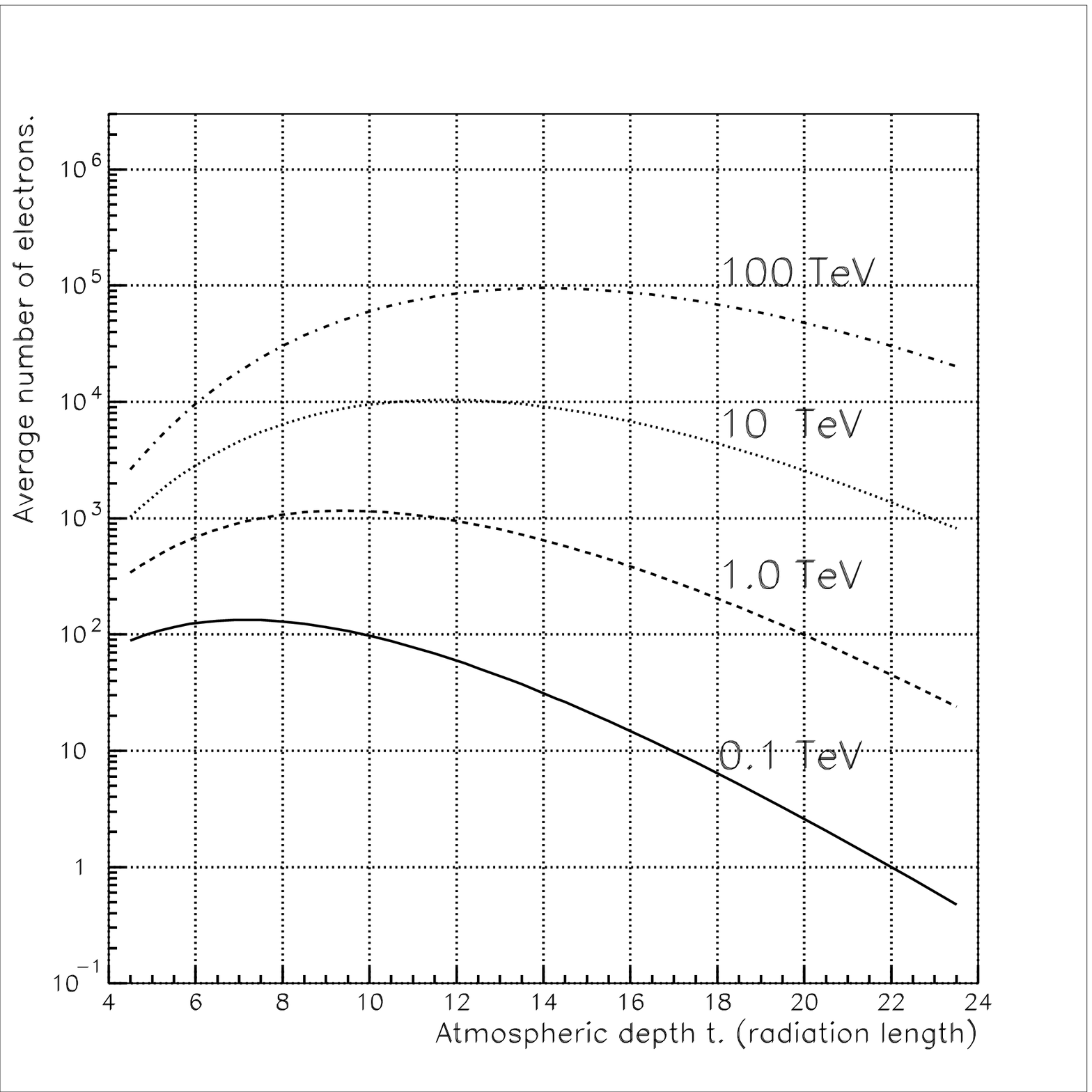}
  }
  \subfigure[Longitudinal development of photon component.]{
    \label{fig:eas:longitude_phtons}
    \includegraphics[width=2.7in]{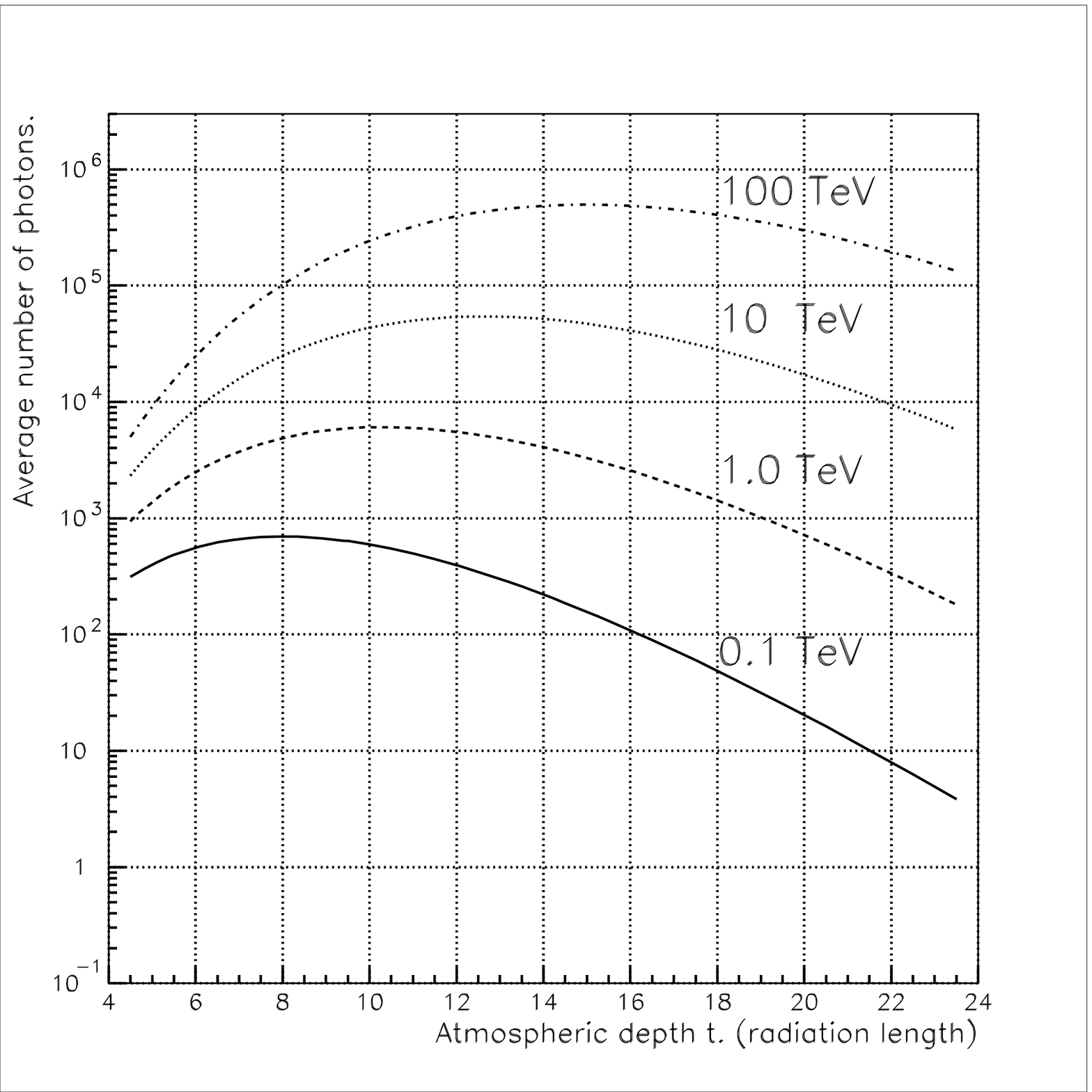}
  }
\caption{Longitudinal development of gamma ray induced showers with
         $E_{th} = 1MeV$ particle detection threshold.}
\label{fig:eas:longitude}
\end{figure}

% ////////////////////////////////////////////////////////////
\section{Lateral Development of Extensive Air Showers.}

As an extensive air shower cascades through the atmosphere
ultrarelativistic particles and photons are produced mainly in the forward
direction. However, because electrons and positrons suffer multiple
Coulomb scattering off the electric fields of nuclei and photons undergo
Compton scattering off atomic electrons, the particles are spread out and
the shower attains a lateral distribution. The density of particles is
greatest near the {\em shower core}, the trajectory that the primary
gamma-ray would have had if it did not interact with the atmosphere. The
average number of photons $\rho(r,t,E_{0})$ per unit area at a distance
$r$ from the shower core agrees well with the Nishimura-Kamata-Greisen
(NKG) formula, with modified depth parameter \cite{Sciascio}:

\begin{equation}
 \rho (r,t,E_{0}) = \frac{N(E_{0},t)}{r_{m}^{2}} f(r/r_{m}, s_{2})
 \label{equation:eas_lateral}
\end{equation}

where
\[ f(r/r_{m}, s_{2}) = \frac{1}{2\pi} \cdot \frac{1}{B(s_{2}, 4.5 - s_{2})}
           \left(\frac{r}{r_{m}} \right)^{s_{2} - 2}
           \left(1 + \frac{r}{r_{m}} \right)^{s_{2} - 4.5} \]

\[ t_{2} = t + b(E_{th}) \]
\[ s_{2} = \frac{3t_{2}}{t_{2}+2y}\]

with $B(x, y)$ being beta function so that $2\pi\int_{0}^{\infty}
f(r/r_{m}, s_{2}) (r/r_{m})d(r/r_{m}) = 1$, $r_{m}$ --- being Moliere
scattering unit, $r_{m}=\frac{E_{s}}{E_{c}}X_{0}= 9.7 g \cdot cm^{-2}$ or
about 110 meters at the elevation of Milagro ($E_{s} =
m_{e}c^{2}\sqrt{4\pi / \alpha} \approx 21MeV$) and $N(E_{0},t)$ is total
number of photons at the depth $t$ \cite{gaisser} given by equation
\ref{equation:eas_longitudal}.

It was also found in \cite{Sciascio} that the lateral density distribution
of electrons is well represented by the same expression if the scattering
unit is substituted by $r_{m}^{\prime} = r_{m}/2$. (This fact that shower
photons are spread farther from the shower axis than the electrons is a
consequence of the fact that photons do not lose energy by ionization and 
can travel larger distances than electrons.) The values of parameters
$b(E_{th})$ are given in the table \ref{table:eas:lateral}.

\begin{table}[htbp]
\begin{center}
\begin{tabular}{|c|c|c|} \hline
  $E_{th}$, & \multicolumn{1}{c|}{electrons} & \multicolumn{1}{c|}{photons} \\ \cline{2-3}
  MeV       &     b                          &    b      \\  \hline
  1         &     0.45                       &    0.83   \\
  5         &    -1.22                       &   -1.49   \\
  10        &    -2.57                       &   -3.45   \\
  20        &    -4.22                       &   -5.51   \\ \hline
\end{tabular}
\end{center}
\caption{ Coefficients $b$ for modified depth calculation \cite{Sciascio} 
          in lateral distribution case, equation
          \ref{equation:eas_lateral}.}
\label{table:eas:lateral}
\end{table}

The average density of photons and electrons per unit area as a function
of distance from the shower core is illustrated on figure
\ref{fig:eas:lateral}

% ////////////////////////////////////////////////////////////
\begin{figure}
\centering
  \subfigure[Lateral distribution of electron component.]{
    \label{fig:eas:lateral_electrons}
    \includegraphics[width=2.7in]{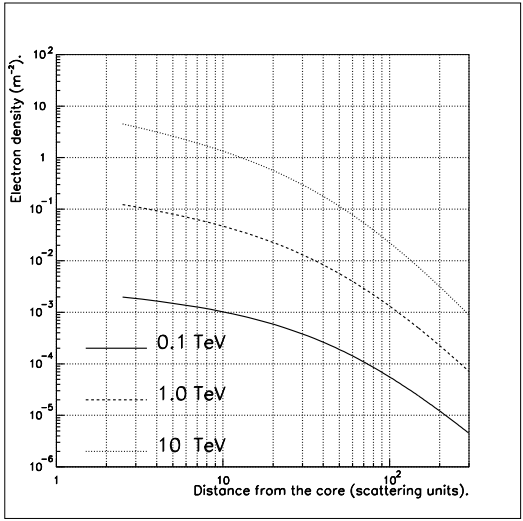}
  }
  \subfigure[Lateral distribution of photon component.]{
    \label{fig:eas:lateral_phtons}
    \includegraphics[width=2.7in]{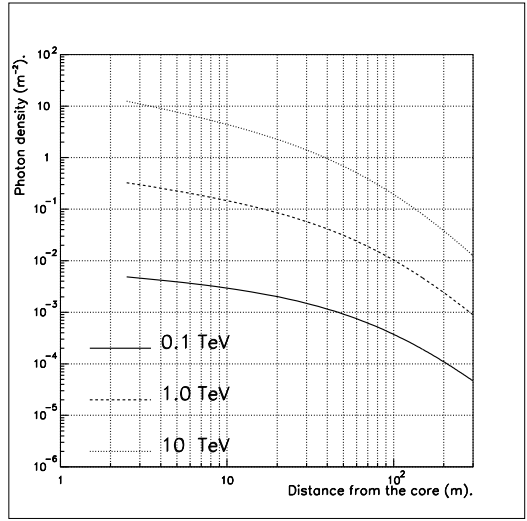}
  }
\caption{Density of particles as a function of core distance for gamma ray
         induced showers with $E_{th} = 1MeV$ particle detection
         threshold. Curves are normalized to the total number of particles
         in the respective components. Plots are made for showers at the
         depth of 20 radiation lengths.}
\label{fig:eas:lateral}
\end{figure}

% ////////////////////////////////////////////////////////////
\section{Temporal Distribution of Extensive Air Shower Particles.
\label{chapter:eas:temporal}}

Once an air shower develops lateral structure, one can speak about the
{\em shower front} --- the forward edge of the advancing cascade. The
arrival time $T$ of the earliest particle hitting a plane perpendicular to
the shower axis at a distance $r$ from the core provides the information
concerning the shape of the front. Since the electron component of the
lateral distribution is attained due to multiple scattering, it is lower
energy electrons and positrons which propagate farther away from the axis.
These travel at lower speeds and one expects them to be delayed with
respect to the energetic one's at the core. The photons are expected to be
more prompt than electrons, but are also delayed with respect to core due
to greater distance traveled. The average arrival time of the first
particle as a function of the core distance for both electron and photon
components is illustrated on figure \ref{fig:eas:timing}.

% ////////////////////////////////////////////////////////////
% ======  Pictures.....===========
\begin{figure}
\centering
\includegraphics[width=5in]{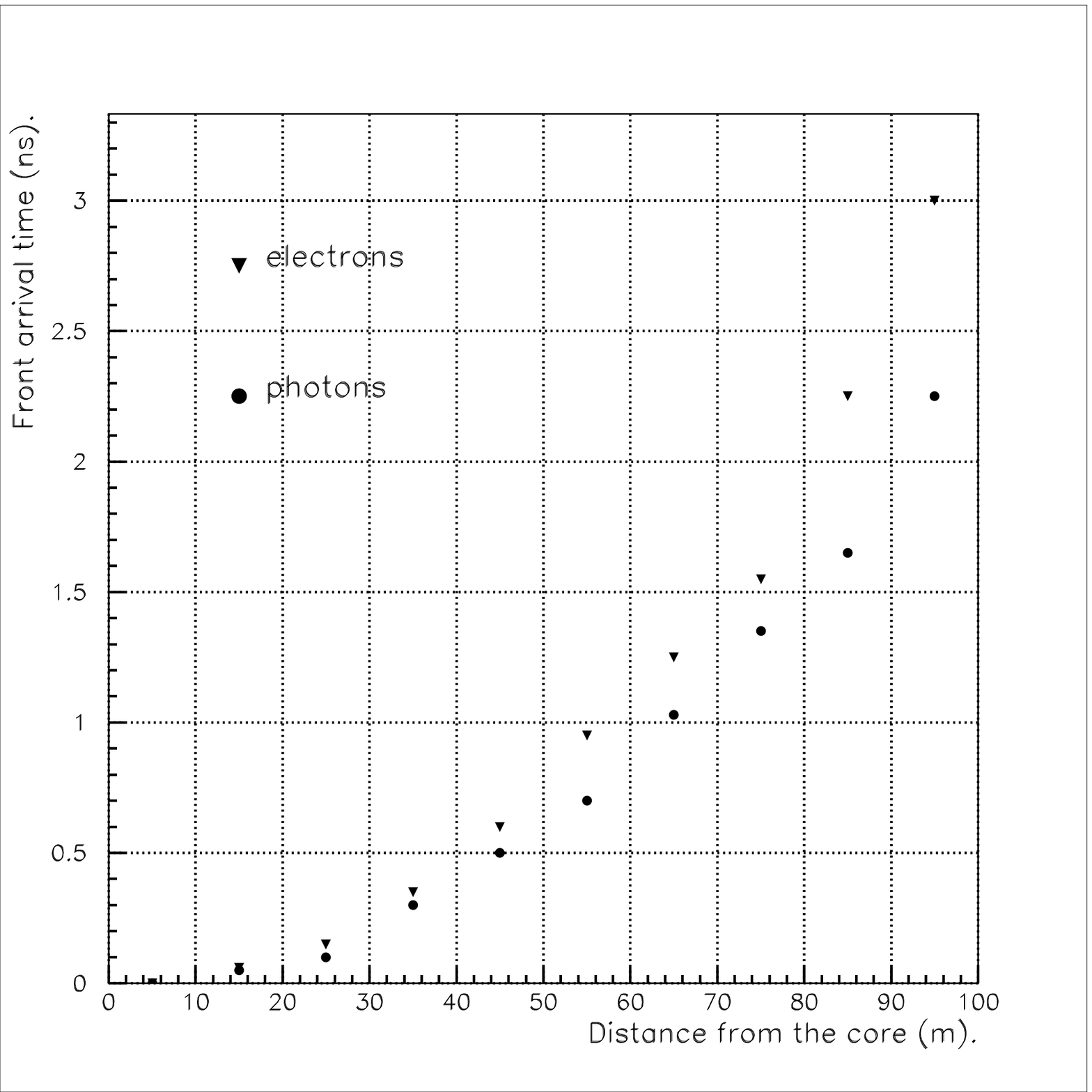}
\caption{Average arrival time of the first particle as a function of the 
core distance.}
\label{fig:eas:timing}
\end{figure}

It appears that the front of each of the components assumes a parabolic
shape. The fluctuations of $T$ around its average appears smaller for the
photon component than for electron one \cite{Sciascio} and, in general,
are quite small. The thickness of the shower is defined by the arrival
time distribution of particles with respect to the front at the given
distance $r$, and is, therefore, due to lower energy electrons and photons
originating later in the shower cascade.

% ////////////////////////////////////////////////////////////
\section{Cherenkov Radiation.}

One of the methods of detection of high energy electrons and positrons and
therefore of the air showers is with the help of Cherenkov radiation.
Cherenkov radiation arises when a charged particle traverses a dielectric
medium with a velocity $v$ which is greater than the speed of light in the
medium $c/n$. Here $n$ is the index of refraction of the medium which, in
general, depends on the wavelength of the emitted radiation and $c$ is the
speed of light in vacuum. The radiation is emitted by atoms and molecules
polarized by the moving particle. According to Huygens' principle, the
partial waves will interfere to create the total wavefront propagating at
an angle $\theta = \arccos \frac{c}{nv}$ with respect to the velocity of
the particle. At lower speeds, some energy of the particle is still lost
due to polarization of the medium, however no Cherenkov radiation results.

The condition for existence of radiation $\frac{c}{nv} = \frac{1}{\beta n}
< 1$ can be expressed in terms of the particle's energy:

\begin{equation}
  E > \frac{m c^{2}}{\sqrt{1-\frac{1}{n^{2}}}} 
\label{equation:cherenkov_threshold}
\end{equation}

where $m$ is the rest mass of the particle. For example, the index of
refraction of water is $n_{water}= 1.35$, and therefore the threshold
energy for electron to produce Cherenkov radiation (and therefore
detection threshold) is equal to 0.759 MeV. For a muon it is 0.157 GeV and
for proton 1.40 GeV.

The number of Cherenkov photons produced per unit path length by such a
particle with charge $Ze$ and per unit wavelength \cite{particle_book}:

\[ \frac{d^{2} N}{dx d \lambda} = \frac{2 \pi \alpha Z^{2}}{\lambda^{2}}
                 \left(1- \frac{1}{\beta^{2} n^{2}(\lambda)} \right) \]

or, in terms of emitted energy:

\[ \frac{d^{2} E}{dx d \lambda} = 
                   \frac{\pi e^{2}Z^{2}}{\epsilon_{0}\lambda^{3}}
              \left( 1 - \frac{1}{\beta^{2} n^{2}(\lambda)} \right)  \]

where $\alpha = \frac{e^{2}}{4 \pi \epsilon_{0} \hbar c} $ is fine
structure constant. Here, we have made explicit the dependence of the
index of refraction on wavelength. Cherenkov radiation is emitted
preferably in the short wave region (blue/violet end of the visible
spectrum).

Note that for ultrarelativistic particles $\beta \approx 1$, the Cherenkov
angle in water is approximately $42^{\circ}$ and the emitted energy
becomes independent of the particle's energy. This fact can be used to
provide absolute energy calibration of photodetectors.

% ////////////////////////////////////////////////////////////
\section{Detection of Extensive Air Showers.}

Air showers produced by high energy photons consist mainly of electrons,
positrons and lower energy photons, and thus hint at methods of detection.
If the energy of the primary photon is greater than about 10 TeV, then
there are enough particles in the cascade reaching the surface of the
Earth to enable the detection by arrays of scintillation counters where
energy of charged particles is converted into flashes of visible light.
These flashes are detected by photoelements. The detectors of this type
are called {\em extensive air shower arrays} (EAS arrays). CASA-MIA and
Tibet are examples of such arrays. The determination of the arrival
direction and of the primary energy of air showers, sampled by a detector
array, makes use of the lateral distribution function over a wide range of
distances: the arrival time of the shower particles is used to determine
the direction, the measurements of the particle density as a function of
core distance together with the direction provide information about the
energy of the primary gamma ray.

Cascades, initiated by a primary photon with energy below several hundred
GeV do not reach the ground, but can be detected by Cherenkov radiation
that charged particles emit in the air as the cascade develops. Such
detectors are termed {\em air Cherenkov telescopes}, such as Whipple and
HEGRA. Detectors of this type, being optical devices, make use of the fact
that most of the Cherenkov light is emitted in the forward direction of
the primary particle and rely on their angular resolution --- truly a
telescope type of a measurement. Arrival direction and energy of the
primary particle is inferred on the basis of the imaged longitudinal
development of the air shower.

Other types of detectors which were used to detect secondary particles are
bubble chambers, spark and ionization chambers. Such methods are rarely
used nowadays. Instead, atmospheric Cherenkov telescopes and extensive air
shower particle detector arrays constitute the two major ground-based
techniques.

% ////////////////////////////////////////////////////////////
\section{Cosmic Rays. Difference Between Cosmic Ray and Gamma Ray
Induced Showers.}

The discussion has been concerned so far with the air showers produced
by primary gamma rays. However, among the particles entering the Earth's
atmosphere gamma rays present a fraction of less than 0.1\%. About 79\% of
the particles are high energy protons and 14\% are alpha particles. The
rest are nuclei of heavier atoms: carbon, nitrogen, oxygen, iron,...

Just as in the case of gamma rays, high energy hadrons also initiate air
showers when they enter the atmosphere. However, the processes in a
hadronic cascade are quite different from the electro-magnetic one.

In such a cascade an incident hadron undergoes strong interactions with
the air nuclei in which protons, neutrons, mesons and hyperons are created
in quantities determined by their relative cross-sections. The most
numerous particles are that of the pion triplet.

The charged pions have a lifetime of about $2.6\cdot10^{-8}$ seconds and,
if they did not interact first, decay into muons and neutrinos:

\[ \pi^{+} \rightarrow \mu^{+} + \nu_{\mu}\]
\[ \pi^{-} \rightarrow \mu^{-} + \bar{\nu}_{\mu}\]

which gives rise to a muonic component of the cascade. If the energy of
charged pions is high, then, due to relativistic time dilatation, they
will have the opportunity to interact with a nucleus rather than decay,
thus producing secondary particles which replenish the hadronic component
of the cascade. Neutral pions, on the other hand, have a much smaller
lifetime of about $10^{-16}$ seconds and decay, dominantly, into photons
($\pi^{0} \rightarrow \gamma + \gamma$). If energy of photons is above
critical, they will initiate electromagnetic cascades. It is due to
presence of the electromagnetic component the hadron induced air showers
are very similar to those initiated by gamma rays. Muons, produced in the
cascade, have a lifetime of $2.2\cdot10^{-6}$ seconds and high energy ones
survive to the sea level because of time dilatation.

Thus, in a hadron cascade, the secondary particles either interact again,
decay or are absorbed as a result of ionization energy loss. The cascade
builds up to a shower maximum as in an electromagnetic cascade after which
the numbers decrease. Because decay muons do not interact by strong
interactions and loose less energy in bremsstrahlung radiation then
electrons, they primarily loose energy by ionization and therefore have a
slower decrease. Because detection of high energy gamma rays relies on
registration of secondary particles, the extensive air showers produced by
cosmic rays constitute background noise for ground-based gamma-ray
detectors. Special techniques and algorithms have to be developed to
suppress this noise in order to increase the sensitivity to photon
primaries. The presence of muons in the hadron induced shower is often
exploited.

% ////////////////////////////////////////////////////////////
\section{Milagro: a Next Generation of EAS Array.}

The design of a detector has to reflect the goals and the obstacles
mentioned above. The great success of air Cherenkov telescopes is due to
their excellent angular resolution and good background rejection. With
advantages come its limitation. Because the energy threshold for electrons
to produce Cherenkov radiation in air is about 21 MeV ($n_{air}=1.000293$,
formula \ref{equation:cherenkov_threshold}), it cuts the number of
detectable particles by a factor of 2 (table \ref{table:eas:longitude},
equation \ref{equation:eas_longitudal}). This bounds the energy of the
primary gamma rays detectable by the technique to be greater than about
100 GeV. Furthermore, the telescopes are narrow-field-of-view optical
devices, therefore, they can observe only a small portion of the sky at a
time during cloudless, moonless nights. This severely impacts on their
ability to detect transitory sources and perform observations of extended
sources.

Extensive air shower arrays, on the other hand, can observe the entire
overhead sky and can operate 24 hours a day regardless of weather
conditions. However, scintillation counters typically cover only a small
fraction of area compared to lateral extent of the shower, and therefore
detect only a small fraction of secondary particles leading to higher
energy threshold, typically above 10 TeV.

% ////////////////////////////////////////////////////////////
% ======  Pictures.....===========
\begin{figure}
\centering
\includegraphics[width=5in]{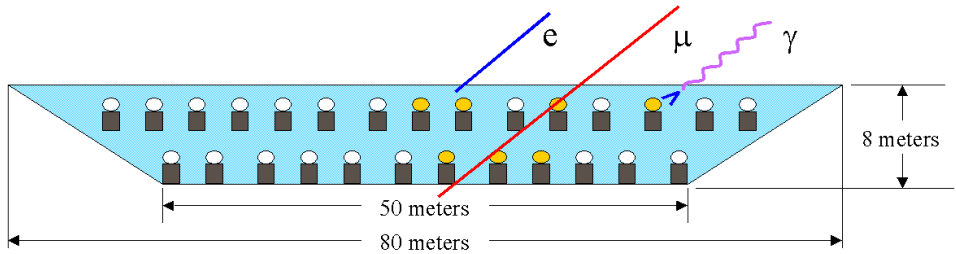}
\caption{ Schematic diagram of the Milagro detector.}
\label{fig:eas:milagro}
\end{figure}

Milagro achieves an energy threshold of 400 GeV by using water as a
detection medium and by locating the detector at relatively high altitude.
Using water allows Milagro to detect not only charged secondary particles
of the air shower, but also secondary photons via Cherenkov radiation of
cascades initiated by them in the water. This is important because
secondary photons present a large fraction of particles reaching the
ground level (table \ref{table:eas:longitude}, figure
\ref{fig:eas:longitude}). Also, the time distribution of the foremost
particle appears to be narrower for the photon component than for the
electron one \cite{Sciascio} rendering the efficient conversion of
secondary photons into Cherenkov radiation by the water medium even more
essential.

Milagro is a covered, light tight pond filled with purified water and
instrumented with photo detectors. Relativistic, charged particles produce
Cherenkov radiation as they traverse the water. The radiation is emitted
in a cone-like pattern with an opening angle of not greater than
$42^{\circ}$. This allows Milagro to instrument a large surface area of
the detector with a sparse array of photo detectors. Methods such as these
have been used for decades in high energy physics \cite{Becker}.

As shown in figure \ref{fig:eas:milagro}, Milagro has been configured with
two planes of photo detectors. The upper layer is used to measure the air
shower front, providing the information needed to reconstruct the primary
particle direction. The lower layer is used to detect penetrating muons or
hadrons and aid in rejecting cosmic ray induced showers.

Milagro fits the classification of extensive air shower
array, shares many principles but improves on detection methodology.

\chapter{The Milagro Detector}

% ////////////////////////////////////////////////////////////
\section{Physical Components of the Detector.}

The Milagro detector is located in the Jemez mountains, about 45 km west
of Los Alamos, New Mexico, at $-106^{\circ} 40' 37''$ East longitude,
$35^{\circ} 52' 43''$ North latitude. The central detector consists of 723
photomultiplier tubes (PMT) deployed in a 60~m~x~80~m~x~8~m pond filled
with water. The elevation of the reservoir above sea level is 2650 meters,
which translates to an atmospheric overburden of 750 $g/cm^{2}$.

The buoyant photomultiplier tubes are held below the water surface by
anchor cords, attached to a grid of weighted PVC pipes. The spacing of
the grid is 2.8 meters. The length of each string was calculated for each
PMT so that the PMTs would all lie in a horizontal plane and form a two
layer structure. Each PMT is floating upright with its photocathode facing
upwards. Each PMT is also surrounded by a conical baffle to block
internally reflected light.

The 450 PMTs of the top layer are deployed under 1.4 meters of water and
are used to measure the arrival direction of the shower. Two hundred
seventy-three additional PMTs are located near the bottom of the pond
under 6 meters of water and are used to distinguish photon- and
hadron-induced air showers. The top layer is called {\em the shower layer}
and the bottom one --- {\em the muon layer}.

The complete set of PMT string attachment points was surveyed after the
grid was installed.Vertical positions of the PMTs were verified after the
pond was filled with water. The final accuracy of PMT coordinates is
estimated to be $\pm 0.03$ m in horizontal and $\pm 0.01$ m in vertical
directions. Coordinates of the PMTs are used for shower reconstruction and
detector calibration. Because PMTs are used to detect Cherenkov light
produced by air shower particles traversing the water, it was necessary to
block the external light from entering the pond with an opaque cover and
to provide for high transparency of the water. The 1 mm thick cover was
made of two black layers of polypropylene with an internal polyester
scrim. The cover can be inflated to allow access into the pond. High
quality of the water was achieved with continuous recirculation and
filtration of the water by a set of carbon filter, $1-\mu$m filter,
ultraviolet lamp and $0.2-\mu$m filter. The attenuation lengths of the
water in the recirculator and that of the pond are indistinguishable and
equal to $13.4 \pm 0.5$ meters \cite{milagro_water}.

A lightning protection system was installed around the experiment as a
safeguard against possible strikes.

% ////////////////////////////////////////////////////////////
\section{The photomultiplier tube.\label{chapter:milagro:pmt}}

The photomultiplier tube is a vacuum device used to transform very faint
light signals into electric ones. It consists of a photocathode, a set of
dynodes and an anode. A photon incident on the photocathode causes the
emission of an electron (called {\em photo-electron (PE)}) into the vacuum
tube via the photoelectric effect. This electron is directed towards the
first dynode by the focusing fields. When it hits the dynode, secondary
electrons are emitted, each of which is guided towards the next dynode.
The dynodes are kept at different electric potentials and thus cause
acceleration of electrons to sufficient energies for secondary emission to
take place. As this cascade develops, the number of electrons increases
exponentially and by the time it reaches the anode the number of electrons
is about $10^{7}$ for every electron emitted at the photocathode.

Besides amplification, the following points motivate the choice of the
Hamamatsu R5912 SEL PMT model:

\begin{itemize}
 \item Cherenkov radiation generated by shower particles in the water is
       emitted mostly at the violet end of the visible spectrum. The
       selected model is sensitive in the 300 - 650 nm wavelength range.

 \item {\em Quantum efficiency,} the ratio of the number of PE produced
       at the cathode to the number of incident photons, is rather high:
       0.2-0.25 for the selected model.

 \item {\em Time resolution.} Time resolution of a PMT is limited by the
       fluctuations in the cascade development starting from the
       photocathode all the way down to the anode. The variance of this
       transit time increases as the input intensity decreases and is 2.7
       ns at the lowest input (1 PE produced).

 \item {\em Late pulsing.} Late pulses on the output of a PMT are believed
       to occur when electrons are being reflected off the first dynode
       and produce secondary electron only upon second re-entry into the
       dynode assembly caused by the focusing fields. The probability of
       late pulses decreases for high light intensities because the
       probability that all electrons produced at the cathode are reflected
       decreases as number of them increases.

 \item {\em Pre-pulsing.} Pre-pulses are considered to be caused by the
       light hitting the first dynode directly, thereby producing an anode
       signal which precedes the main one induced by the photo-electron.
       The Probability of pre-pulsing is higher for larger light
       intensities.

 \item {\em After-pulsing.} After pulses are thought to be caused by
       residual gas molecules in the tube being ionized by electrons and
       subsequently hitting the photocathode and producing electrons. The
       signal induced by these electrons follows the main anode one.

 \item {\em Saturation.} Saturation is the effect of decrease of the
       amplification coefficient of the PMT for high light intensities.
       This is caused by the inability of the dynodes to accelerate the
       increased number of the secondary electrons to sufficiently high
       energy.

\end{itemize}

Overall, the characteristics of the selected PMT model were found suitable
for the application.

% ////////////////////////////////////////////////////////////
\section{Electronics. Time Over Threshold.}

Each PMT is connected to its own electronic channel with a single cable
used to supply high voltage for the tube and to deliver a signal pulse
from it. The signal carries two pieces of information about the Cherenkov
light hit: the time when the hit occurred and it strength.

% ======  Pictures.....===========
\begin{figure}
\centering
\includegraphics[width=3.0in]{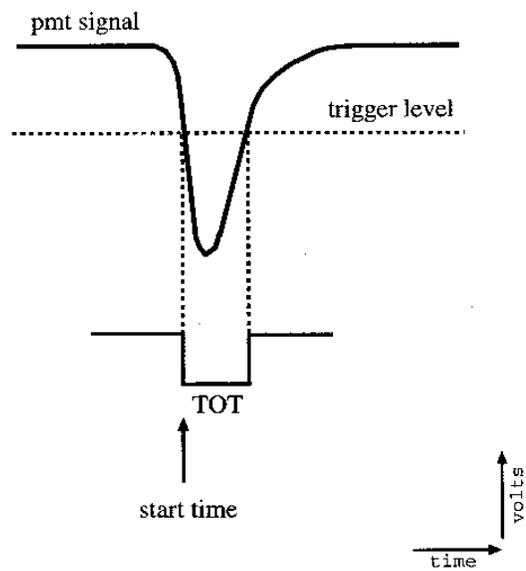}
\caption{Time over threshold.}
\label{fig:tot}
\end{figure}

The time of the hit is determined by the time when the amplified signal
crosses the discriminator with the preset threshold. This time is termed
{\em start time}. A time-to-digital converter (TDC) is used to digitize
the start time and to pass it to the data acquisition computer. Except for
the saturation, the output anode charge of the PMT is proportional to the
number of Cherenkov photons which struck the tube, therefore, the output
charge is the measure of the hit strength.

Conventionally, analog-to-digital converters (ADC) are utilized to
digitize the charge information. Because of cost, speed and dynamic range
limitations of typical ADC's, a time-over-threshold (TOT) method was
adopted instead.  The idea of the method is simple (figure~\ref{fig:tot}).
A capacitor $C$ is quickly charged by the PMT signal with total charge $Q$
and then is slowly discharged via a resistor $R$. The voltage on the
capacitor as a function of time is given by:

\[ V(t) = \frac{Q}{C} e^{-t/RC} \]

If the time during which the voltage exceeds the preset threshold is
measured, then it is seen that

\[  time \; over \; threshold \; \sim \ln Q \]

Thus, time over threshold is related to the initial charge $Q$ and
provides a way to measure the signal strength. Implementation of this
method allows the use of a single multi-hit TDC module to digitize both
start time and the time over threshold counterparts.

In practice, two discriminators with different threshold levels (low and
high) are used to guard against pre- or after-pulsing of the PMTs. These
accompanying pulses are usually small and do not cross the high threshold.
Therefore, the use of high threshold information is preferred where
available. The low threshold was set to about 1/4 of the average signal
produced by a single photoelectron and high threshold to about 7 PE.

Each shower layer PMT signal crossing the low discriminator also
participated in the trigger formation process by generating a 25 mV, 300
ns long pulse \cite{milagrito:nim}. The analog sum of all such pulses is
sent to a discriminator with a threshold corresponding to 60 PMT being
hit. If the discriminator is fired the trigger pulse is generated. Setting
the trigger threshold much higher than 60 PMT would increase the energy
threshold of the detector. Making it much lower would enable single muons
to trigger the detector. The choice of 300 ns window is motivated by the
duration of near horizontal shower propagation through the detector. The
trigger provides a common stop for all TDC's, all start times are
referenced to it.

The absolute time of each triggered event is provided by a Global
Positioning System clock. A schematic diagram of a PMT channel is
presented in figure~\ref{fig:channel}.

% ======  Pictures.....===========
\begin{figure}
\centering
\includegraphics[width=5.5in]{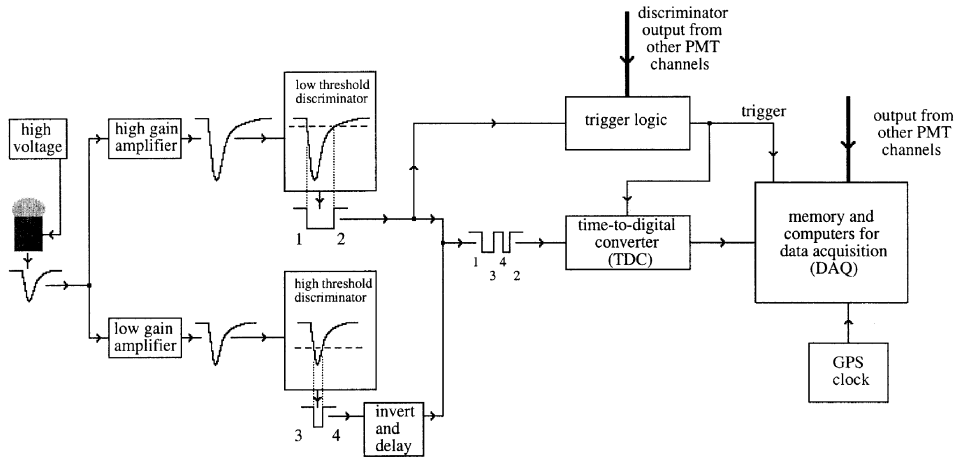}
\caption{Schematic diagram of a PMT channel.}
\label{fig:channel}
\end{figure}

% ////////////////////////////////////////////////////////////
\section{Timing Edges.}

Time-over-threshold pulses generated by both discriminators (if crossed) 
are multiplexed into a series of time edges. (Figure~\ref{fig:timeEdges})
It is the polarity \footnote{Polarity is the direction in which the
threshold was crossed:  going up or down.} and time of these edges which
is being recorded as raw data from a PMT. Strong hits cross both
thresholds and result in a 4-edge event, weak ones cross low threshold
only and produce 2 timing edges. 

% ======  Pictures.....===========
\begin{figure}
\centering
\includegraphics[width=4.5in]{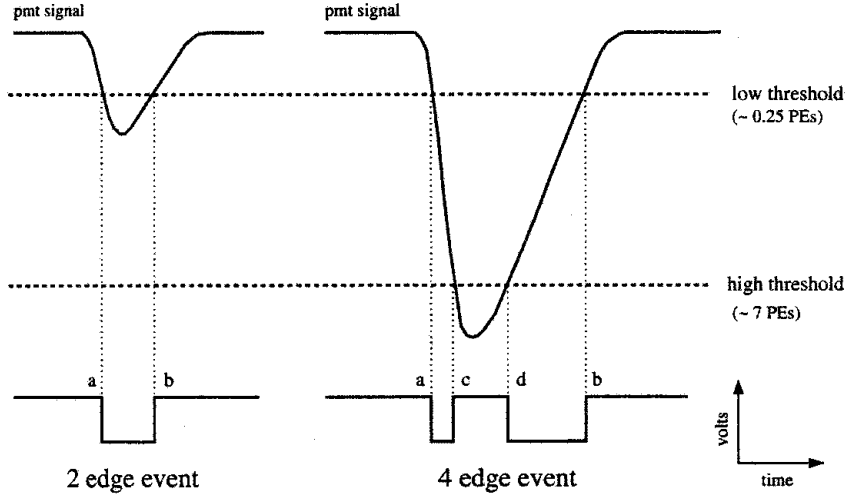}
\caption{Timing edges.}
\label{fig:timeEdges}
\end{figure}

In practice many edge events can be observed. One edge events can result
when the other edge (or edges) is truncated by the trigger time
window. Four edge events can be a sequence of two 2-edge events. Multi
edge hits are also possible. A filtering algorithm had been designed to
clean the hits. It bases its decisions on the polarity of hits and their
relative time separation. On average, about 8\% of PMT hits are rejected as
having invalid edge sequences. The accepted PMT hits are characterized by
start time (LoStart) and time over low threshold (LoTOT) and their high
threshold counterparts (HiStart, HiTOT) if available.

% ////////////////////////////////////////////////////////////
\section{Monte Carlo Simulation of the Detector.}

The simulation of the detector response is done in two steps: (1) initial
interaction of the primary particle (proton or photon) with the atmosphere
and the subsequent generation of secondary particles, and (2) detector
response to the secondary particles reaching the detector level.  The
CORSIKA~\cite{corsika_web} air-shower simulation code provides a
sophisticated simulation of secondary particle development in the Earth's
atmosphere induced by a primary particle with energy up to $10^{20}$ eV.
Within CORSIKA, the VENUS code is used to treat hadron-nucleus and
nucleus-nucleus collisions at high energies whereas GHEISHA is used at low
energies ($<80$ GeV). Electromagnetic interactions are simulated using EGS
4 code. The atmosphere adopted in CORSIKA consists of nitrogen, oxigen and
argon in volume proportions of 78.1\%, 21.0\% and 0.9\% respectively. The
density variation of the atmosphere with altitude is modeled by 5 layers.
In CORSIKA a flat atmosphere is adopted which is a good approximation up
to zenith angles of about $70^{\circ}$ where discrepancy with the
shperical one reaches one radiation length.

The simulation of the detector itself is based on GEANT~\cite{geant_web}.
All of the secondary particles in a shower cascade reaching the Milagro
are used as input to the GEANT with uniform random placement of the shower
core around the detector. GEANT simulates the electromagentic and hadronic
interactions of particles in the pond and results in the set of simulated
times and pulse heights at each PMT. This information is saved in the same
format as the real data and is used to establish properties of the
detector. The detailed simulation of the electronics and phototubes
continues to be the area of active reseach. Further improvements and
understanding are expected.

\chapter{Calibration.}

% ////////////////////////////////////////////////////////////
\section{System Setup and Goals.}

The calibration system has been designed to reflect the physics goals of
the detector and is used to obtain parameters needed to transform the raw
counts to physically meaningful arrival times and light intensities which
then can be used for event reconstruction. Despite the considerable effort
that has been made to construct all PMT channels of the detector as
uniformly as possible, in order to achieve the high precision required for
the event reconstruction the remaining variations between channels have to
be compensated for. A separate set of calibration parameters is determined
for each PMT channel.

The desire to determine the positions of events on the Celestial Sphere
with systematic errors much less than the expected angular resolution
(which is about $1^{\circ}$) dictates that the locations of the
photo-tubes be known to about 10~cm accuracy in horizontal direction and
3~cm in vertical, and PMT timing accuracy to about 1~ns. Photographic and
theodolite surveys were used to ensure accurate PMT position
determinations.

To achieve the stated time accuracy it is important to calibrate the TDC
conversion factors, compensate for pulse amplitude dependence of TDC
measurements (known as slewing correction) and synchronize all TDCs (find 
TDC time off-sets) to the required accuracy. Time over threshold to
photo-electron conversion must be determined to convert all PMT amplitude
measurements to a common unit for each event. All of the above is achieved
with the help of the laser calibration system.

The calibration system is based on the laser---fiber-optic--diffusing ball
concept used in other water Cherenkov detectors~\cite{IMB_calibration}. A
computer operated motion controller drives a neutral density filter wheel
to attenuate a 300 picosecond pulsed nitrogen dye laser beam. The selected
dye emitted light at 500 nm. The beam is directed to one of the thirty
diffusing laser balls through the fiber-optic switch (see
figure~\ref{fig:calibration:setup:system}). Part of the laser beam is sent
to a photo-diode. When triggered by the photo-diode, the pulse-delay
generator sends a trigger pulse to the data acquisition system. A laser
fire command is issued by the motion controller, providing full automation
of the calibration process. The balls are floating in the pond so that
almost every PMT can be illuminated by more than one light source.
Complete description, operation, analysis, problems and suggestions are
described in \cite{Milagro_calibration_manual}.

% ======  Pictures.....===========
\begin{center}
\begin{figure}
\includegraphics[width=5.5in]{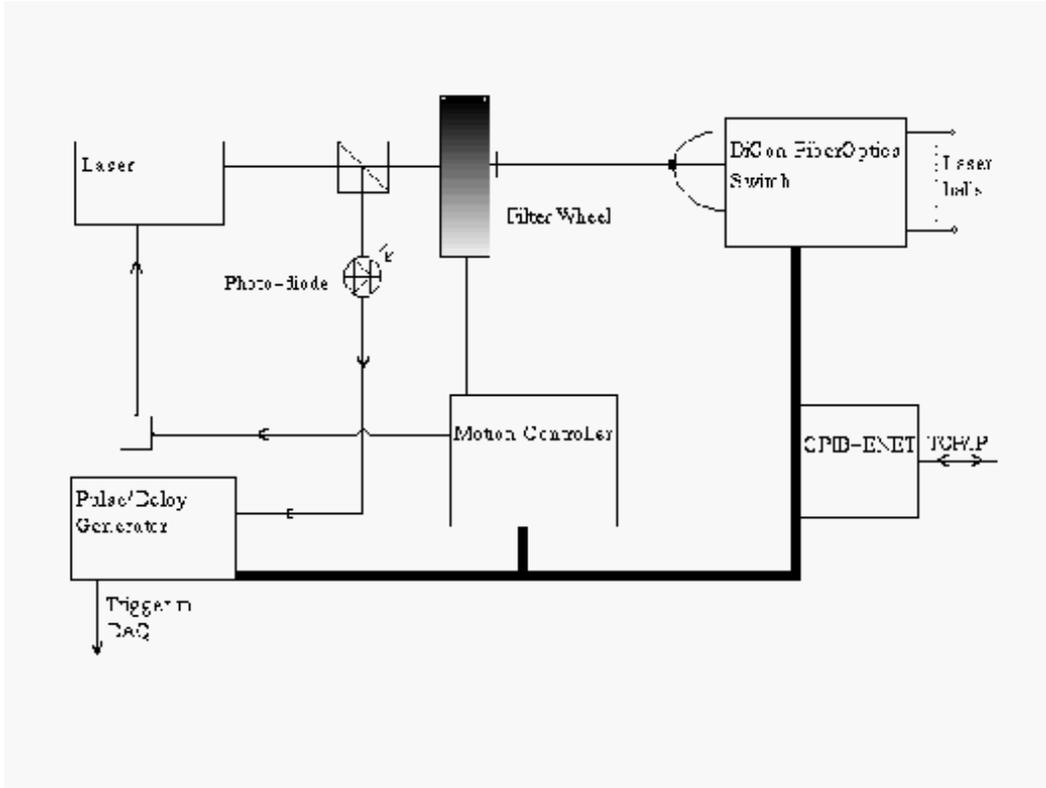}
\caption{Calibration system setup.}
\label{fig:calibration:setup:system}
\end{figure}
\end{center}

% ////////////////////////////////////////////////////////////
\section{Timing Calibration.}
\subsection{Slewing Calibration.}

Time slewing is the dependence of the reaction time of a PMT together with
its electronics on the intensity of the incident light. The amount of
light in the calibration system is regulated by the filter wheel whose
transmission property ranges from completely opaque to completely
transparent thus enabling the study of the effect. The ideal situation of
no slewing should reveal itself as independence of PMT registration times
with respect to photo-diode when intensity of light in the calibration
system is changed.\footnote{ The photo-diode is thought as having no
slewing problem because intensity of light incident on it is constant.}

In practice, it takes longer for weak pulses to cross a discriminator
threshold, and thus results in a delay. This is illustrated on the
figure~\ref{fig:calibration:slewing:tot}.

% ======  Pictures.....===========
\begin{figure}
\centering
\includegraphics[width=3.0in]{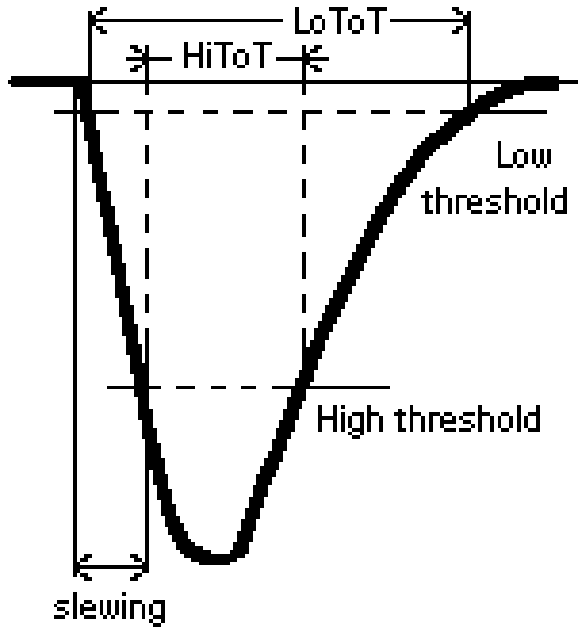}
\caption{Slewing and Time over Threshold. The weaker the pulse (or,
         equivalently, the higher the discriminator threshold) the latter
         the registration time is. Infinitely large pulse will cross the
         threshold without any delay, or zero slewing.}
\label{fig:calibration:slewing:tot}
\end{figure}

Because the light intensity is characterized by time over threshold, the
amount of slewing is studied with respect to this variable. The method of
generation of such a slewing curve is explained with the help of the
figure \ref{fig:calibration:slewing:concept}.

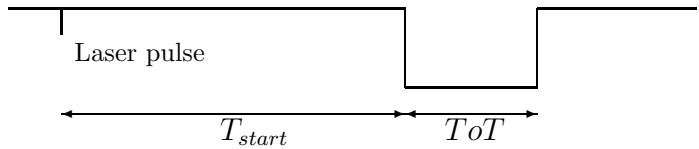
\begin{figure}
\centering
%\shadowbox{
\begin{picture}(300,100)(0,15)
\put(20,90){\line(1,0){150}}
\put(170,60){\line(1,0){50}}
\put(220,90){\line(1,0){60}}
\put(170,90){\line(0,-1){30}}      % start of ToT
\put(220,90){\line(0,-1){30}}      % end of ToT
\put(40,90){\line(0,-1){10}}       % laser pulse

\put(60,50){\vector(1,0){110}}     % T-start
\put(60,50){\vector(-1,0){20}}     % T-start
\put(195,50){\vector(1,0){25}}     % ToT
\put(195,50){\vector(-1,0){25}}    % ToT

\put(100,40){$T_{start}$}   
\put(185,40){$ToT$}
\put(45,70){\footnotesize Laser pulse}
\end{picture}

%}
\caption{Conceptual drawing of a measurement performed for timing
calibration.}
\label{fig:calibration:slewing:concept}
\end{figure}

This diagram is high/low threshold independent, the scheme is applied
independently for both threshold levels. The slewing curve is a plot of
$T_{start}$ vs $ToT$. Note, that $T_{start}$ includes the propagation time
of pulses in the water and in the optical fiber, delays associated with
the details of the particular PMT channel and common detector trigger
delays.  Because common offsets are irrelevant, after correcting for
relative fiber delays and water propagation times which effectively shifts
the curve up or down, this will become a complete timing calibration. The
slewing correcting curve is found by fitting a polynomial to $T_{start}$
vs $ToT$. If a PMT gets a slewing curve from more than one laser ball, the
curve resulting from the maximal light illumination is chosen. An example
of a slewing curve is presented in
figure~\ref{fig:calibration:slewing:curve}.  The TDC's used in Milagro
employ a common stop, thus larger values of $T_{start}$ correspond to
earlier times.

\begin{figure}
\centering
  \subfigure[Example of the $T_{start}$ vs $ToT$ data points.]{
    \label{fig:calibration:slewing:curve::1}
    \includegraphics[width=2.5in]{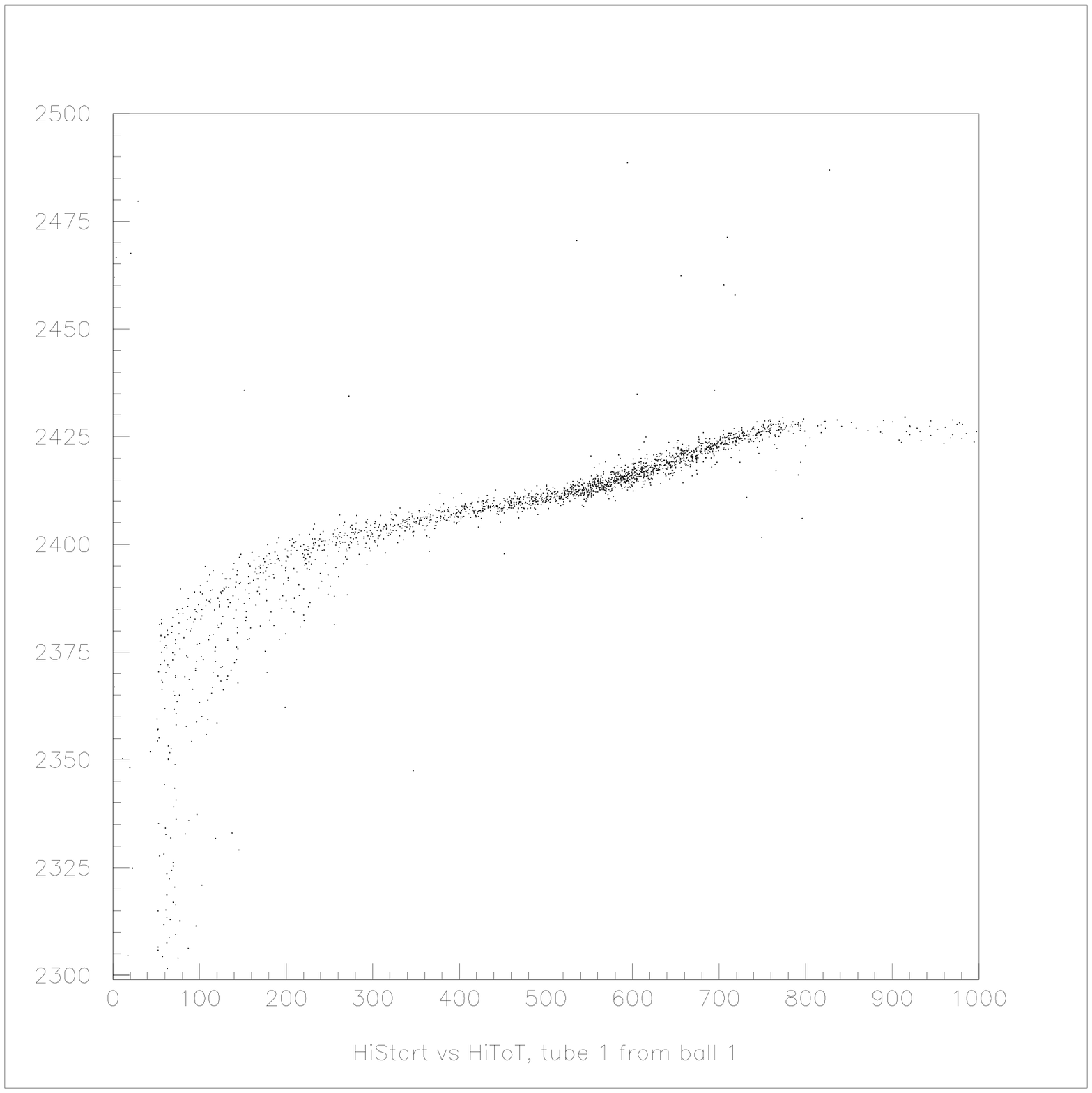}
  }
  \subfigure[Example of the polynomial fit. The values in each bin are the
             means of the $T_{start}$ obtained by a fit of a Gaussian
             distribution in the corresponding $ToT$ bin.]{
    \label{fig:calibration:slewing:curve:2}
    \includegraphics[width=2.5in]{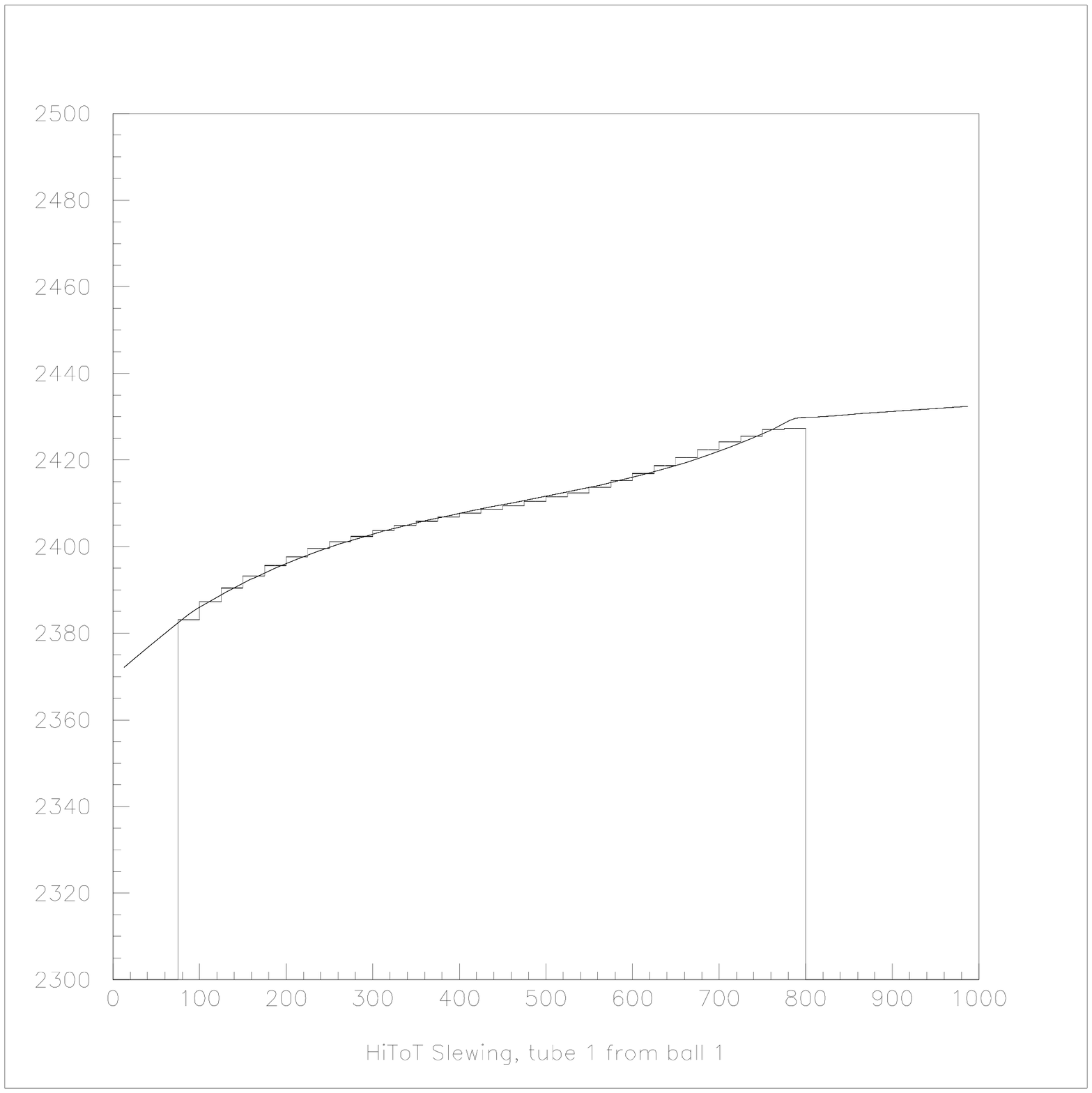}
  }
\caption{Plots showing a typical data obtained from the calibration 
         system for the purpose of timing calibration. Units of both axis
         are TDC counts.}
\label{fig:calibration:slewing:curve}
\end{figure}

% ////////////////////////////////////////////////////////////
\subsection{TDC Conversion Verification.}

The time to digital converters measure time in the units of ``counts''.
According to LeCroy 1887 FASTBUS TDC specifications one count corresponds
to 0.5 nanoseconds. This was verified by insertion of known variable delay
in the photo-diode trigger chain with the help of DG535 digital
delay/pulse generator by Stanford Research Systems. The shift in the start
times ($T_{start}$) of all PMTs was found to be very consistent with the 2
count per 1 ns conversion. This assured that all TDC clock speeds are the
same and meaningful interpretation of time across all PMT channels is
available.

% ////////////////////////////////////////////////////////////
\subsection{Fiber-Optic Delays. Speed of Light in Water.}

The last step of timing calibration is correcting the slewing curve for
both the light propagation time in water between the PMT and the laser
ball used to generate the curve and the relative delay in the fiber
attached to that ball. In order to correct for the propagation time,
coordinates of PMTs and balls and the speed of light in water have to be
given. PMT and laser ball coordinates are known from the survey. (If laser
ball coordinates are deemed to be inaccurate, they can be found from the
calibration data: see \cite{tped_memo,tped_icrc}.) Because only a typical
index of refraction of water is found in reference tables and because
fiber-optic delays vary from ball to ball, these parameters have to be
evaluated on the basis of the calibration data itself.

The method to solve the problems uses the ability to cross calibrate a PMT
using several laser balls. If $T_{1}$ and $T_{2}$ are slewing corrected
times registered by a PMT from two laser balls (1 and 2), then the
difference $(T_{1} - T_{2})$ does not depend on delays in the PMT
electronics (common to both measurements), but instead reflects the
difference in propagation times from the corresponding balls and the
relative fiber delays. Consider the following quantity:

\[ \tau = T_{1} - T_{2} + \Delta_{fiber} + \Delta_{propagation}\]

If speed of light and fiber delays are correct, then $\tau$ is zero within
errors of measurement. When for given laser ball pair many PMTs with their
$\tau$'s are considered, one sees that for PMTs located approximately half
way between the balls $\Delta_{propagation} \approx 0$ and deviation of
average $\tau$ from zero is due to fiber delay difference. On contrary,
PMTs located in the close proximity of either ball provide information
about the speed of light: the usage of incorrect speed of light will widen
the distribution of $\tau$'s. Thus the analysis of all $\tau$'s for all
PMTs and laser ball pairs enables the determination of speed of light in
water and relative fiber delays.

Slewing curves are shifted by the appropriate propagation time and fiber
delay to yield the final timing calibration for each PMT.

% ////////////////////////////////////////////////////////////
\section{Photo-Electron Calibration.}
\subsection{Occupancy Method.}

Time over threshold to photo-electron calibration is based on the
well-known occupancy method, which was one of the methods of PE
calibration of Milagrito \cite{isabel_memo} and other water Cherenkov
detectors~\cite{IMB_calibration}. The data used for PE calibration is the
same as for the timing calibration which is obtained with the laser light
passing through the filter wheel with different transparency settings
(which are changed by rotating the wheel). The main task of the PE
calibration is determination of the number of photo-electrons for a given
incident light intensity (ToT). The basic method consists of two steps: 1.
at low light levels calibrate using the occupancy method, and 2. extend to
higher light levels using given attenuation properties of the filter
wheel.

\begin{enumerate}
\item Calibration at low light levels.

At low light levels it is possible to measure the number of
photo-electrons as function of ToT directly. The method is founded on the
following assumptions:

\begin{itemize}
 \item The number of photo-electrons produced at PMT's photocathode obeys
       a Poisson distribution:

       \[ P_{\lambda}(n) = \frac{\lambda^{n}}{n!}e^{-\lambda}  \]

       where $\lambda$ is the mean number of the photo-electrons produced.
       (This is justified because the probability of emitting a single
        photo-electron does not depend on the fact of possible emission of
        other photo-electrons.)

 \item A constant light intensity source is used in calibration.
\end{itemize}

Then, the probability $\eta$ that at least one photo-electron was produced
while the photocathode was illuminated (the probability that the PMT
``saw'' the illumination) is called occupancy and is given by: 

\[  \eta = P(n>0) = 1 - P_{\lambda}(n=0) = 1 - e^{-\lambda}\]

Occupancy can be easily measured if the PMT is illuminated several times
with the same pulse intensity:

\[  \eta = \frac{number \; of \; observed \; pulses}
                {total \; number \;  of \; pulses}\]

Therefore, if $\eta$ is known, then $\lambda = -\ln(1-\eta)$. This method
is applicable at relatively low light levels (when $\lambda < 2$) because
at high levels the occupancy saturates to 1 and a small measurement error
in $\eta$ will lead to a big error in $\lambda$:

\[ \Delta \lambda = \frac{1}{1-\eta} \Delta \eta = e^{\lambda} \Delta \eta\]

\item Calibration at high light levels.

Calibration at high light levels is achieved by noting that in the absence
of PMT/electronics saturation the number of photo-electrons is
proportional to the light intensity at the photocathode. If the
transmittance of the filter wheel $T$ is known, then:

\[  \lambda = a T \]

where $a$ is some coefficient which is different for each laser ball PMT
pair. This coefficient can be found at low light levels where $\lambda$ is
also known. Error of this method will grow linearly with light intensity,
not exponentially as in low light level case. Thus, given transmittance
$T$ of the filter wheel, ToT-to-PE conversion can be achieved at any light
level. When, in reality, saturation is present prescribing the number of
photo-electrons using this as a requirement allows protection against the
effect. Traditional amplitude-to-digital conversion methods which make use
of the air shower data are susceptible to such non-linearity.

\end{enumerate}

% ////////////////////////////////////////////////////////////
\subsection{In Situ Filter Wheel Calibration.}

Traditionally, filter wheel calibration is obtained either from the
manufacturer or by a separate measurement. In Milagro, in situ filter
calibration method was developed which uses the same calibration data. The
idea is grounded on the supposition that for any two sufficiently close
levels of transmittance of the filter wheel ($T_{1}$ and $T_{2}$) there
exists a PMT in the pond for which the occupancy method is valid on both
light intensities. If $\lambda_{1}$ and $\lambda_{2}$ are corresponding
photo-electron measurements, then: 

\[     \frac{T_{2}}{T_{1}} = \frac{\lambda_{2}}{\lambda_{1}}\]

This line of arguments is used to relate $T_{3}$ to $T_{2}$, $T_{4}$ to
$T_{3}$ and so on, leading to restoration of the transmittance levels for
the entire wheel. Because absolute transmittance of the wheel is
irrelevant one can always set $T_{0}=1$.

% ////////////////////////////////////////////////////////////
\subsection{Dynamic Noise Suppression.}

Small amounts of radioactive elements present in the water, ambient light,
Cherenkov light produced by the shower particles or thermo-electron
emission cause signals on the output of PMT channels which constitute
noise hits. The presence of these hits will lead to overestimation of
occupancy implied by the laser light level and thus will damage the
accuracy of the calibration. Dynamic noise suppression is a technique
allowing the solution of this problem on the tube by tube basis.

An event, registered by a PMT could have come from a laser pulse or from a
noise hit which are independent, therefore the probability of observing a
hit $P(any)$ is given by:

\[  P(any) = P(laser+noise) = P(laser) + P(noise) - P(laser) \cdot P(noise)\]

where $P(laser)$ is probability of observing a pulse due to laser (true
occupancy), $P(noise)$ is probability of observing a pulse due to noise.
$P(any)$ is measured by sending laser pulses into the pond, $P(noise)$ is
obtained by sending ``random'' (no laser light) triggers to the data
acquisition system, then the occupancy is given by:

\[  \eta = P(laser) = \frac{P(any) - P(noise)}{1- P(noise)}  \]

This is indeed how occupancy was determined for the filter wheel
calibration and for the ToT-to-PE conversion. To demonstrate the
importance of the suppression refer to
figure~\ref{fig:calibration:wheel:noise} which shows filter wheel
calibration curves obtained with and without noise suppression. 

% ======  Pictures.....===========
\begin{figure}
\centering
\includegraphics[width=5.5in]{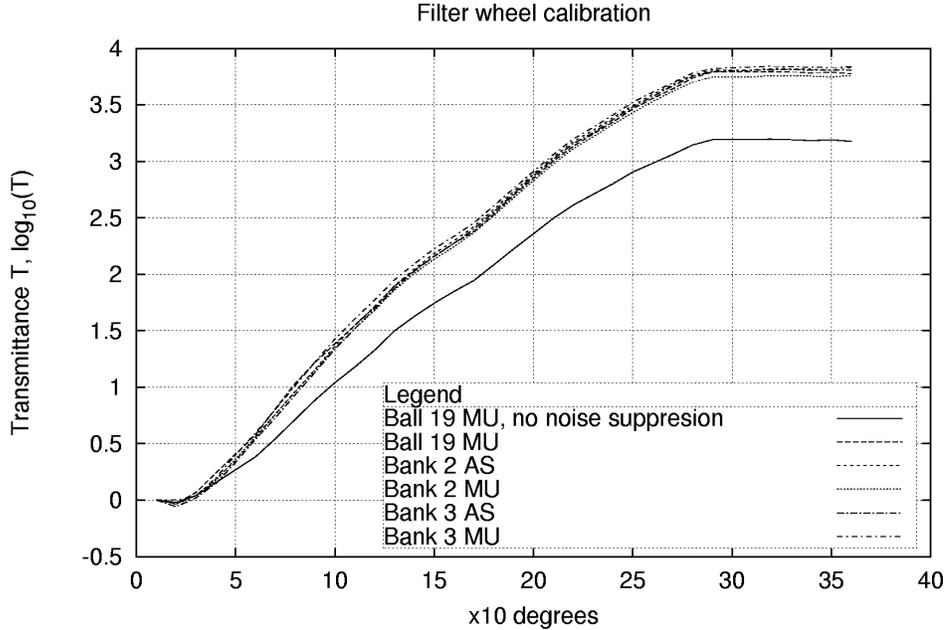}
\caption{Filter wheel calibration curves obtained with different laser
ball --- PMT combinations are presented. Bank 2,3 represent laser balls
11-20 and 21-30 respectively. AS, MU --- PMT's of air shower and muon layers
respectively. Calibrations with ball 19 and muon layer PMT's are also
shown with and without the noise suppression to illustrate its
importance.}
\label{fig:calibration:wheel:noise}
\end{figure}

% ////////////////////////////////////////////////////////////
\section{Slewing Extrapolation.}

The maximum light level at which a PMT was calibrated depends on the
distances to the laser balls, laser light output and the calibration
system optics alignment. The range of $ToT$ within which a tube is
calibrated may be too narrow to allow interpretation of the strongest hits
generated by the shower particles. The desire to make use of these hits
requires extrapolation of the calibration curves beyond the measured
points by some plausible functional form. The functional shape depends on
the PMT discriminator threshold level, amplification coefficients and so
on. Instead of trying to make physical model of the channel a statistical
approach is followed.

It is believed that all PMT channels were manufactured to meet common
characteristics. Therefore, the study of the channels' responses
(calibration) can be viewed as a multiple (about 700 times, broken PMTs do
not count) measurement of a single function of $ToT$. The fact that the
calibration curves obtained for different channels are slightly different
can be attributed to the ``manufacturing imperfections'' (such as
unavoidable uncontrollable spread of characteristics of electronic
components) and can be treated as such.Thus, the curves for all PMTs can
be viewed as particular realizations of the same random function
\cite{random_invitation}.

Inasmuch as a random function is a mathematical concept of great
complexity and in most general case it can be regarded as a
non-denumerable set of scalar random variables, it is natural to seek the
expression of the random function in terms of simpler random concepts:
ordinary scalar random variables. Therefore, the following representation
of the calibration curve $X(t)$ is tried (here $t$ or $t'$ denote $ToT$):

\[
  X(t) = m(t) + \sum_{\nu=1}^{n} V_{\nu} \phi_{\nu}
\]

Here, $\{ \phi_{\nu}, \nu=1,..,n \}$ are some non-random functions, $\{
V_{\nu} \}$ are random variables. This representation is said to be
canonical if $m(t)$ is the mean function ($m(t)= M[X(t)]$) and $V_{\nu}$'s
are uncorrelated with zero mean \cite{random_pugachev}:

\[
  M[V_{\nu}] = 0, \;\;\; D[V_{\nu}] = M[(V_{\nu})^{2}] =D_{\nu}
\]
\[
  M[V_{\nu} \cdot V_{\mu}] = 0 \;\;\; \mu \ne \nu
\]

When such a representation is found, each particular calibration curve
$x(t)$ can be extrapolated using the expansion and the set of coefficients
$v_{\nu}$ can be calculated in the range where calibration data is
available. The construction of the canonical representation makes use of
the average function $m(t)$ and the correlation function $K(t,t')=
M[X^{\circ}(t) \cdot X^{\circ}(t')]$, where $X^{\circ}(t) = X(t) - m(t)$.
It can be shown that

\[
  \phi_{\mu}(t) = \frac{1}{D_{\mu}} \sum_{k=1}^{m} a_{\mu k} K(t,t_{k})
\]

\[
  V_{\mu} = \sum_{k=1}^{m} a_{\mu k} X^{\circ}(t_{k})
\]

satisfy the conditions of the canonical representation if coefficients
$a_{\mu k}$ are chosen such that

\[
  \sum_{k,l=1}^{m} a_{\mu k}a_{\nu l} K(t_{k},t_{l}) = 0 \;\; \mu \ne \nu
\]

which is always possible if the number $m$ of sample points $\{t_{k}\}$ is
large enough ($m > (n-1)/2$) \cite{slewing_extrapolation}. The root mean
square deviation of the representation from the realization is given by:

\[
  rms(t) = \sqrt{K(t,t) - \sum_{\mu=1}^{n} D_{\mu} (\phi_{\mu}(t))^{2}}
\]

This scheme is used for extrapolation of slewing curves of timing
calibration. It is implemented to the first order of the canonical
expansion \cite{slewing_extrapolation}. The only sample point $t_{1}$ is
the highest value of $ToT$ available from the calibration data. Therefore,
the extrapolation is obtained by:

\[
  x(t) = m(t) + \frac{x(t_{1}) - m(t_{1})}{K(t_{1},t_{1})} K(t,t_{1})
\]

The typical extrapolation range needed to interpret shower data is less
than 100 ns in $ToT$ leading to the error of the extrapolation being less
than 0.7 ns \cite{slewing_extrapolation}. Comparison of the extrapolated
curves and measured ones obtained in different calibration runs yielded
the measured extrapolation error of 0.55 ns \cite{stability_study} in good
agreement with the expectations.

\chapter{Event Reconstruction.}

% ////////////////////////////////////////////////////////////
\section{Primary Particle Direction Reconstruction.}

The data registered in each event is used to infer the characteristics of
the primary particle that created the shower: its direction of incidence,
energy and type.\footnote{The PMT registration times and corresponding
light intensities deduced from the raw counts and calibration parameters
present the event data.} In order to be able to make the inference about
the properties of the primary particle based on the observed quantities
one has to assume the relation between the two. One of these relations is
expressed by equation \ref{equation:eas_lateral}, the lateral distribution
of the shower. If the arrival direction is known, then the depth $t$ of
the atmosphere traversed by the shower can be determined from geometrical
consideration and then, equation \ref{equation:eas_longitudal} together
with equation \ref{equation:eas_lateral} relate the measured lateral
distribution $\rho(r)$ with energy of the primary particle and shower core
position. From the practical point of view, the fluctuations in the shower
development, small size of the detector and fluctuations in its response
make this program infeasible \cite{Gaurang:energy_memo}. The detector is
currently being upgraded with the array of water tanks which will aid in
core location, allowing achievement of 30\% energy resolution
\cite{Gaurang:energy_memo}.

The task of determination of the primary particle direction is much
simpler and is of highest importance (from the gamma ray astronomy point
of view). The basic idea is as follows. Let us assume that the shower
front is a plane, perpendicular to the direction of motion of the primary
particle. Then, as in any air shower array, each detector $i$ in the set
records the local arrival time $T_{i}$ of the shower front (see figure
\ref{fig:reconstruct:idea}) which has to be contrasted with the model: a
plane defined by a normal vector $\vec{n}=(a,b,c)$, the direction of the
primary particle:

%///////////////////////////////
% picture
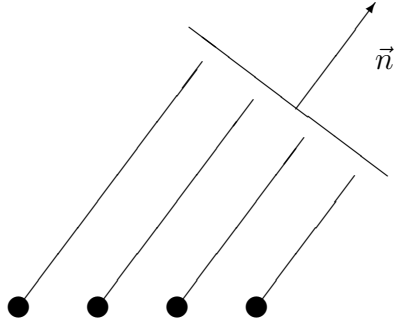
\begin{figure}
\centering
\setlength{\unitlength}{1.5pt}
\begin{picture}(150,100)(0,-10)

%\put(0,0){\circle*{5}}
%\put(150,100){\circle*{5}}

\put(10,0){\circle*{5}}
\put(10,0){\line(3,4){46.4}}

\put(30,0){\circle*{5}}
\put(30,0){\line(3,4){39.2}}
\put(50,0){\circle*{5}}
\put(50,0){\line(3,4){32}}
\put(70,0){\circle*{5}}
\put(70,0){\line(3,4){24.8}}

\put(103.07,33.07){\line(-4,3){50}}
\put(80,50.07){\vector(3,4){20}}
\put(100,60.07){$\vec{n}$}

\end{picture}

\caption{Schematic diagram of a shower array detectors and how timing
information can be used to determine the direction of the primary particle
$\vec{n}$. Lines represent measured arrival times, filled circles --- the
locations of detectors.}

\label{fig:reconstruct:idea}
\end{figure}

\[ T_{i} =\frac{ax_{i}+by_{i}+cz_{i}}{v_{0}\sqrt{a^{2}+b^{2}+c^{2}}}
           + T_{0}  \]

where, $(x_{i},y_{i},z_{i})$ are coordinates of PMT $i$, $v_{0}$ --- is
the speed of the shower front in air, $T_{0}$ --- is a common time offset
for the whole event. Noting that Milagro is a flat array, thus $z_{i}
\equiv 0, \forall i$, and introducing notations:

\[ A = \frac{a}{v_{0}\sqrt{a^{2}+b^{2}+c^{2}}}, \;\;\;
   B = \frac{b}{v_{0}\sqrt{a^{2}+b^{2}+c^{2}}}  \]

we obtain the final model

\[ T_{i} = Ax_{i} + By_{i} + T_{0}  \]

When coefficients $A$ and $B$ are found, the direction of the incoming
shower is reconstructed as

\[ \left\{ \begin{array}{rcl}
        \cos(\phi) & = & \frac{A}{\sqrt{A^{2}+B^{2}}} \\
        \sin(\phi) & = & \frac{B}{\sqrt{A^{2}+B^{2}}} \\
            \theta & = & \arcsin(v_{0}\sqrt{A^{2}+B^{2}})
           \end{array}
   \right.
\]

where $(\theta, \phi)$ are zenith and azimuth angles, and $v_{0}$, the
speed of the shower front in air, is approximately equal to the speed of
light.

The method of determination of coefficients $A$ and $B$ from the measured
times $T_{i}$ is a $\chi^{2}$ fit where weights are chosen based on the
recorded light intensity of the PMTs. This is because more energetic
particles appear to suffer less fluctuations due to the combined effect of
the shower development and PMT time resolution and thus are given higher
weight. The weights were derived by studying the distributions of
$T_{i}$'s \cite{Joe_Bussy}. These distributions are not Gaussian, as
assumed by the $\chi^{2}$ fit, therefore, several iterations are made. The
experimental points are also occasionally just way off (such as times of
incidental hits). Points like this are called {\em outliers}. They can
easily turn a $\chi^{2}$ fit on otherwise adequate data into nonsense.
Their probability of occurrence in the assumed Gaussian model is so small
that the estimator is willing to distort the whole curve to try to bring
them, mistakenly, into line. Therefore, on the first iteration only PMTs
with hits of greater than 2 PE are used. On the next iterations the PE
restriction is gradually relaxed so that PMTs with weaker signals are
allowed to participate. At the same time, the definition of outliers
becomes more stringent. After the first iteration points with distance of
more than two standard deviations away from the plane are removed from the
fit on the second pass. Before the last, fifth, pass outliers are defined
as all hits with distance of more than 0.5 standard deviations away from
the plane. The results of the last iteration are recorded as the
reconstructed direction of the primary particle. The number of PMTs that
participated in the last pass are also noted as $N_{fit}$.

More sophisticated algorithms were also attempted, they, however, required
a significantly higher computer power without improving angular resolution
by a noticeable amount \cite{andy:likelihood}

% ////////////////////////////////////////////////////////////
\section{Core Location.}

The inability to make precise determination of the shower core impacts not
only energy estimation but also the direction reconstruction. Indeed, as
was mentioned in section \ref{chapter:eas:temporal}, the shower front is
not flat, but rather presents a paraboloid surface. Therefore, in order to
be able to use the described plane fit algorithm, the shower front has to
be ``unfolded'' into the plane. The amount of this curvature correction is
certainly core distance dependent. Failure to apply curvature correction
will result in a systematic error in the direction reconstruction which
increases with core displacement from the pond. Because there are many
showers with different core positions, this will lead to degradation of
average angular resolution.

Thus, for the purposes of angle determination, the following core locating
algorithm was adopted \cite{Greg:corefitter}. First the direction to the
core from the center of the pond is determined by calculating the location
of the provisional core. The provisional core position is found in two
iterations. Iteration one: the average PMT coordinate is calculated where
the weight is taken to be $\sqrt{PE}$ of the corresponding PMT. The second
iteration is the same as the first, but where only PMTs within $\pm
45^{\circ}$ sector of the direction found on the iteration one are used.
After the provisional core is found, the decision is made whether the
shower core is on the pond or it is outside of it. The decision is made on
the basis of profile of the average PE versus distance from the center of
the pond for PMTs within $\pm 60^{\circ}$ sector of the provisional core
direction. If it is declining sufficiently fast towards the edge of the
pond, the core is adopted to be on the pond. Otherwise, it is off. On-pond
cores are found as $PE$ weighted average of PMT coordinates in the circle
of 8 meters around provisional core. Off-pond cores are placed at a
distance of 50 meters from the center of the pond in the direction defined
by the provisional core. Only PMTs of the shower layer are used in this
algorithm. The value of 50 meters is chosen because simulations indicate
that this is the most probable core displacement for showers triggering
the detector.

In the early part of the detector running, a more primitive version of the
algorithm was used.

% ////////////////////////////////////////////////////////////
\section{Curvature Correction.}

The origin of the curvature correction is traced back to the desire of
using the plane model of the shower front in the direction reconstruction
and to the details of the shower development underlined in section
\ref{chapter:eas:temporal}: the shower front is parabolic. It was shown
that electron and photon components of the shower have different shapes of
the corresponding fronts (figure \ref{fig:eas:timing}). This means that
the shape of the front, as sampled by the array, depends on energy
threshold of the detectors and on the relative efficiency for photon
conversion. For this reasons, methods of determination of the curvature
correction from the real data were developed \cite{Joe_Bussy}. A
particular representation of the correction was chosen and then parameters
were optimized in a multistage iteration process. The correction was shown
to improve angular resolution compared to a plane fit.

% ////////////////////////////////////////////////////////////
\section{Sampling Correction.}

The curvature correction and core finding algorithms provide for primary
particle direction reconstruction given that the times of the arrival of
the shower front are measured by the detectors. However, as was mentioned
in section \ref{chapter:milagro:pmt}, each PMT has quantum efficiency of
about $0.2-0.25$. This means that the first particle, carrying information
about the shower front, is detected with some probability $p$. If this
particle is missed, the tube is presented a chance of detecting the second
particle in the shower and so on up to the total number of particles
$\rho(r)dxdy$ which make up the thickness of the front at the specific
core distance $r$. Thus, if $f_{n}(T,r)$ is the time distribution of the
$n$-th particle, $f_{1}(T,r)$ being the shower front, then the
distribution $g(T,r)$ of the detected arrival times is given by:

\[  g(T,r) = pf_{1}(T,r) +(1-p)pf_{2}(T,r)+\dots +
(1-p)^{n-1}pf_{n}(T,r)+\dots \]

If nothing else is known, the $g(T,r)$ has to be interpreted as the shower
front instead of $f_{1}(T,r)$. This would certainly lead to poorer angular
resolution.\footnote{Particles following one after another have their
arrival times ordered: $T_{1} < T_{2} < \dots < T_{n} < \dots$. Among
these times, $T_{1}$ obeys distribution $f_{1}$, $T_{2}$ obeys $f_{2}$ and
so on where corresponding variances are believed to increase with $n$.
Therefore the variance of $g(T,r)$ satisfies inequality $var(g(T,r)) \geq
var (f_{1}(T,r))$ with equal sign if and only if $g(T,r) \equiv
f_{1}(T,r)$ --- 100 \% detection efficiency. } The Milagro analysis tries
to exploit the correlation that particles trailing the front are of
monotonously decreasing energy, making it possible, effectively, to
estimate the order number $n$ of the detected particle. With regard to the
core distance dependence, each of the $f_{n}(T,r)$ is narrower where
density of particles $\rho(r)$ is higher, that is towards the core. Thus,
the sampling correction is aimed at inferring the arrival time of the
front needed for direction reconstruction, it is due to finite detection
efficiency of the PMTs and it is a function of core distance and light
intensity registered by the PMT. Such estimated front arrival times are
subject to higher fluctuations compared to direct measurements and
therefore are given less weight in the direction reconstruction algorithm
\cite{Joe_Bussy}.

% ////////////////////////////////////////////////////////////
\section{Cosmic Ray Rejection.}

The identification of the type of the primary particle (photon or cosmic
ray) in Milagro makes use of the presence of muons and hadrons in
cosmic-ray initiated cascades. In the top layer of the detector their
presence is obscured by the general illumination by the shower particles,
in the bottom layer, however, they may be noticed. Penetration of one of
these particles to the muon layer should lead to clusters of high light
intensities. In the case of a primary photon, the illumination of the
bottom layer should be uniform. It was found with the help of Monte Carlo
simulations that the parameter $X_{2}$ defined as

\[ X_{2} = \frac{N_{2}}{PEmax}\]

is sensitive to the type of the primary particle. Here $N_{2}$ is the
number of PMTs in the muon layer with registered light intensities greater
or equal to 2 PE, PEmax --- is the maximum intensity detected in the event
by the muon layer (\cite{Gaurang:gamma_proton} and
\cite{Gus:gamma_proton}). The simulated distributions of parameter $X_{2}$
for gamma- and proton- induced showers are presented on figure
\ref{fig:x2_distribution}. The $X_{2}$ distribution for the proton
dominated data (also shown on figure \ref{fig:x2_distribution}) agrees
with the simulated proton distribution well. This match is a check of the
Milagro Monte Carlo simulations.

% ////////////////////////////////////////////////////////////
% ======  Pictures.....===========
\begin{figure}   
\centering
\includegraphics[width=5.0in]{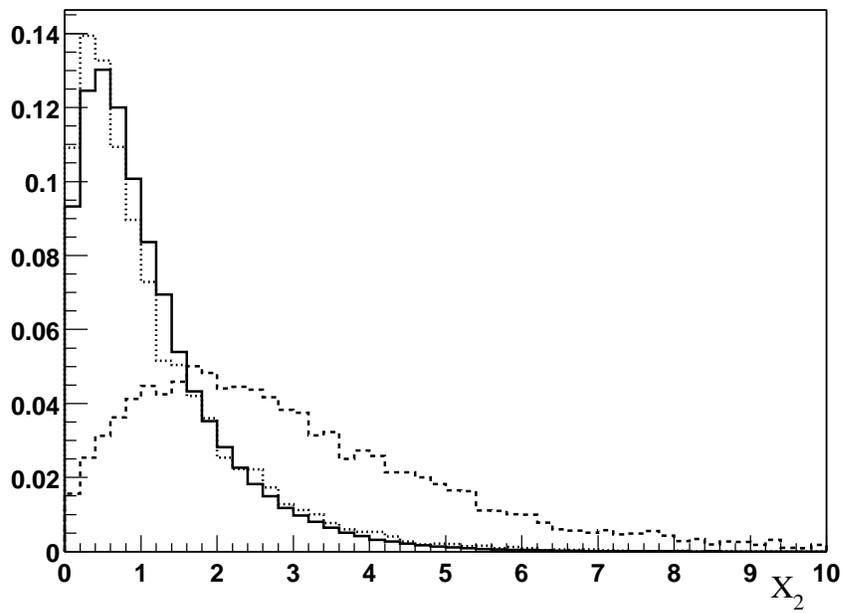}
\caption{Distributions of parameter $X_{2}$. Dashed line is Monte Carlo
simulations for photon induced showers, dotted line is simulation for
proton induced showers, solid line is the data. Each distribution is
normalized to 1.}
\label{fig:x2_distribution} 
\end{figure}

It is clear that events with large values of the parameter are more likely
to be gamma ray induced. It was determined that the optimal cut for gamma
rays is accepting events with $X_{2} > 2.5$. At this cut, about 90\% of
protons are rejected while about 50\% of gamma rays are accepted.

Attempts to develop a more complicated criterion have not yet yielded a
significant improvement over the present scheme.

\chapter{Performance of the Milagro Detector.}

% ////////////////////////////////////////////////////////////
\section{Operation of the Milagro.}

The Milagro detector operates largely unattended in a reliable and stable
manner. Automatic alerts are generated under serious error conditions such
as loss of electrical power, an abnormal event rate, or overheating of the
electronics. The nature of the error, time of the day and weather
conditions determine the response time and restart of the data taking.
Less serious problems can be corrected remotely. There are also scheduled
down times to accommodate repair activities and calibrations. Other than
that, data are acquired continuously. During operation of the detector
some PMT-electronics channels cease to work and have to be turned off.
This problem has typically been traced to water leakage of the under-water
connectors linking coaxial electric cable with the PMT. Repairs of these
connectors performed once a year are a major part of the scheduled
maintenance down time. The repair time is chosen to be September when the
pond water is the warmest to facilitate the scuba diving to retrieve PMTs.

% ////////////////////////////////////////////////////////////
\section{Angular Resolution.}

The cosmic ray shadow of the Moon has been used to measure the angular
resolution of an EAS array above 50 TeV \cite{cygnus:moon}. At TeV
energies, the geomagnetic field will displace the Moon cosmic ray shadow.
Another estimate of the angular resolution is obtained by studying
$\Delta_{eo}$ --- the difference between directions reconstructed by the
same ``color'' PMTs if the detector is imagined to be painted in the white
and black squares of a checkerboard. (The PMT numbering scheme was
designed in such a way that the color of the PMT is defined by the parity
(even/odd) of its number.) The quantity $\Delta_{eo}$ is not sensitive to
certain systematic effects, common to the both parts of the detector,
however, in the absence of these effects the dispersion of $\Delta_{eo}$
is twice the overall angular resolution \cite{cygnus:deleo}. From the
studies of the Moon shadow, the $\Delta_{eo}$ and the Monte Carlo
simulations, the estimated angular resolution of the instrument including
pointing effects is $0.75^{\circ}$ for all events with $N_{fit} > 20$.

% ////////////////////////////////////////////////////////////
\section{Absolute Energy Scale.}

The absolute energy scale can be determined by examining the magnetic
displacement of the shadow of the Moon. A preliminary estimate of the
median energy of the detected events is $640\pm70$ GeV which is in
excellent agreement with simulations which predict median energy of 690
GeV.

% ////////////////////////////////////////////////////////////
\section{Cosmic Ray Rejection at Work.}

All observations to date indicate that the flux of gamma rays from the
Crab nebula is constant which makes it a very useful source for testing
the sensitivity of different instruments. The cosmic ray rejection method
was tested on the Crab. If the data is analyzed without application of the
$X_{2}$ cut, the Crab is observed with significance of 1.4. Analysis of
the data passing the rejection criterion yields significance of 5.4. (The
notion of significance is discussed in section
\ref{chapter:technique:sinificance}.) The data set used in this study
covers the period between June 8, 1999 and April 1, 2002
\cite{milagro_crab}.

\chapter{Data Analysis Technique.\label{chapter:technique}}

%////////////////////////////////////////////////
\section{Coordinate Systems on the Celestial Sphere.}

%////////////////////////////////////////////////
\subsection{Equatorial Coordinate System.}

Inasmuch as distances to the majority of astronomical objects are much
bigger than the size of the Earth and its orbit, it is sufficient to
specify directions to the objects regarding distances as equal and
infinitely large. Within this framework, the stars appear to be located on
the surface of an imaginary sphere with the observer at its center. This
sphere is called the {\em Celestial sphere}. Various coordinate systems on
the sphere are used, all defined by the corresponding choices of reference
circles. {\em Horizon} and {\em equatorial} coordinate systems are
important examples employed in this work \cite{sky_coordinate_definiton}.
Figure \ref{fig:coords:definition} illustrates the following definitions.

% ////////////////////////////////////////////////////////////
% ======  Pictures.....===========
\begin{figure}
\centering
\includegraphics[width=5in]{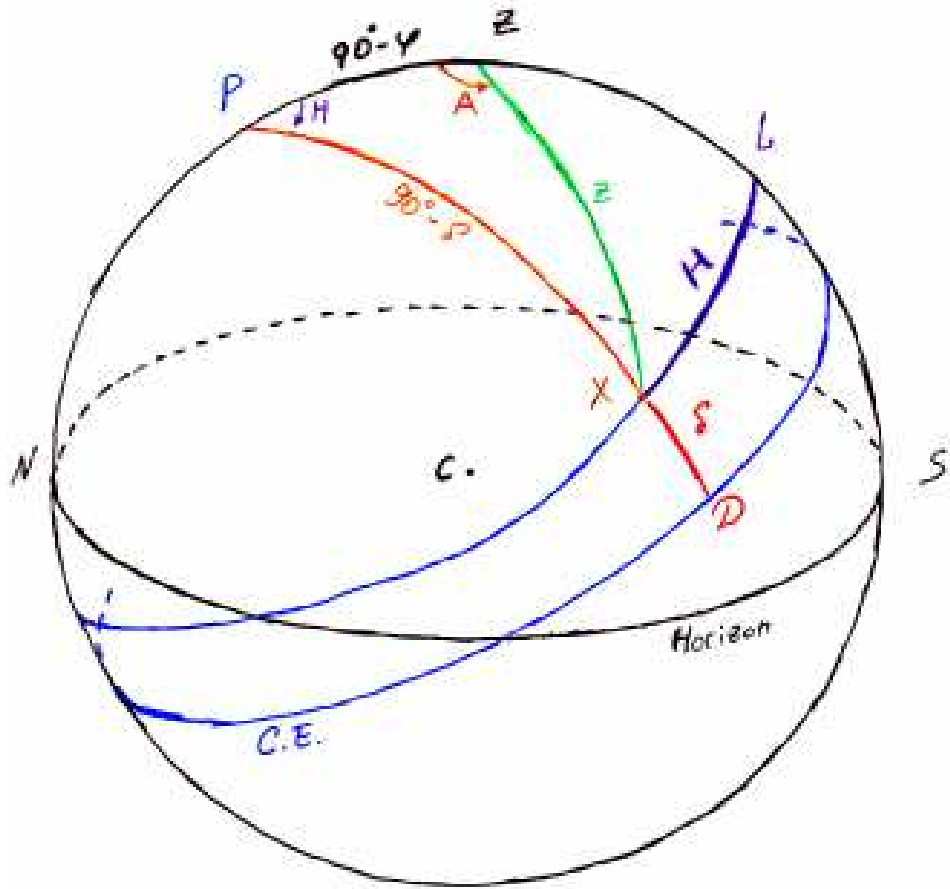}
\caption{Definitions of horizon and equatorial coordinate systems.}
\label{fig:coords:definition}
\end{figure}

\begin{description}

\item[$C$]  - the observer.   

\vspace{-0.1in}
\item[$CP$] - line parallel to the axis of rotation of the Earth.

\vspace{-0.1in}
\item[$P$] - North celestial pole.

\vspace{-0.1in}
\item[{\em Celestial Equator}] - is defined as intersection of a plane
perpendicular to $CP$ at point $C$ and the celestial sphere. (This is the
projection of Earth's equator.)

\vspace{-0.1in}
\item[$Z$] - the zenith, intersection of the celestial sphere with the
outward continuation of the plumb line at the observer's location.

\vspace{-0.1in}
\item[$\angle PCZ = 90^{\circ} - \phi$] - definition of $\phi$. Angle
$\phi$ is astronomical latitude of the observer on the Earth.

\vspace{-0.1in}
\item[{\em Horizon}]- is defined as intersection of a plane perpendicular
to $CZ$ at point $C$ and the celestial sphere.

\vspace{-0.1in}
\item[$N,S$] - are North and South of the {\em Horizon}. $N$ and $S$ are
defined as the intersection of a great circle $PZ$ centered at observer
$C$ with the {\em Horizon}. Arc $PZLS$ is called {\em local reference
celestial meridian}.

\vspace{-0.1in}
\item[$X$] - is a star.

\vspace{-0.1in}
\item[$\angle ZCX=\stackrel{\smile}{ZX}=z$] - zenith distance of star $X$.

%\vspace{-0.1in}
\item[$A$] - azimuth, is a dihedral angle between reference meridian and 
$ZXC$ plane.

\vspace{-0.1in}
\item[$\angle LCX=\stackrel{\smile}{LX}=H$] - hour angle, is a dihedral
angle between reference celestial meridian and that of the star $PXD$.

\vspace{-0.1in}
\item[$\angle XCD=\stackrel{\smile}{XD}=\delta$] - is Declination of a
star (Dec.).

\end{description}

The equatorial coordinate system of hour angle and declination
$(H,\delta)$ is built around the axis of rotation of the Earth whereas the
horizon one of zenith and azimuth $(A,z)$ uses a plumb line as the
reference. The law of cosines for trihedral angles applied to the
spherical triangle $XPZ$ (figure \ref{fig:coords:definition}) two times
yields the relation between the systems:

\[ \cos ( 90^{\circ}-\delta ) = \cos z \cos (90^{\circ}-\phi) + \sin z
\sin (90^{\circ}-\phi) \cos A  \]
\[ \cos z  = \cos (90^{\circ}-\delta) \cos (90^{\circ}-\phi) + \sin
(90^{\circ}-\delta) \sin (90^{\circ}-\phi) \cos H  \]

or,

\[ \sin \delta = \sin \phi \cos z + \cos \phi \sin z \cos A  \]
\[ \tan H = \frac{\sin z \sin A}{\cos \phi \cos z - \sin z \sin \phi \cos
A} \]

The value of observer's latitude $\phi$ defined as the complement of the
angle $\angle PCZ$ is assumed to be known. Both of these coordinate
systems are {\em local} coordinate systems in the sense that both of them
revolve with the Earth. If the object $X$ is stationary in space, due to
the Earth's rotation it will appear to be moving and its local coordinates
will be changing. With this motion, the declination $\delta$ of the
stationary object in the equatorial system will remain constant, while its
hour angle $H$ will change incrementing by $360^{\circ}$ when the Earth
makes one full revolution around its axis $CP$. The time required to
complete such a revolution is called {\em sidereal day}. In contrast, the
{\em solar day} or {\em universal day} is defined as the time between two
appearances of the Sun on the local reference celestial meridian. The
universal time is different from the sidereal time due to Earth's orbital
motion around the Sun.

{\em Equatorial celestial} coordinate system is defined as declination
$\delta$ and right ascension $\alpha$ of the object, where $\alpha =
H_{\Upsilon} - H$, $H_{\Upsilon}$ is the hour angle of the
vernal equinox. This coordinate system is independent of the Earth's
rotation and observer's longitude. {\em Local sidereal time} is defined as
$H_{\Upsilon}$ expressed in the units of time.

Local sidereal time as well as the geodesic latitude of Milagro are
provided by the GPS system clock. In general, there is no exact relation
between astronomical latitude defined above and the geodesic one, but
according to indications in literature \cite{geodesy} the difference is of
the order of $15''$ in magnitude, much smaller than the angular resolution
of the detector and thus can be neglected.

%////////////////////////////////////////////////
\subsection{Galactic Coordinate system.}

% ======  Pictures.....===========
\begin{figure}
\centering
\includegraphics[width=2.9in]{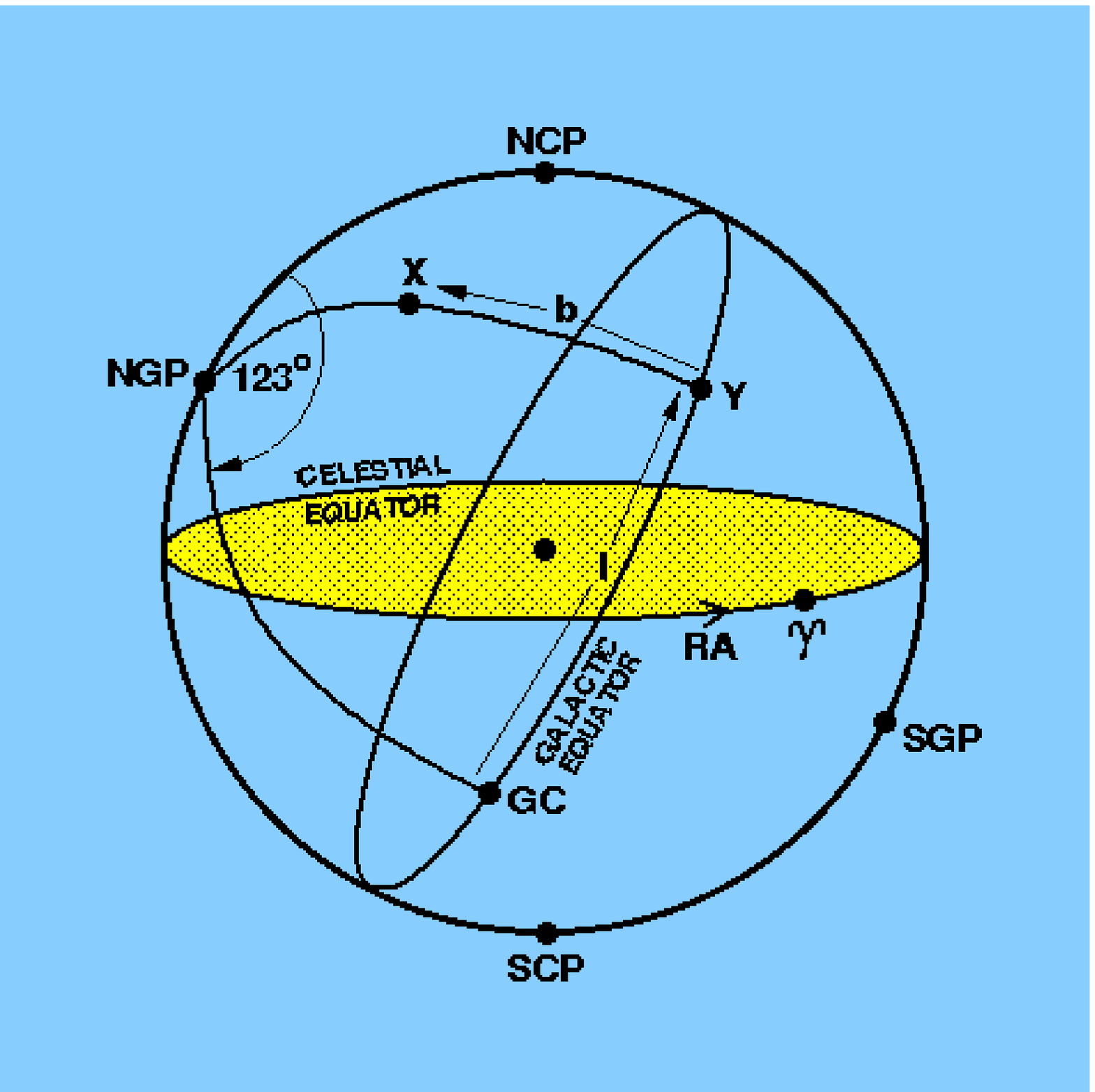}
\caption{Definition of Galactic coordinate system in its relation to the
equatorial coordinate system.}
\label{fig:galactic_defintions}
\end{figure}

% ////////////////////////////////////////////////////////////
\begin{figure}
\centering
  \subfigure[View of the Galaxy from the North Galactic pole. Galactic
             longitude grid is shown. The Galaxy, at the Earth's position,
             is rotating toward the direction of $l=90^{\circ}$.]{
    \label{fig:lagactic:longitude}
    \includegraphics[width=2.7in]{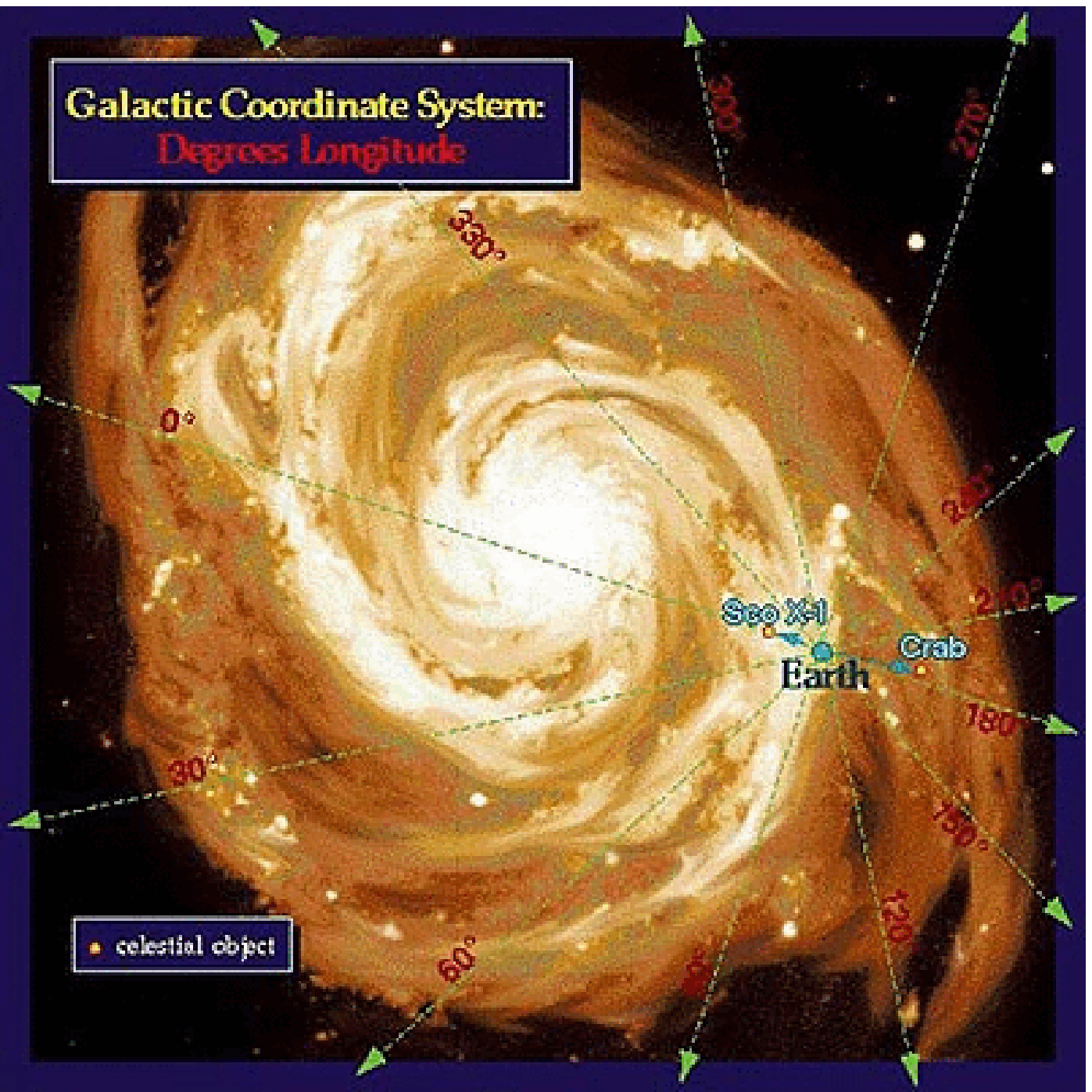}
  }
  \subfigure[Side view of the Galaxy. Galactic latitude grid is shown. The
             arrow indicates the direction of increment of Galactic
             longitude which is opposite to the Galactic rotation.]{
    \label{fig:lagactic:latitude}
    \includegraphics[width=2.7in]{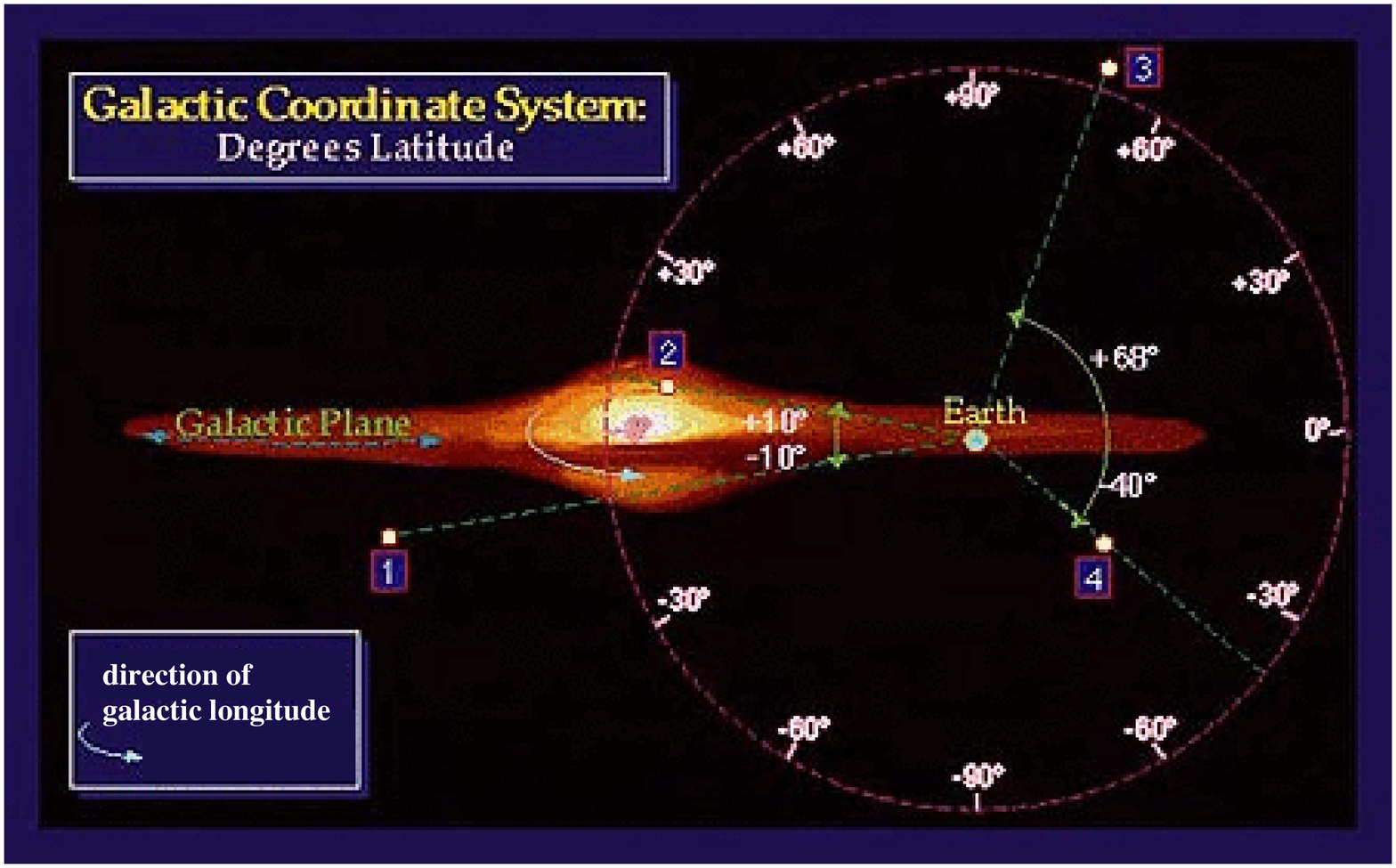}
  }
\caption{Artist view of the Milky Way Galaxy from aside.} 
\label{fig:lagactic}
\end{figure}

The reference plane of the galactic coordinate system is the disc of our
Galaxy (i.e. the Milky Way) and the intersection of this plane with the
celestial sphere is known as the galactic equator, which is inclined by
about $63^{\circ}$ to the celestial equator. Galactic latitude, $b$, is
analogous to declination, but measures distance north or south of the
galactic equator, attaining $+90^{\circ}$ at the north galactic pole (NGP)
and $-90^{\circ}$ at the south galactic pole (SGP). The galactic latitude
of the star X on the figure is arc YX and is north (figure
\ref{fig:galactic_defintions}).

Galactic longitude, $l$, is analogous to right ascension and is measured
along the galactic equator in the same direction as right
ascension.\footnote{This sense of rotation, however, is opposite to the
sense of rotation of our Galaxy. The Sun, together with the whole Solar
System, is orbiting the Galactic Center at the distance 28,000 light
years, on a nearly circular orbit, moving at about 250 km/sec. It takes
about 220 million years to complete one orbit (so the Solar System has
orbited the Galactic Center about 20 to 21 times since its formation about
4.6 billion years ago). Considering the sense of rotation, the Galaxy, at
the Sun's position, is rotating toward the direction of $l=90^{\circ}$,
$b=0^{\circ}$. Therefore, the galactic north pole, defined by the galactic
coordinate system, coincides with the rotational south pole of our Galaxy,
and vice versa.} The zero-point of galactic longitude is in the direction
of the Galactic Center (GC), in the constellation of Sagittarius; it is
defined precisely by taking the galactic longitude of the north celestial
pole to be exactly $123^{\circ}$. The galactic longitude of the star X on
the figure is given by the angle between GC and Y.

The galactic north pole is at RA = 12:51.4, Dec = +27:07 (2000.0), the
galactic center at RA = 17:45.6, Dec = -28:56 (2000.0). The inclination of
the galactic equator to Earth's equator is thus $62.9^{\circ}$. The
intersection, or node line of the two equators is at RA = 18:51.4, Dec =
0:00 (2000.0), and at $l = 33^{\circ}$, $b=0^{\circ}$
\cite{galactic_coefficients}. Transformation between celestial coordinate
system and galactic one are given, therefore, by
\cite{handbook_space_astro}:

\vspace{+0.1in}
\begin{tabular}{rcl}

$\cos b \cos (l-33^{\circ})$  & = & $ \cos \delta \cos(\alpha - 282.85^{\circ})$ \\

$\cos b \sin (l-33^{\circ})$  & = & $\cos \delta \sin(\alpha - 282.85^{\circ}) \cos 62.9^{\circ}
                                    + \sin \delta \sin 62.9^{\circ}$ \\
$\sin b $                     & = & $\sin \delta \cos 62.9^{\circ} 
                                    - \cos \delta \sin(\alpha - 282.85^{\circ}) \sin 62.9^{\circ}$ \\
\end{tabular}

%////////////////////////////////////////////////
\section{Significance of a Measurement.\label{chapter:technique:sinificance}}

According to its definition \cite{neyman_pearson}, a {\em statistical
hypothesis} is a theoretical prediction or assertion about the
distribution of one or several measurable quantities. It is therefore
desired to devise a rule of checking whether the data are consistent with
the hypothesis, called {\em null hypothesis} $H_{0}$. Such a rule,
however, can not be hoped to tell whether the hypothesis is true or false,
it can only insure that in the long run of experience wrong propositions
will not be accepted too often. A naive method of testing is to calculate
the probability that a certain character $x$ of the observed data would
arise if $H_{0}$ were true. If the probability is small, this is
considered as an indication that the hypothesis is false and it is
rejected. However, if $x$ is a continuous variable, then probability of
any value of $x$ is zero. Therefore, tests are based on the notion of
critical region. The {\em critical region} $w$ is the subset of space $W$
of all possible values of $x$ such that $x$ falling into $w$ leads to the
rejection of the hypothesis. The hypothesis is accepted or remains in
doubt in all other cases. It follows then, that if there are two
alternative tests for the same hypothesis, the difference between them
consists in the difference in critical regions. It also follows that
$H_{0}$ can be rejected when, in fact, it is true ({\em error of the first
kind}); or it can be accepted when some other {\em alternative hypothesis}
$H_{1}$ is true ({\em error of the second kind}). (Admittance of the
existence of an alternative hypothesis is clear, otherwise the null
hypothesis would not be questioned.) The probabilities of errors of these
two kinds depend on the choice of the critical region. The critical region
is said to be the best for testing hypothesis $H_{0}$ with regard to
$H_{1}$ if it is the one which minimizes the probability of the error of
the second kind among all regions which give the same fixed value of the
probability of the error of the first kind. The construction of the best
critical region, resulting in the most efficient test of $H_{0}$ with
regard to $H_{1}$ was considered by Neyman and Pearson in
\cite{neyman_pearson}. The problem is solved for the general case of
simple hypotheses. A hypothesis is said to be {\em simple} if it
completely specifies the probability of the event; it is {\em composite}
if the probability is given only up to some unspecified parameters.

Critical regions $w(\epsilon)$ corresponding to different probabilities
$\epsilon$ of errors of the first kind are engineered before the test is
performed. When experimental data $x$ is obtained, the smallest $\epsilon$
is found such that $x \in w(\epsilon)$. It is then said, that based on the
experimental data, the null hypothesis can be rejected with {\em
significance} $\epsilon$.

\vspace{+0.5in}

A typical astronomical experiment consists of two independent observations
of some regions of the sky yielding two measured counts $N_{1}$ and
$N_{2}$ made during time periods $t_{1}$ and $t_{2}$ respectively with all
other conditions being equal. The null hypothesis being tested is that
$N_{1}$ and $N_{2}$ constitute a sample of size 2 from a single Poisson
distribution (adjusted for the duration of observations) which is due to
common illumination. The alternative is that the two measurements are due
to Poisson distributions with unrelated parameters. In other words, the
alternative hypothesis $H_{1}$ is of the form:

\[ H_{1}: \; \; p_{1}(N_{1}, N_{2}) = 
           \frac{(\lambda_{1}t_{1})^{N_{1}}}{N_{1}!}
                   e^{-\lambda_{1} t_{1}}
           \frac{(\lambda_{2} t_{2})^{N_{2}}}{N_{2}!}
                        e^{-\lambda_{2} t_{2}}
\]

while the null hypothesis is also of the composite type, obtained from the
$H_{1}$ by setting $\lambda_{1}=\lambda_{2}$:

\[ H_{0}: \; \; p_{0}(N_{1}, N_{2}) =
                              \frac{(\lambda t_{1})^{N_{1}}}{N_{1}!}
                              \frac{(\lambda t_{2})^{N_{2}}}{N_{2}!}
                              e^{-\lambda (t_{1}+t_{2})}
\]

that both measurements are due to the same unknown count rate
$\lambda=\lambda_{1}=\lambda_{2}$.

Despite the fact that $p_{0}(N_{1}, N_{2})$ admits the existence of
regions similar to the sample space $W$ with regard to the parameter
\footnote{ According to publication \cite{neyman_pearson}, the region $w$
is called {\em region of size $\epsilon$ similar to sample space $W$ with
regard to parameter $\beta$} if

\[ \int_{w} p_{0}(x, \beta) dx = \epsilon = const \]

for all values of parameter $\beta$. The hypothesis $p_{0}(N_{1}, N_{2})$
satisfies the special case considered in \cite{neyman_pearson}, for which

\[ \frac{\partial \phi}{\partial \beta} = A + B \phi \]

where $\phi = \frac{1}{p_{0}} \frac{\partial p_{0}}{\partial \beta}$,
coefficients $A$ and $B$ may depend on $\beta$, but not on $x$. For the
case of interest, $\beta = \lambda$, $\phi = \frac{N_{1}+N_{2}}{\lambda} 
- 2$, $A= -\frac{2}{\lambda}$, $ B= -\frac{1}{\lambda}$.
} 
$\lambda$, no common best critical region with the respect to every
alternative admissible $(\lambda_{1} \ne \lambda_{2})$ hypothesis
$p_{1}(N_{1}, N_{2})$ can be found.

Therefore, the following practical method is adopted which is to consider

\begin{equation}
 S = \frac{N_{1} - \alpha N_{2}}{\sqrt{N_{1}+\alpha^{2} N_{2}}},
   \; \;\;\;\;  \alpha = t_{1}/t_{2}
\label{equation:significance}
\end{equation}

Arguments can be made (appendix \ref{appendix:poisson_arguments}) that for
large $N_{1}$ and $N_{2}$, under the null hypothesis, $S$ obeys the
standard normal distribution in the vicinity of zero. The bounds of this
approximation $-S_{0} < S < S_{0}$ enlarge when smallest of the $N$'s
increases (equation \ref{equation:lima_bound}). The critical regions
$w(\epsilon)$ with respect to the alternative hypothesis $\lambda_{1} >
\lambda_{2}$ are, therefore, defined as $S > S(\epsilon)$, where:

\[
   \epsilon = \frac{1}{\sqrt{2\pi}} 
              \int_{S(\epsilon)}^{\infty}e^{-\frac{x^{2}}{2}} dx 
\]

 and provide the way to calculate the significance of the measurement
$\epsilon$. (The critical regions $w(\epsilon)$ with respect to the
alternative hypothesis $\lambda_{1} < \lambda_{2}$ are defined as $S <
S(\epsilon)$, $\epsilon = \frac{1}{\sqrt{2\pi}}
            \int_{-\infty}^{S(\epsilon)}e^{-\frac{x^{2}}{2}} dx$.) Because
of the one-to-one correspondence between $\epsilon$ and $S(\epsilon)$, the
significance of a measurement can be quoted in the units of $S$ (see 
table \ref{table:significance_probability}). The error
on the value of $\epsilon$ due to this approximation is less than $
\frac{1}{\sqrt{2\pi}}
              \int_{S_{0}}^{\infty}e^{-\frac{x^{2}}{2}} dx$.

\begin{table}[htbp]
\begin{center}
\begin{tabular}{|c|c|} \hline
 $|S(\epsilon)|$ &    $\epsilon$    \\ \hline
       1.0       &   0.15866 \\
       2.0       &   0.02275 \\
       2.5       &   $6.21  \cdot 10^{-3}$  \\
       3.0       &   $1.35  \cdot 10^{-3}$  \\
       3.5       &   $2.326 \cdot 10^{-4}$  \\
       4.0       &   $3.17  \cdot 10^{-5}$  \\ \hline

\end{tabular}
\end{center}
\caption{The correspondence between $|S|$ and $\epsilon$ for the
         alternative hypothesis $\lambda_{1} > \lambda_{2}$ or
         $\lambda_{1} < \lambda_{2}$ \cite{korn_korn}.}
\label{table:significance_probability}
\end{table}

%////////////////////////////////////////////////
\section{Effective Area.}

Considering the detector and the atmosphere together as parts of a giant
apparatus, it is obvious that not all particles entering the atmosphere
will trigger the detector. (Particles on the other side of the Earth will 
not be registered by the detector.) Thus, it is meaningful to speak about
detection probability, or more precisely, about conditional probability
for a particle to be detected given its type, its direction of incidence,
its energy and displacement of its trajectory from the detector's center
(core position). Then, the number of detected events is given by:

\[ N_{obs} = \sum_{k} \int P(detect|k, E, \Theta, x,y) \;
                           F(k, E, \Theta) \; dE \; d\Theta \; dx \; dy \]

where $F(k, E, \Theta)$ is the signal function, the total number of
particles of $k$-th kind emitted with energies between $E$ and $E+dE$ in
the directions $\Theta \div \Theta + d\Theta$ of local sky and landing in
the area $dxdy$ during the measurement period. Note, that the signal
function does not depend on the core position $(x,y)$, which is a
reasonable assumption: landing coordinates on the Earth have nothing to do
with the emission process in the sky. Therefore, the integration over core
positions can be performed and the result is called the {\em effective
area}:

\begin{equation}
\int_{-\infty}^{+\infty} P(detect|k, E, \Theta, x,y) dxdy =
 A_{eff}(k, E, \Theta)
\label{equation:effective_area:definition}
\end{equation}

\begin{equation}
N_{obs}  = \sum_{k} \int A_{eff}(k, E, \Theta) F(k, E, \Theta)
                             \; dE \; d\Theta
\label{equation:effective_area:Nobserved}
\end{equation}

The term effective area is used because $A_{eff}(k, E, \Theta)$ has units
of area and can be interpreted as the geometrical area of a fictitious
detector with detection efficiency of 100\% independent of a particle's
core position over its extent and zero outside. Because air showers may
have lateral extent of 300 meters or more, they may trigger the detector
even if the core is very far from the detector. That is why effective area
can be much larger than the geometrical area of a real detector. Due to
the fact that the signal function $F(k, E, \Theta)$ and the number of
observed events $N_{obs}$ are related via equation
\ref{equation:effective_area:Nobserved}, effective area becomes an
important characteristic of a detector. Detectors of relatively small
sizes can be calibrated using test beams of particles of interest and the
effective areas can be measured experimentally. For ground based detectors
such as Milagro where the Earth's atmosphere is a component of the
detector the effective area has to be computed with the help of computer
simulations. Definition \ref{equation:effective_area:definition} of
effective area provides a direct method for its determination: Monte Carlo
integration. To do this, $N_{total}$ showers are simulated with core
distances chosen uniformly over a sufficiently large area $A_{0}$ and
those which trigger the detector are counted.

\[ 
  A_{eff}(k, E, \Theta) = \frac{N_{trigg}(k, E, \Theta)}{N_{total}(k, E, \Theta)} A_{0}
\]

Such Monte Carlo integrations were performed for photon and proton
primaries with energies in the range of 0.1TeV-100TeV and zenith angles
between $0^{\circ}$ and $45^{\circ}$. Appropriate tables are saved and
used in the analysis. (Dependence of the effective area on the azimuth
angle is disregarded in this calculation, therefore, in what follows the
effective area is denoted as $A_{\gamma, cr}(z,E)$.) It is important to
note that because the atmosphere can be considered as one of the
constituents of the apparatus, changes of its conditions can lead to
variations of the effective area.

%////////////////////////////////////////////////
\section{Determination of Gamma Ray Flux from a Source.}

When the presence of a source is established by means of the hypothesis
test of section \ref{chapter:technique:sinificance}, it is desirable to
measure the features of the source, the parameters of its signal function.

In the case of a point source, the signal function has the form

\[
 F(E,z) = \int_{observation \; period} \Phi(E,t) \delta(z-z(t)) dt
\]

where $\Phi(E,t)$ is the differential flux of the emitted particles
\linebreak $([\Phi(E,t)]=\frac{1}{m^{2}\cdot sec \cdot eV})$ and $z(t)$ is
the trajectory of the source in the local sky ($z$ is the zenith
coordinate). If the source is steady, $\Phi$ does not depend on time and
the signal function becomes:

\[ 
 F(E,z) = \Phi(E) \cdot T(z)
\]

where $T(z)$ is the effective amount of time spent by the source in the
zenith region between $z$ and $z+dz$. The case of an extended source may
be reduced to a collection of point sources, and if it is uniform, it will
be absorbed by $T(z)$. The physical processes responsible for the gamma
ray emission from the Galactic plane, as well as those of the cosmic rays
are believed to be steady and are believed to produce power law
differential flux type spectrum:

\[ 
    \Phi_{k}(E) \sim E^{-\alpha_{k}}
\]

\[
 \Phi_{k}(E) = \frac{\alpha_{k}-1}{E_{0}^{-\alpha_{k}+1}}
               F_{k}(>E_{0}) \cdot E^{-\alpha_{k}}
\]

leading to the signal function prototype:

\begin{equation}
 F_{k}(E,z) = \frac{\alpha_{k}-1}{E_{0}^{-\alpha_{k}+1}}
              F_{k}(>E_{0}) \cdot E^{-\alpha_{k}} \cdot T(z)
 \label{equation:source_function_prototype}
\end{equation}

where $F_{k}(>E_{0})$ is the integral flux of particles with energies
greater than $E_{0}$, $F_{k}(>E_{0}) =
\int_{E_{0}}^{\infty}\Phi_{k}(E)dE$. Then, the number of gamma ray and
cosmic ray induced events from the source region can be calculated with
the help of equation \ref{equation:effective_area:Nobserved}:

\[ N_{\gamma} = \frac{\alpha_{\gamma}-1}{E_{0}^{-\alpha_{\gamma}+1}}
                F_{\gamma}(>E_{0}) \int \int
                A_{\gamma}(z,E) \; E^{-\alpha_{\gamma}} \; T(z) \; dz \; dE
\]

\[ N_{cr} = \frac{\alpha_{cr}-1}{E_{0}^{-\alpha_{cr}+1}}
                F_{cr}(>E_{0}) \int \int
                A_{cr}(z,E) \; E^{-\alpha_{cr}} \; T(z) \; dz \; dE
\]

Then the ratio of the integral gamma and cosmic ray fluxes from the source
is related to the ratio $ \frac{N_{\gamma}}{N_{cr}}$, the
experimentally available quantity by:

\begin{equation}
  \frac{F_{\gamma}(>E_{0})}{F_{cr}(>E_{0})} =
   \frac{1}{\eta(\alpha_{\gamma}, \alpha_{cr})}  \;
   \frac{N_{\gamma}}{N_{cr}}
\label{equation:flux_ratio}
\end{equation}

where 

\[
  \eta(\alpha_{\gamma}, \alpha_{cr}) =
                             \frac{\alpha_{\gamma}-1}{\alpha_{cr}-1} \;
                             E_{0}^{\alpha_{\gamma}-\alpha_{cr}} \; 
       \frac{\int\int A_{\gamma}(z,E)\;E^{-\alpha_{\gamma}}\;T(z)\;dz\;dE}
            {\int\int A_{cr}(z,E)\;E^{-\alpha_{cr}}\;T(z)\;dz\;dE}
\]

The dimensionless coefficient $\eta(\alpha_{\gamma}, \alpha_{cr})$ is the
energy and transit averaged ratio of effective areas for the given source.
The effective areas were obtained with the Monte Carlo simulation,
however, the ratio of them $\eta$ is believed to be less sensitive to
possible imperfections of the simulations than effective areas themselves.

Formula (\ref{equation:flux_ratio}) is the principal formula for
interpretation of measurements for the assumed signal function
(\ref{equation:source_function_prototype}).

\chapter{Background Estimation. Null Hypothesis.}

%/////////////////////////////////////////////////
\section{Time Swapping Method.}

Most air showers detected are produced by charged cosmic rays that form a
background (chapter \ref{chapter:eas}) to the search for gamma initiated
showers from a source. Because of their charge and because of the presence
of random magnetic fields in the interstellar medium, the cosmic ray
particles lose all memory of their initial directions and sites of
production, and can be regarded as forming isotropic radiation. Emission
from a gamma ray source would appear as an excess number of events coming
from the direction of the source. Therefore, a search for emission is a
statistical test with the null hypothesis that there is no source
conducted by means of two observations.  Indeed, one of the measurements
could be direct counting of the events in the angular bin in celestial
coordinates containing the source and the other observation, called
off-source, providing information about background level in the
neighborhood of the source bin. If according to the results of the test
the two measurements are inconsistent with each other, it is said that the
presence of the gamma ray emission is established. Otherwise, the
measurements do not contradict to the null hypothesis that the
observations are due to isotropic background and no source detection can
be claimed. This, however, does not preclude its existence. In other
words, chapter \ref{chapter:technique} is the navigation map for source
detection: identify the candidate source by its coordinates on the sky,
perform measurements and calculate significance, if significant, estimate
of the source function parameters can be tried.

Special consideration must be given to the significance test (section
\ref{chapter:technique:sinificance}) because of the ambiguous statement
that ``two independent observations ... with all other conditions being
equal''.  The off-source observation can be performed at the same time as
the on-source one utilizing the wide field of view of the detector, or it
can be performed at a different time making measurement in the same
directions of the field of view. (Due to the Earth's rotation, the
off-source bin may present itself in the directions of local coordinates,
previously pointed at by the source bin.) Both of these stipulations
contradict to the conditions of ''being equal'': if observations are done
at the same time, then non-uniformity in the acceptance of the array to
air showers due to detector geometry must be compensated for; if
observations are done at different times, then uniform operation of the
detector must be assured. The mechanism of such an equalization is called
{\em background estimation}.

The widely accepted method of background estimation
\cite{cygnus_methods,flyeye_methods} follows the second path by
recognizing that no major changes in the detector configuration are
usually made on the short time scale and by taking advantage of the
rotation of the Earth. Therefore, and if the cosmic ray background is due
to isotropic radiation, the number of detected events as a function of
local coordinates $x$ and time $t$ can be written in the form:

\begin{equation}
 d N(x,t) = G(x) \cdot R(t) \; dx dt
\label{equation:sloshing:stability_assumption}
\end{equation}

Here $R(t)$ is overall event rate, $G(x)$ --- acceptance of the array such
that $\int_{field \; of \; view} G(x) \; dx = 1$. The number of background
events in the source bin, is then given by

\begin{equation}
 N_{b} = \int \int (1-\phi(x,t)) \; G(x) \; R(t) \; dx dt
\label{equation:sloshing:background}
\end{equation}

where $\phi(x,t)$ is equal to zero if $x$ and $t$ are such that it
translates into inside of the source bin, and is one otherwise.

In the context of section \ref{chapter:technique:sinificance}, the number
of events $N_{s}$ obtained from the observation of the source celestial
bin can be considered as $N_{1}$ and number $N_{b}$ of estimated
background events can be considered as $\alpha N_{2}$. Then equation
\ref{equation:significance} can be written as:

\begin{equation}
  S = \frac{N_{s} - N_{b}}{\sqrt{N_{s} + \alpha N_{b}}}
\label{equation:sloshing:significance}
\end{equation}

where $\alpha$ is the ratio of effective exposures of the two measurement
as before.

The task now is to evaluate integral in
(\ref{equation:sloshing:background}). Recall \cite{numerical_recipes} that
Monte Carlo integration of $\int_{V} g(x)f(x)dx$ can be performed by
generating $x$ uniformly distributed over the area of integration $V$ and
then calculating the sum:

\[
  \int_{V} g(x)f(x)dx = \frac{V}{N} \sum_{i=1}^{N} g(x_{i})f(x_{i})
\]

where $N$ is the sample size generated. However, if $f(x)$ can be
interpreted as probability density function, then integration can be
performed by generating $x$ according to distribution $f(x)$ and
calculating sum:

\[  
  \int_{V} g(x)f(x)dx = \frac{1}{N} \sum_{i=1}^{N} g(x_{i})
\]

The proof is evident if the integration variable is changed to
$y=\int_{-\infty}^{x}f(x)dx$, $y \in [0,1]$. Therefore, if $N_{0}$ is
total number of events detected during integration time of equation
\ref{equation:sloshing:background}, $N_{0} = \int R(t) dt$, then,
introducing $r(t)= R(t)/N_{0}$, both $G(x)$ and $r(t)$ can be interpreted
as probability density functions in corresponding spaces and integration
of (\ref{equation:sloshing:background}) can be done by means of Monte
Carlo:

\begin{equation}
  N_{b}= \frac{N_{0}}{N} \sum_{i=1}^{N} (1-\phi(x_{i},t_{i}))
\label{equation:sloshing:montecarlo}
\end{equation}

where $(x,t)$ are distributed according to joint probability density
$G(x)r(t)$. A list of all coordinates of the detected events is regarded
as a sample from the $G(x)$ distribution, while list of all times --- as
the one from $r(t)$. Therefore, sample from $G(x)r(t)$ distribution can be
generated from the data by randomly associating an event's local
coordinate $x$ with an event's time $t$ among the pool of detected events.
The so created coordinate-time pair is called a {\em generated event}. The
accuracy of Monte Carlo integration increases as the square root of the
number of generated events.

In practice, Monte Carlo integration is performed by substituting each
real event's arrival time by a new time from the list of registered times
of collected events in a finite time window. That is why the method is
referred to as {\em time swapping method}. The swapping is repeated
$\beta$ times per each real event, $\beta$ typically being around 10.
Then, $N_{b}$ is estimated as number of generated events in the source bin
divided by $\beta$ (equation \ref{equation:sloshing:montecarlo}). The
method entails that the second, off-source, celestial bin is defined by
the regions of the sky which have the opportunity to present themselves
into the same set of local coordinates as the source bin does due to the
Earth's rotation during the time period of integration.

%/////////////////////////////////////////////////
\section{Accounting for Signal Events During Swapping.
                        \label{section:background:signal_events}}

Despite the fact that the time swapping method possesses the desired
equalization properties, namely, the generated events all obey correct
local angle distribution and correct timing distribution, even if the
detector stability assumption (equation
\ref{equation:sloshing:stability_assumption}) holds, the application of
the method introduces difficulties. Indeed, equation
(\ref{equation:sloshing:significance}) for significance of a measurement
was obtained with the assumption of independence of $N_{1}$ and $N_{2}$.
The time swapping realization of background estimation as described
includes on-source events $N_{s}$ in the calculation of $N_{b}$. The
problem can be brought to an extreme as in the case of sighting of the
Polar Star. In this case, the source does not present any apparent motion
in local coordinates because it lies on the axis of rotation of the Earth.
The off-source bin does not exist and measurement can not be performed. In
the framework of the time swapping method, the background $N_{b}$ will be
estimated to be exactly equal to $N_{s}$ reporting zero significance. This
is clearly unsatisfactory. There are two directions to proceed: one ---
modify the significance calculation to reflect the dependence of
measurements, two --- modify the time swapping method to regain
independence while keeping all other properties intact. The latter one is
followed in the current analysis.

To regain independence of the two measurements, the events from the source
bin should not participate in the background estimation. However, simply
removing these events from the swapping procedure will destroy its
foundation that the list of local coordinates and times represent samples
from $G(x)$ and $R(t)$ respectively. This problem is solved by the
following algorithm \cite{zoshka}.

Denote by $N_{out}(x,t)$ the number of detected events originated outside
of the source bin, and $R_{out}(t)$ their total event rate, then it is
readily seen that

\[
  d N_{out}(x,t) = \phi(x,t) G(x) R(t) \; dx d t
\]

Integrating this equation with respect to $t$ and $x$ the system of
equations on unknown $G(x)$ and $R(t)$ is obtained:

\[
  \left\{ 
  \begin{array}{lcl}
   N_{out}(x) & = & G(x) \int \phi(x,t') \cdot R(t') \; dt'  \\
   R_{out}(t) & = & R(t) \int \phi(x',t) \cdot G(x') \; dx'
  \end{array}
  \right|
\]

The numerical solution of these integral equations (see appendix
\ref{chapter:background_equations}) provides $R(t)$ and $G(x)$ based on
data $N_{out}(x)$ and $R_{out}(t)$ from the outside of the source bin. One
can arrive at these equations by invoking the maximum likelihood principle
based on the Poisson distribution of detected events with average
$\mu(x,t) = \phi(x,t) G(x) R(t) dtdx$.

The event rate $R(t)$ is considered to be constant on the very short time
scale (24 sidereal seconds amount to rotation of the Earth by
$0.1^{\circ}$, much smaller than angular resolution of the detector which
defines ``very short'') and therefore is saved as a histogram. This is the
distribution of detected times estimated using off-source events only.
Generated event times are drawn from it. The task now is to generate a
sample from $G(x)$ using off-source events. This sample should contain
$N(x) = G(x) \int R(t)dt$ events with given local coordinates $x$,
however, the number of off-source events available is $N_{out}(x) = G(x)
\int \phi(x,t)R(t)dt$. Therefore, missing events are created by swapping
available ones

\[
 1+\alpha '(x) = \frac{G(x) \int R(t)dt}{G(x) \int \phi(x,t)R(t)dt} =
                 \frac{G(x) \int R(t)dt}{N_{out}(x)} 
\]

number of times. This can be done by choosing actual number of swaps from
a Poisson distribution with parameter $(1+\alpha '(x))$. The described
method is the Monte Carlo generation scheme applied to the equation
(\ref{equation:sloshing:montecarlo}) where only events from the off-source
region are used.

Two remarks are in order. First, the value $\alpha '(x)$ depends on local
coordinate $x$. Second, $\alpha '$ has a similar meaning as $\alpha$ in
equations (\ref{equation:sloshing:significance}) and
(\ref{equation:significance}): ratio of on and off source exposures.
However, if the region of interest is a subregion of the excluded bin,
then $\alpha$ in equation (\ref{equation:sloshing:significance}) pertains
to the subregion, whereas, $\alpha '$ corresponds to the excluded bin.

The importance of this modification is demonstrated on figure
\ref{fig:galactic_latitide:mc} where results of Monte Carlo simulations
are presented. The figure shows excess number of events ($N_{\gamma}
\equiv N_{s} - N_{b}$) as a function of Galactic latitude before (figure
\ref{fig:galactic_latitide:mc:before}) and after (figure
\ref{fig:galactic_latitide:mc:after}) the change. The 25\% signal loss is
recovered by the modification.

% ====== Pictures.....=========== 
\begin{figure}
\centering
  \subfigure[Excess number of events $N_{\gamma}$ as a function of
             Galactic Latitude. Source region is not excluded.]{
    \label{fig:galactic_latitide:mc:before}
    \includegraphics[width=2.7in]{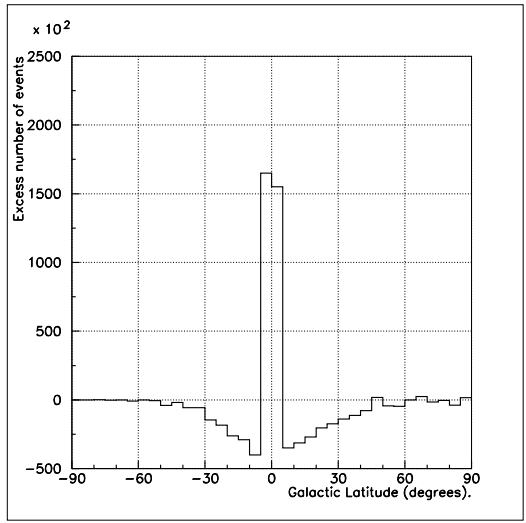}
  }
  \subfigure[Excess number of events $N_{\gamma}$ as a function of
             Galactic Latitude. The region of $\pm 7^{\circ}$ around
             Galactic equator is excluded. Modified time swapping method
             is used.]{
    \label{fig:galactic_latitide:mc:after}
    \includegraphics[width=2.7in]{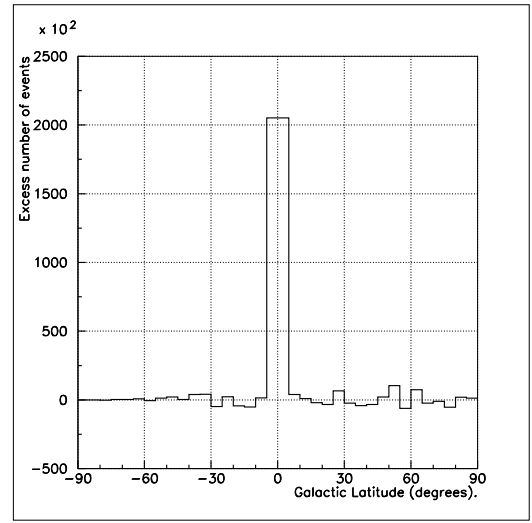}
  }

\caption{Plots showing the results of Monte Carlo simulations with uniform
         Galactic signal flux being 0.0088 that of background in the
         region of $\pm 5^{\circ}$ around the Galactic equator. The
         expected bin content is about 205000.}
\label{fig:galactic_latitide:mc}
\end{figure}

%/////////////////////////////////////////////////
\section{Significance of a Measurement and the Time Swapping Method.}

Significance calculation according to equation
(\ref{equation:sloshing:significance}) relies on the integral evaluation
(\ref{equation:sloshing:background}) of the estimated number of background
events $N_{b}$. The time swapping method employs Monte Carlo integration
and thus introduces additional fluctuations in the estimate of $N_{b}$.
The significance calculation has to reflect this fact and has to reduce to
the equation (\ref{equation:sloshing:significance}) in the limit of
infinite number of generated events, $\beta \rightarrow \infty$. Also, as
was mentioned earlier, it is the time swapping method which defines the
off-source region, therefore, the ratio of on- and off-source exposures
$\alpha$ has to be calculated within the framework of the method and can
not be supplied externally. In what follows, the explicit dependence on
local coordinates $x$ will be omitted and restored at the end. Also, new
definitions are introduced, old ones are as before.

Let $N$ be the number of events available for swapping in the off-source
region. $N$ is obtained as a result of a measurement and therefore is a
sample of size one from a Poisson distribution with some parameter $\mu$:

\[
   P_{\mu}(N) = \frac{\mu^{N}}{N!} e^{-\mu}
\]

These events participate in the swapping procedure, during which the total 
number of generated events $M$ is drawn from a Poisson distribution with
parameter $(1+\alpha ') \beta N$:

\[
  P_{(1+\alpha ') \beta N}(M) = \frac{((1+\alpha ') \beta N)^{M}}{M!}
                                e^{-(1+\alpha ') \beta N}
\]

During this Monte Carlo integration, $m$ events out of $M$ end up in the
source region of interest which obeys a binomial distribution with some
parameter $p$:

\[
  B_{p}(m,M) = C_{M}^{m} p^{m}(1-p)^{M-m}
\]

Combining all of the above, the probability distribution of $m$ is given
by:

\[
  P(m) = \sum_{M=m}^{\infty} B_{p}(m,M) \sum_{N=0}^{\infty} 
                        P_{(1+\alpha ') \beta N}(M) \cdot P_{\mu}(N)
\]

Calculation of dispersion $D(m)$ of $m$ makes use of the following facts:

\[
   D(m) = E[m^{2}] - E[m]^{2}
\]

\[
 N \sim P_{\mu}(N) : \;\;\; E[N] = \mu, \; E[N^{2}] = \mu + \mu^{2}
\]

\[ 
  m \sim B_{p}(m,M) : \;\;\; E[m] = pM, \; E[m^{2}] = pM(1-p) + p^{2}M^{2}
\]

and results

\[
  E[m] = (1+\alpha ')\beta p\mu
\]

\[
  E[m^{2}] = (1+\alpha ')\beta p \mu + 
             (1+\alpha ')^{2}\beta^{2} p^{2} \mu(1+\mu)
\]

Therefore,

\[
   D(m) = (1+\alpha ')\beta p \mu \left( 1 + (1+\alpha ')\beta p \right)
\]

The unknown probability $p(x)$ is estimated in such a way that the average
$E[m]$ is equal to the obtained: $m=(1+\alpha ')\beta p \mu$. The
parameter $\mu$ is estimated to be equal to the number of events available
for swapping $N_{out}(x)$. Thus, the dispersion of the number of
generated events in the source region is estimated as:

\[
  D(m) = m(x)\left(1+ \frac{m(x)}{N_{out}(x)}\right)
\]

Therefore, the dispersion of the number of background events $N_{b} =
\frac{1}{\beta}m$ is given by:

\[
  D(N_{b}) = \frac{N_{b}(x)}{N_{out}(x)}N_{b}(x)  + \frac{1}{\beta} N_{b}(x)
\]

The ratio $\frac{N_{b}(x)}{N_{out}(x)}$ is the measurement of $\alpha$,
the second term vanishes as $\beta$ increases. Thus, finally, the
significance of a measurement within framework of time swapping is given
by

\begin{equation}
  S(x)= \frac{N_{s}(x) - N_{b}(x)}{\sqrt{N_{s}(x) +
        \alpha(x) N_{b}(x)  + \frac{1}{\beta} N_{b}(x)}}, \;\;\;\;
  \alpha(x) = \frac{N_{b}(x)}{N_{out}(x)}
\label{equation:sloshing:significance_corrected}
\end{equation}

%/////////////////////////////////////////////////
\section{Compounding Independent Experiments.
         \label{section:compounding_experiments}}

When two independent observations of a source are made, a question of
reporting total significance arises. The problem is addressed by noting
that the two observations imply existence of two sets of numbers
($N_{s}^{(1)}, N_{b}^{(1)}, \alpha^{(1)}$) and ($N_{s}^{(2)}, N_{b}^{(2)},
\alpha^{(2)}$) needed for significance calculation in each of the
experiments according to equation
(\ref{equation:sloshing:significance}).\footnote{The discussion is of the
general character, therefore, correction due to time swapping (equation
\ref{equation:sloshing:significance_corrected}) is ignored in this
section.} The combined significance is obtained by

\[
  S_{compound} = \frac{(N_{s}^{(1)}-N_{b}^{(1)}) +
                    (N_{s}^{(2)}-N_{b}^{(2)}) }
                    {\sqrt{
                     (N_{s}^{(1)}+\alpha^{(1)}N_{b}^{(1)}) + 
                     (N_{s}^{(2)}+\alpha^{(2)}N_{b}^{(2)})
                    }}
\]

This expression is a direct extension of equation
(\ref{equation:sloshing:significance}) and bears all the properties of
significance. Examples of such a compounding are combination of
independent experiments conducted at different times, or combination of
independent but concurrent sightings of different fragments of a source.

% ////////////////////////////////////////////////////////////
% ======  Pictures.....===========
\begin{figure}
\centering
\includegraphics[width=5in]{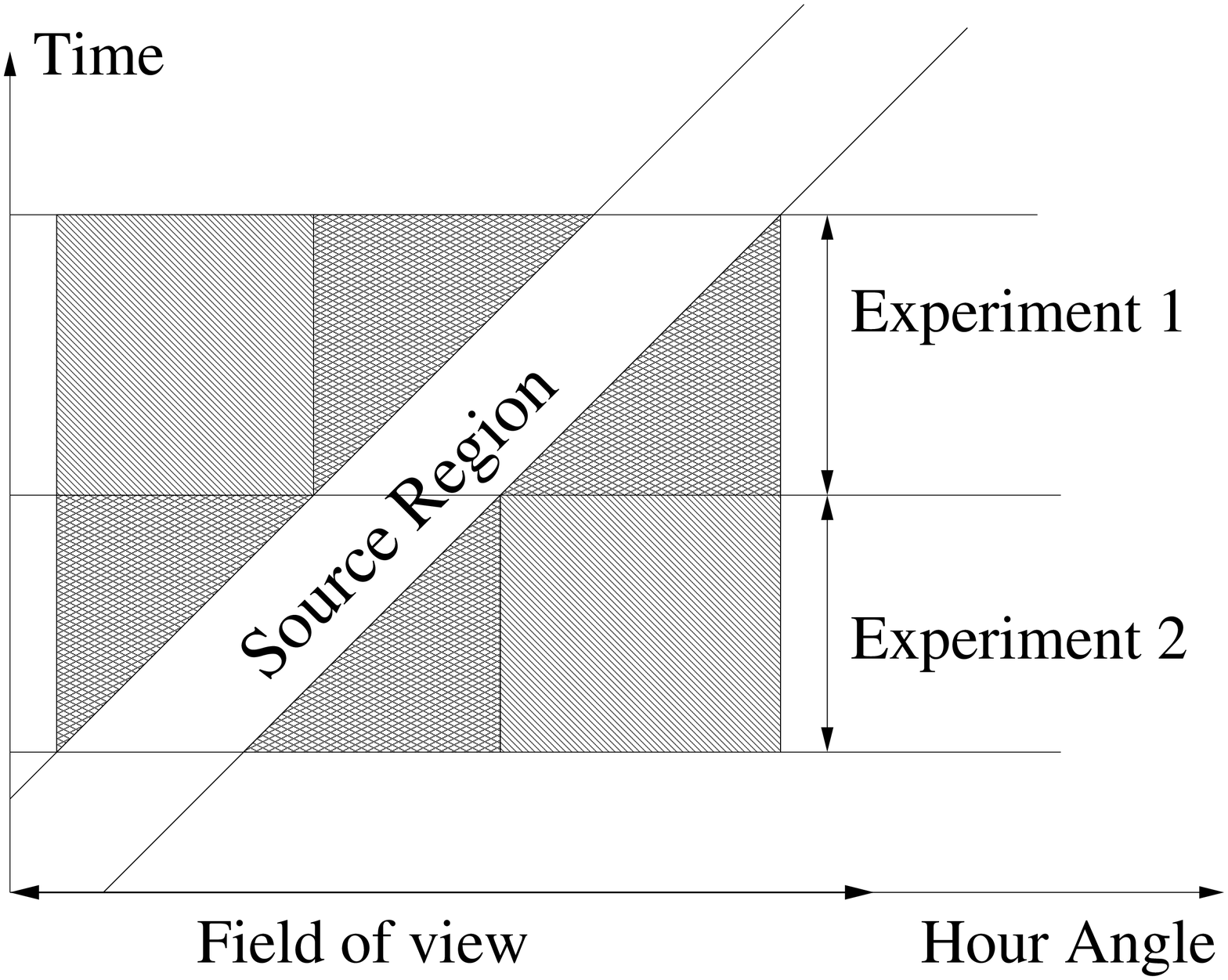}
\caption{Illustration of difference between union and compounding of
experiments.}
\label{fig:background:unite}
\end{figure}

However, the two experiments can be combined in a different way, if the
stability assumption (\ref{equation:sloshing:stability_assumption}) is
common for both observations, namely, the two data sets can be united and
the time swapping procedure can be applied to this common data set. Such a
compounding may lead to increased total significance if otherwise useless
information obtained during experiment 1 can prove to be valuable for
experiment 2 and vise versa. The situation is illustrated on the figure
\ref{fig:background:unite} where two contiguous experiments are presented.
The shaded area is the united off-source region, whereas, the heavily
shaded is the combination of the independently useful off-source regions
of the two experiments. It is seen that in united data set
$N_{s}^{(union)} \equiv N_{s}^{(1)}+N_{s}^{(2)}$, $N_{b}^{(union)} \simeq
N_{b}^{(1)}+N_{b}^{(2)}$ (statistically equal), but $\alpha^{(union)} \leq
\alpha^{(1,2)}$ because $\alpha$'s are ratios of exposures of on- and
off-source regions, the latter one being expanded by the unification.
Therefore, significance

\[
 S_{union} = \frac{N_{s}^{(union)} - N_{b}^{(union)}}
             {\sqrt{ N_{s}^{(union)} + \alpha^{(union)}N_{b}^{(union)}}}
\]

is statistically higher than $S_{compound}$ because of its lower
denominator. The increased sensitivity provides an incentive to broaden
the limits of integration (\ref{equation:sloshing:background}) for as long
as stability assumption (\ref{equation:sloshing:stability_assumption})
allows.

%/////////////////////////////////////////////////
\section{Detector Stability Assumption Test. Diurnal Modulations.}

Despite the fact that no reconfigurations to the detector on the short
time scale are made, the acceptance of the array $G(x)$ depends on
transmission properties of the atmosphere and therefore, verification of
the stability assumption (\ref{equation:sloshing:stability_assumption}) is
still warranted \cite{todd:systematics}.

The test of stability would be a comparison of two acceptances $G_{1}(x)$
and $G_{2}(x)$ measured at different times $t_{1}$ and $t_{2}$. On
physical grounds, the detector possesses a certain degree of azimuthal
symmetry, so does the atmosphere, therefore acceptance is considered as
function of zenith and azimuth angles $G(z,A)$. A measurement is a
generation of histograms $G_{1}(z,A)$ and $G_{2}(z,A)$ from the data for a
certain duration of time around $t_{1}$ and $t_{2}$. (This interval was
chosen to be 30 solar minutes.) These histograms are then to be compared
using a significance test. It has to be recognized that presence of
sources on the sky will mimic instability (section
\ref{chapter:background:anisotropy}), therefore, zenith and azimuth angle
distributions alone are compared instead. The test is based on ideas of
compounding probabilities from independent significance tests
\cite{fisher} and is implemented as a series of $\chi^{2}$ tests of
$G_{i}(x)$ and $G_{j}(x)$ (yielding $\chi^{2}(t_{i},t_{j})$) and then
combined $\chi^{2}_{total}(\Delta t)$ for time separation $\Delta t =
t_{i} - t_{j}$ is calculated:

\[ 
   \chi^{2}_{total}(\Delta t) = \sum_{\Delta t = t_{i} - t_{j}}
                                \chi^{2}(t_{i},t_{j})
\]

% ////////////////////////////////////////////////////////////
% ======  Pictures.....===========
\begin{figure}
\centering
\includegraphics[width=5in]{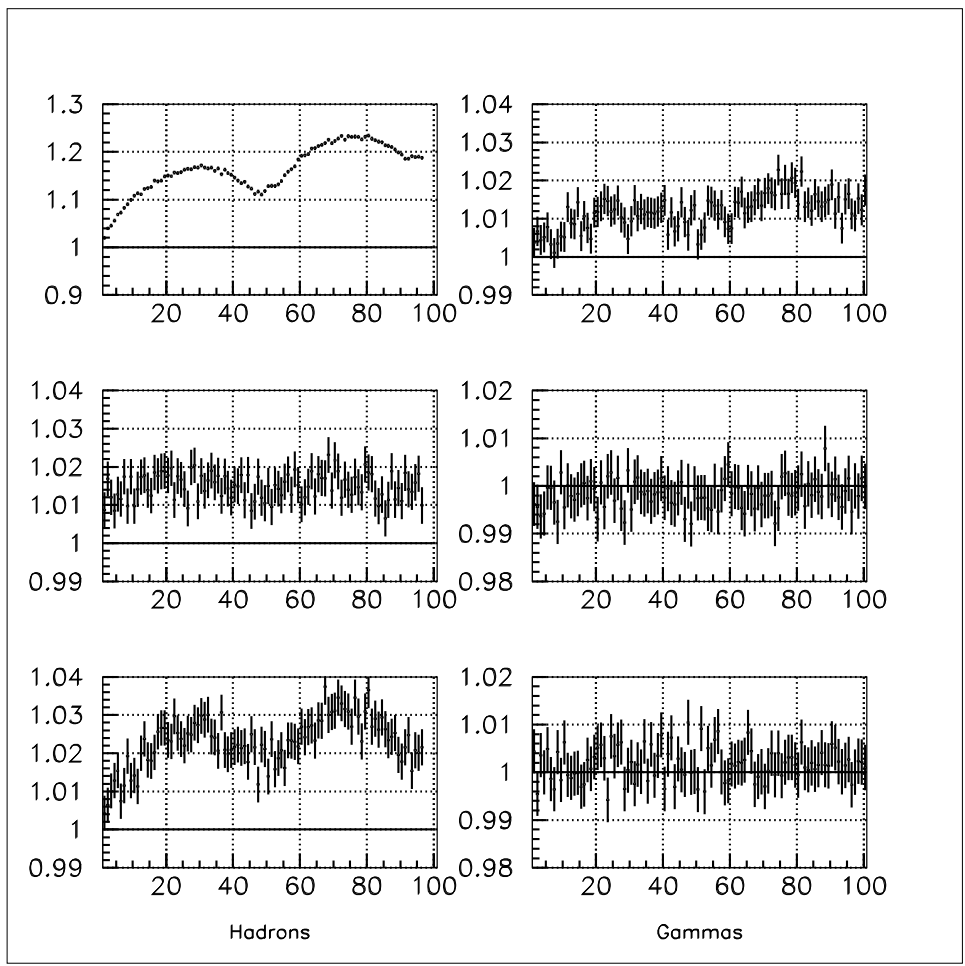}
\caption{
         Results of stability test on data collected between July 19, 2000
         and September 17, 2000 are presented. Horizontal axis is time
         separation $\Delta t$ in 30 minute units, vertical axis is
         corresponding $\frac{\chi^{2}_{total}(\Delta t)}{m_{total}(\Delta
         t)}$. Top plots are the test of zenith distributions, middle and
         bottom ones are the test of azimuth distributions for $0^{\circ} 
         < z < 30^{\circ}$ and $30^{\circ} < z < 50^{\circ}$
         respectively. On the left all data were used, on the right, only 
         events passing $X_{2}>2.5$ cut. Solid horizontal line is the 
         expected value of one if the stability assumption holds.
} \label{fig:background:stability_test}
\end{figure}

The test statistic $\chi^{2}_{total}(\Delta t)$ so obtained follows a
$\chi^{2}$ distribution with $m_{total} = \sum_{\Delta t = t_{i} - t_{j}}
m(t_{i},t_{j})$ degrees of freedom if observed differences are of random
nature only. The average of $\chi^{2}_{total}$ is equal to $m_{total}$
while its variance is equal to $2m_{total}$. The results of such a test
are presented on figure \ref{fig:background:stability_test} as the plot of
$\frac{\chi^{2}_{total}(\Delta t)}{m_{total}(\Delta t)}$. The closer the
plotted value to the expected one, the better the assumption of stability
(\ref{equation:sloshing:stability_assumption}) holds. It is seen from the
plots that the degree of violation of the assumption grows with time
separation $\Delta t$ as might be expected, but then it drops before
growing again. This is interpreted as presence of a periodic component
which insured that two acceptances $G_{1}(z)$ and $G_{2}(z)$ separated by
24 solar hours are ``closer'' to each other than, say, those separated by
only 12. Azimuthal distributions show violation of the stability
assumption to a lesser degree. This statement is considered to be true
after application of the hadron rejection cut as well. Thus, despite the
fact that no human intervention on the short time scale is made, the
acceptance of the detector changes. As was seen in section
\ref{section:compounding_experiments}, the duration of the validity of the
assumption (\ref{equation:sloshing:stability_assumption}) defines the time
integration limits in equation (\ref{equation:sloshing:background}) which,
in turn, defines the off-source region. Inasmuch as search for very weak
signal requires sensitivity pushed to its limits, care is taken to improve
the stability assumption in particular with regard to zenith angle
dependence.

The investigation of changes in zenith distribution (simplified by the
above noted periodicity) was performed in the following way
\cite{diurnal_memo}. Every half hour, a zenith distribution is generated
from registered events. These distributions are compared to the average
distribution accumulated over a one day period. Half hour distributions
and the average one are normalized and then subtracted to give a plot of
distribution's shape change. It was observed that the shape of this change
is approximately constant with amplitude varying from half hour to half
hour. Therefore, the improved stability assumption is chosen to be of the
form:

\begin{equation}
  N(x,t) = G(x) R(t) e^{ \theta (t)  K(z) + q(\theta (t)) } \; dt
 \label{equation:sloshing:zenith_modulation}
\end{equation}

where $\theta (t)$ is the amplitude of the correction at time $t$, $K(z)$
is the polynomial correction function coefficients of which are obtained
from the above study (see appendix \ref{chapter:zenith_correction}),
$q(\theta (t))$ is the normalization factor so that $\int G(x) e^{ \theta
(t)  K(z) + q(\theta (t)) } dx = 1$.

The choice (\ref{equation:sloshing:zenith_modulation}) of the form of time
dependence of a distribution is motivated by the Darmois theorem
\cite{Eadie}. According to this theorem, exponential class
(\ref{equation:sloshing:zenith_modulation}) of probability density
functions is the only class which admits number of sufficient statistics
for unknown parameter $\theta$ independent of the number of observations.
In particular, if $(Z_{1}, Z_{2}, ..., Z_{N})$ is a random sample of size
$N$ from the distribution (\ref{equation:sloshing:zenith_modulation}),
then

\[ Y = \sum^{N}_{i=1} K(Z_{i}) \]

is sufficient statistics for $\theta$. The importance of sufficiency is   
explained by the fact that $Y$ exhausts all the information about $\theta$
that is contained in the sample.

By differentiating

\[
  \int G(z) e^{ \theta  K(z) + q(\theta)} dz =1
\]

twice with respect to $\theta$ we obtain the following two relations for  
expectation and variance:

\[ E[K(Z)]   = - q\prime(\theta) \]
\[ var[K(Z)] = - q\prime\prime(\theta) \]

Both of these properties allow estimating parameter $\theta$ from the
data.  Indeed, because $\theta = 0$ corresponds $q=0$ (no correction to
average distribution), both $\theta$ and $q$ can be determined as

\[  \left\{ \begin{array}{rcl}
           \theta & = & 2\frac{q\prime(\theta) - q\prime(0)}
                              {q\prime\prime(\theta) + q\prime\prime(0)}
                        + O(\theta^{2})\\
         q(\theta)& = & \frac{1}{2}(q\prime(\theta) + q\prime(0))\theta
                        + O(\theta^{2})
           \end{array}
    \right.
\]

Expectation and variance and therefore $q\prime$ and $q\prime\prime$ are
determined from the sample and thus are known. The example of the average
daily amplitude dependence is shown on figure
\ref{fig:backgroun:zenith_amplitude}. The value of the amplitude is
typically within $\pm 4 \cdot 10^{-5}$ range. The plot can also be used to
justify the choice of half hour intervals for the amplitude measurement.
(This, however, is already seen from the figure
\ref{fig:background:stability_test}.)

% ////////////////////////////////////////////////////////////
% ======  Pictures.....===========
\begin{figure}
\centering
\includegraphics[width=5in]{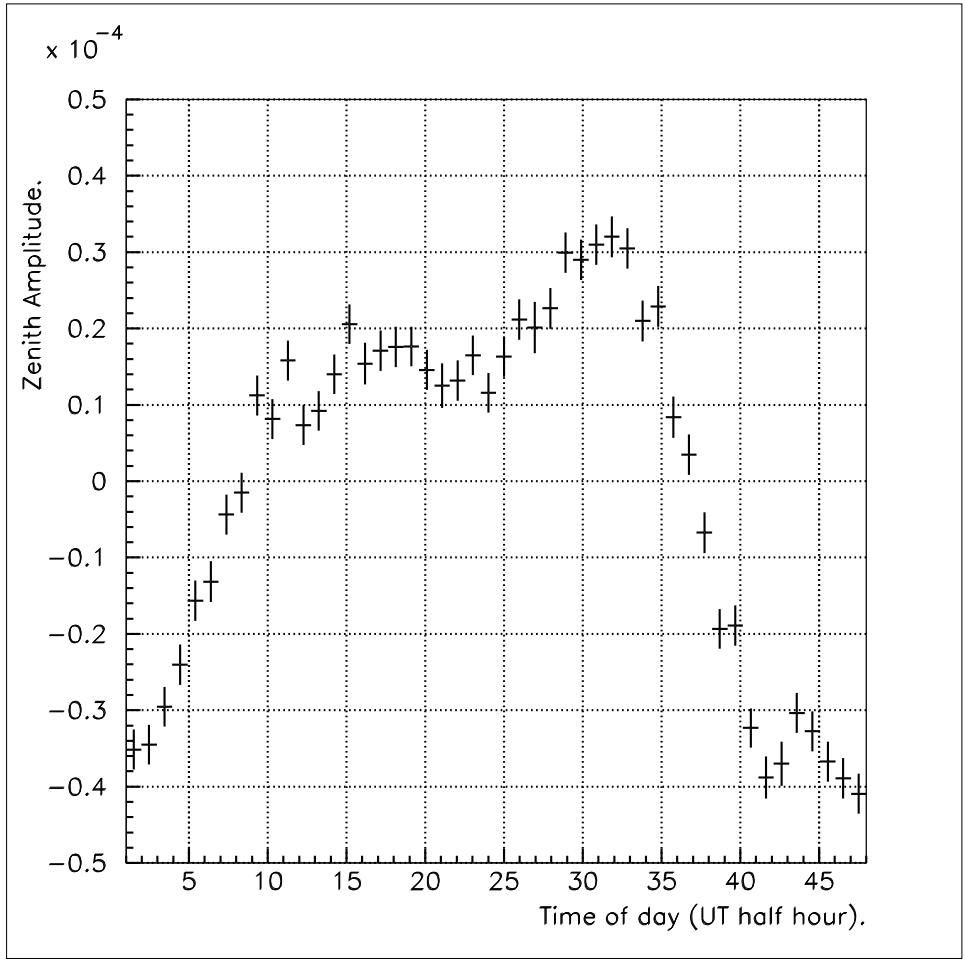}
\caption{Average daily dependence of the zenith correction amplitude
         $\theta$ derived from July 19 - September 17, 2000  Milagro data
         after application of the hadron rejection cut.
} \label{fig:backgroun:zenith_amplitude}
\end{figure}

%/////////////////////////////////////////////////
\section{Accounting for Modulations in the Framework of Time Swapping
Method.}

Incorporation of the improved stability assumption
(\ref{equation:sloshing:zenith_modulation}) into the framework of time
swapping is equivalent to modification of Monte Carlo integration where
generated events are distributed according to $G(x) R(t) e^{ \theta (t)  
K(z) + q(\theta (t))}$ instead of $G(x) R(t)$. This is accomplished by
means of the rejection method \cite{numerical_recipes}. In this method, an
event is drawn from the $G(x) R(t)$ distribution as before. But now, the
candidate event time is accepted only if a random variable $p$ distributed
uniformly between 0 and $P(z) = \max_{\{t\}} e^{ \theta (t)  K(z) +
q(\theta (t))}$ is such that $p < e^{ \theta (t)  K(z) + q(\theta (t))}$.
(Parameters $\theta$ and $q(\theta)$ are assumed to be known at this
stage.) If this inequality is violated, another candidate time is sought
and then tried. This is continued until time is accepted, resulting in
generation of event. Inasmuch as generation of the events is based on the
improved stability assumption, the rest of the data analysis algorithm is
preserved.

% ====== Pictures.....=========== 
\begin{figure}
\centering
  \subfigure[Distribution of $\chi^{2}$ for 90 bin histogram. Solid line 
             is the best fit of $\chi^{2}$ distribution function with
             84.11 degrees of freedom.]{
%    \label{fig:chi_squared:mc:1}
    \includegraphics[width=2.7in]{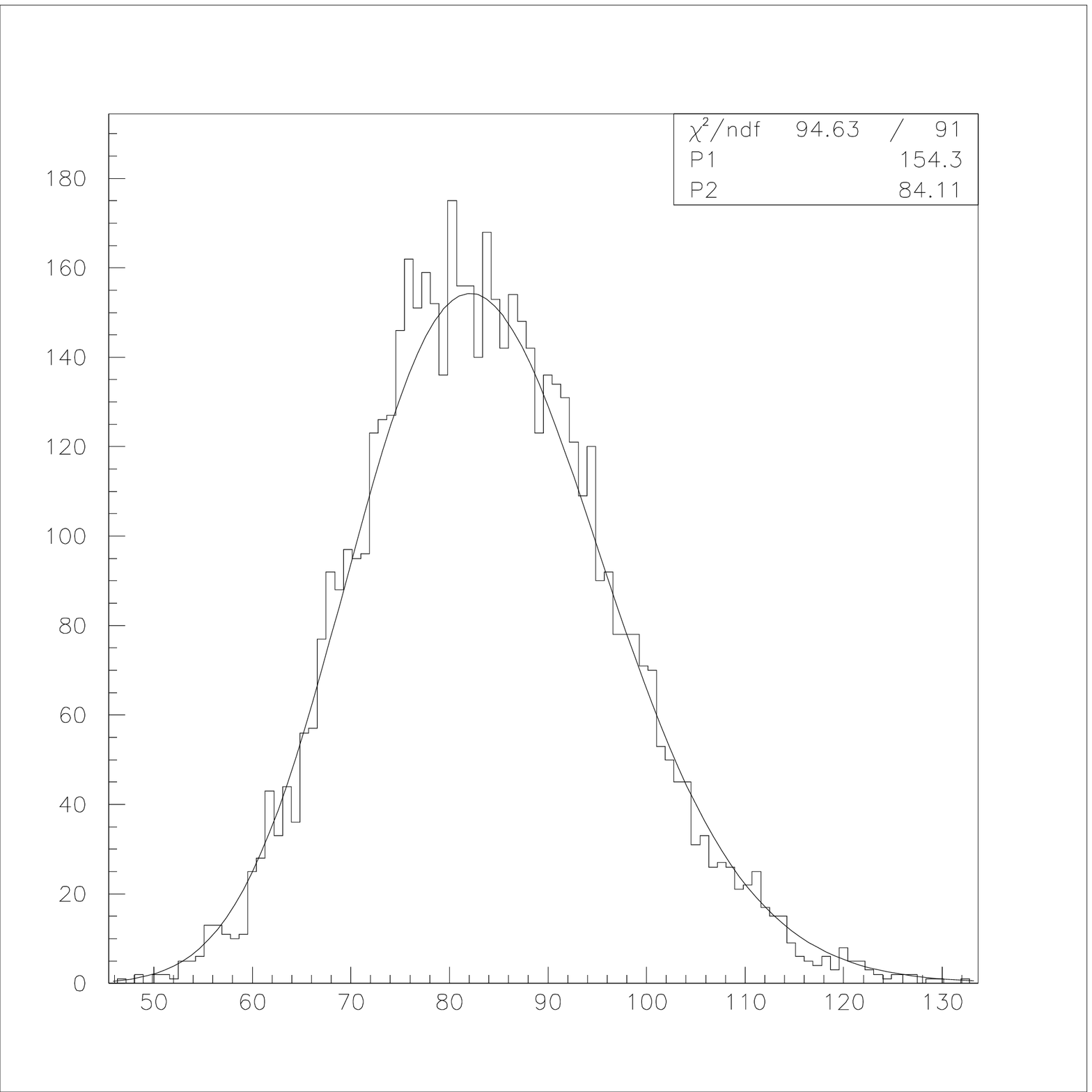}
  }
  \subfigure[Distribution of $\chi^{2}$ for 360 bin histogram. Solid line  
             is the best fit of $\chi^{2}$ distribution function with
             337.0 degrees of freedom.]{
%    \label{fig:chi_squared:mc:2}
    \includegraphics[width=2.7in]{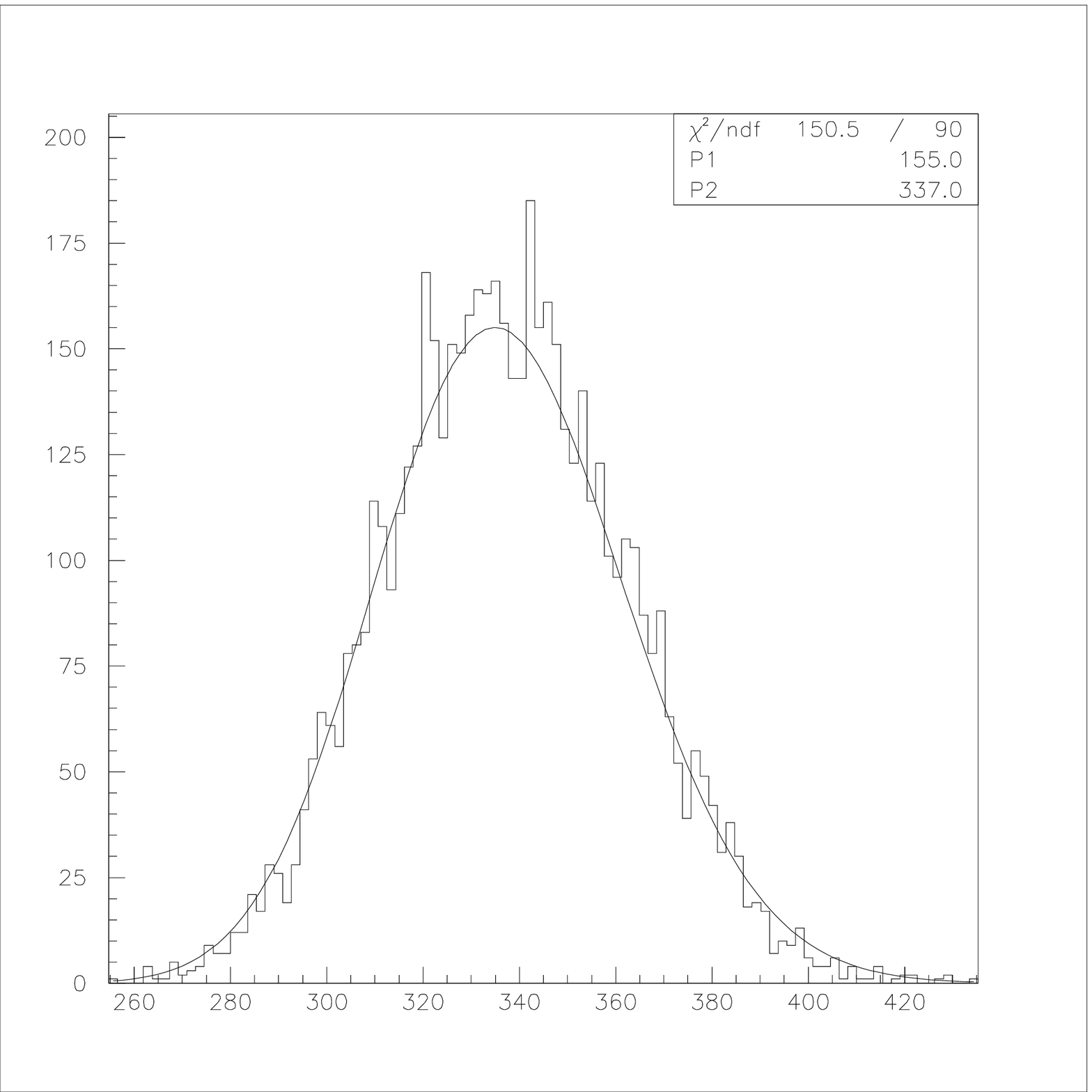}
  }
\caption{Plots showing Monte Carlo simulated distributions of
         $\chi^{2}$ of the difference of histograms with 90 and 360
         bins. Both histograms have the same number of entries (5713).}
\label{fig:chi_squared:mc}
\end{figure}

% ====== Pictures.....=========== 
\begin{figure}
\centering
  \subfigure[Distribution of $\chi^{2}$ for standard stability
             assumption (equation \ref{equation:sloshing:stability_assumption}).
             Solid line is the best fit of $\chi^{2}$ distribution
             function with 93.84 degrees of freedom. The goodness of this
             fit is characterized by $\chi^{2}/ndf = 506.2/50$.]{
%    \label{fig:chi_squared:zenith:2}
    \includegraphics[width=2.7in]{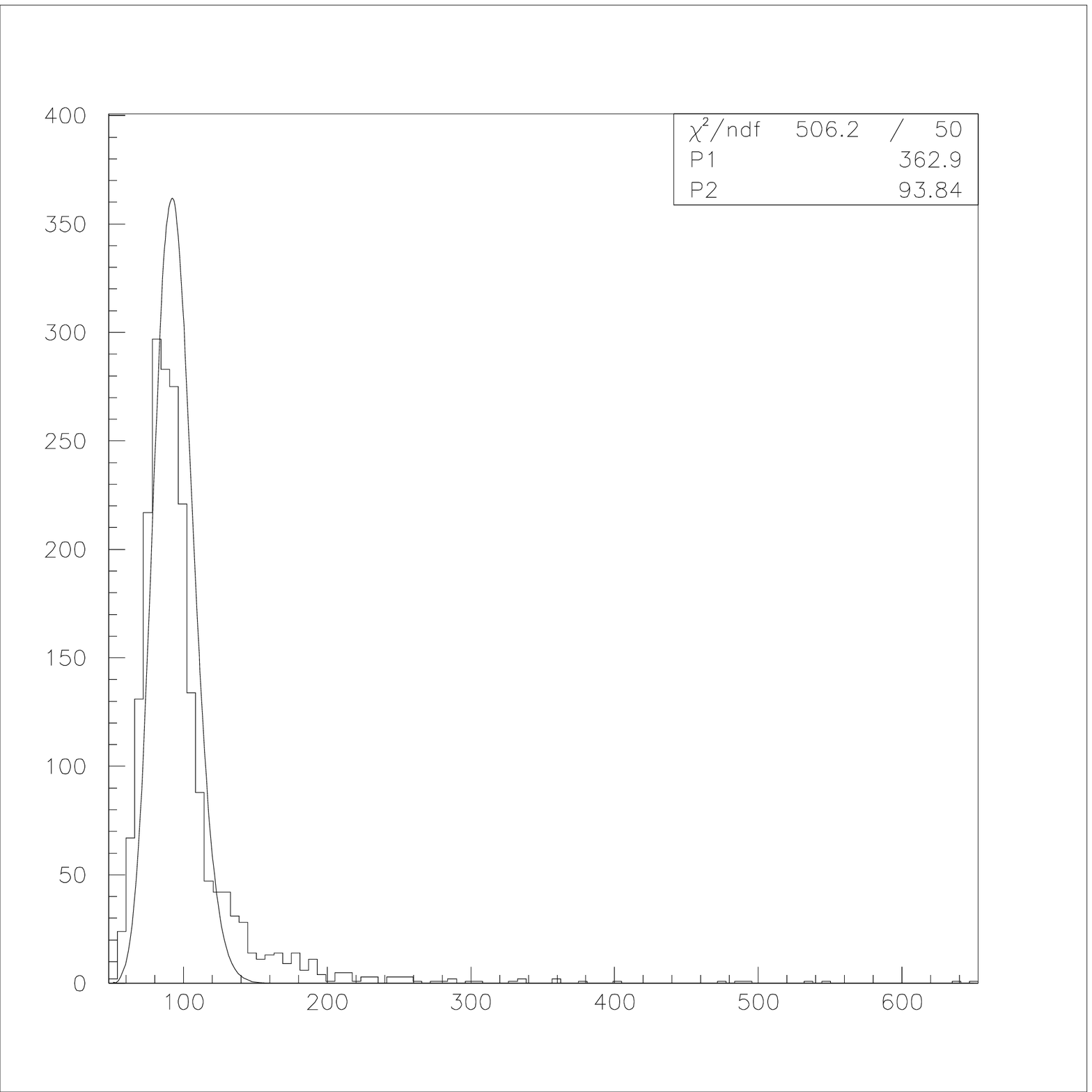}
  }
  \subfigure[Distribution of $\chi^{2}$ for improved stability 
             assumption (equation \ref{equation:sloshing:zenith_modulation}).
             Solid line is the best fit of $\chi^{2}$ distribution
             function with 84.10 degrees of freedom. The goodness of this
             fit is characterized by $\chi^{2}/ndf = 91.34/86$.]{
%    \label{fig:chi_squared:zenith:1}
    \includegraphics[width=2.7in]{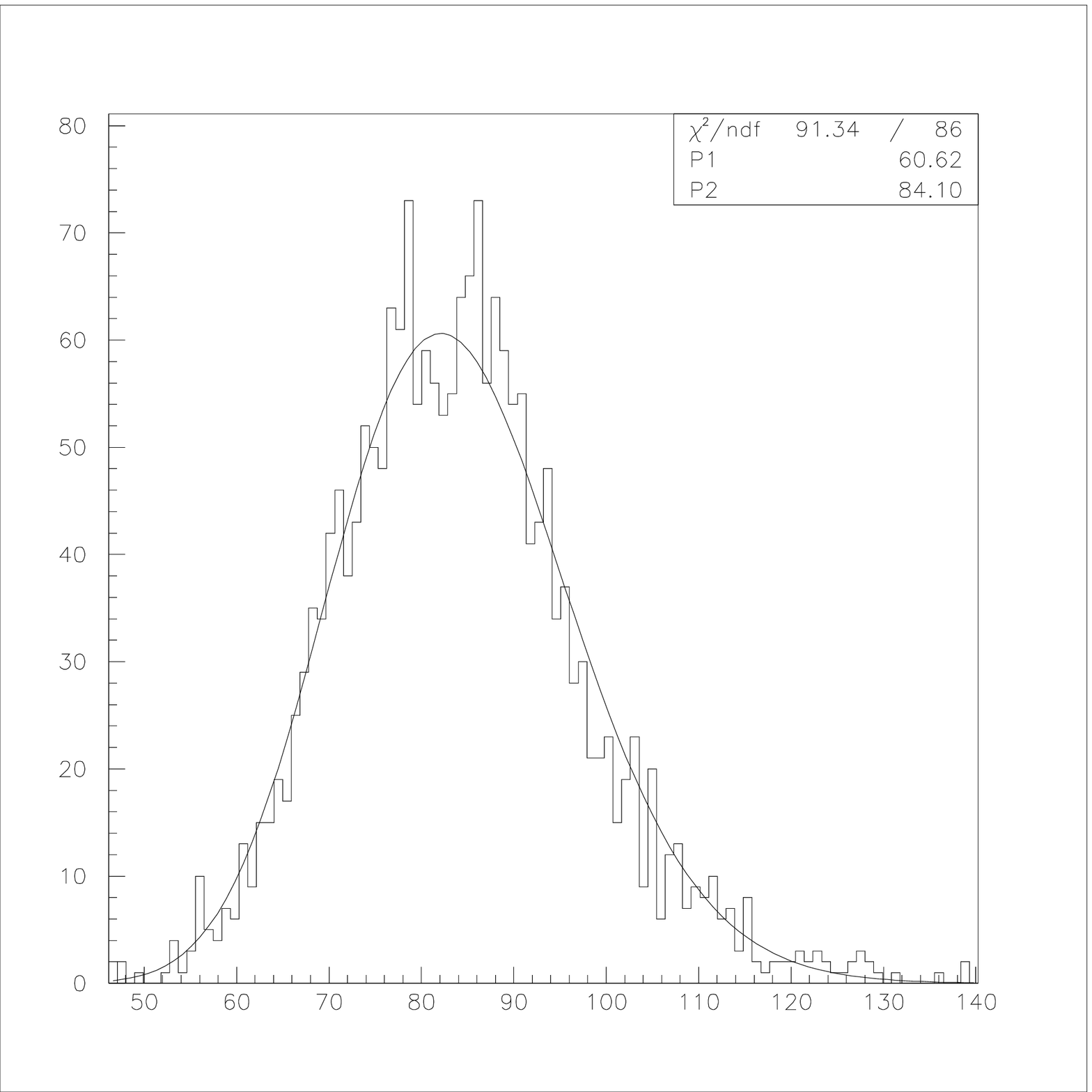}
  }
\caption{Plots showing distributions of $\chi^{2}$ of the difference of
         simulated and real zenith angle distributions for every half hour
         with and without correction. The expected number of degrees of
         freedom is 84.25. Both histograms have the same number of entries
         (2073).}
\label{fig:chi_squared:zenith}
\end{figure}

% ====== Pictures.....=========== 
\begin{figure}
\centering
  \subfigure[Distribution of $\chi^{2}$ for standard stability
             assumption (equation \ref{equation:sloshing:stability_assumption}).
             Solid line is the best fit of $\chi^{2}$ distribution
             function with 339.6 degrees of freedom. The goodness of this
             fit is characterized by $\chi^{2}/ndf = 69.82/83$.]{
%    \label{fig:chi_squared:azimuth:2}
    \includegraphics[width=2.7in]{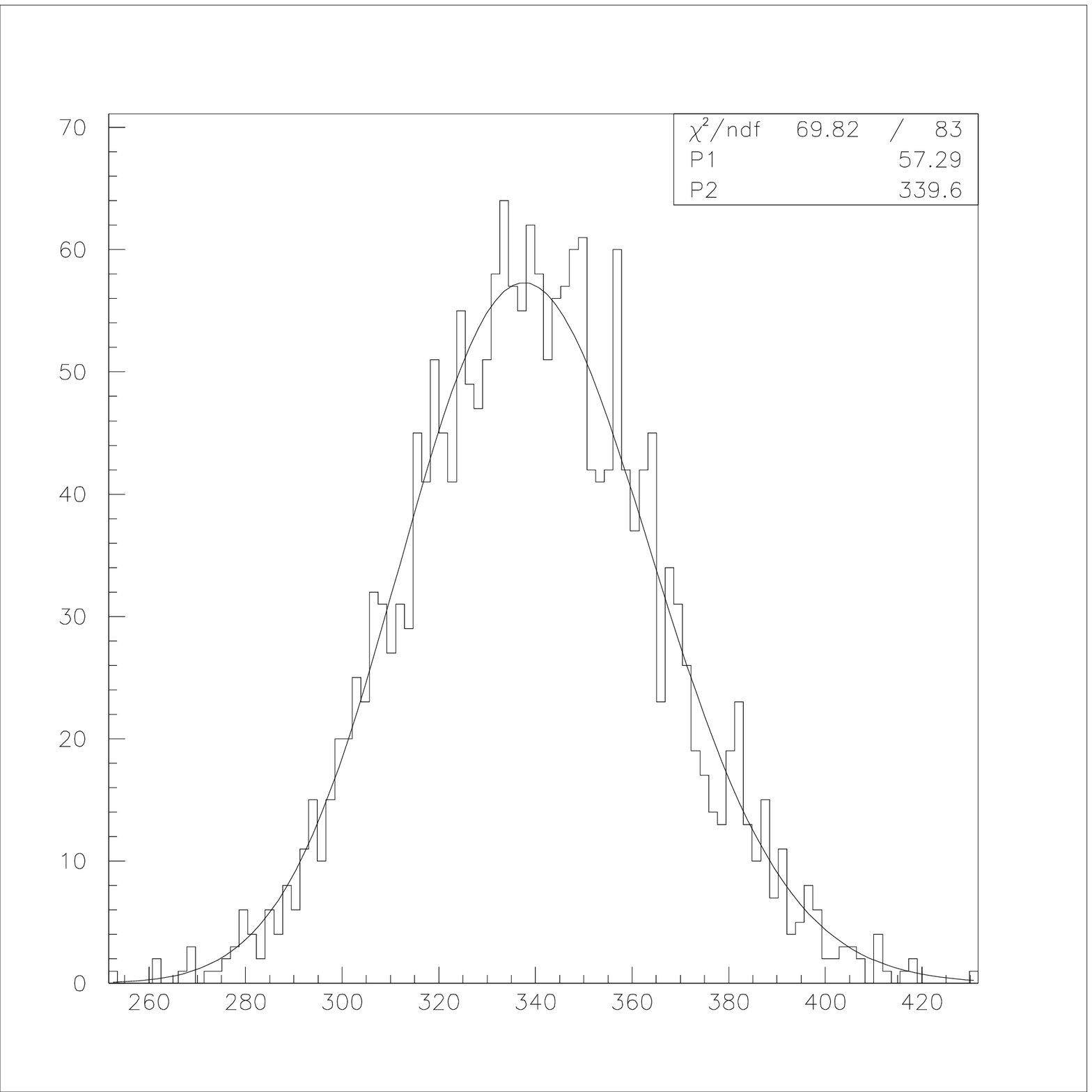}
  }
  \subfigure[Distribution of $\chi^{2}$ for improved stability
             assumption (equation \ref{equation:sloshing:zenith_modulation}).
             Solid line is the best fit of $\chi^{2}$ distribution
             function with 337.6 degrees of freedom. The goodness of this
             fit is characterized by $\chi^{2}/ndf = 92.33/92$.]{
%    \label{fig:chi_squared:azimuth:1}
    \includegraphics[width=2.7in]{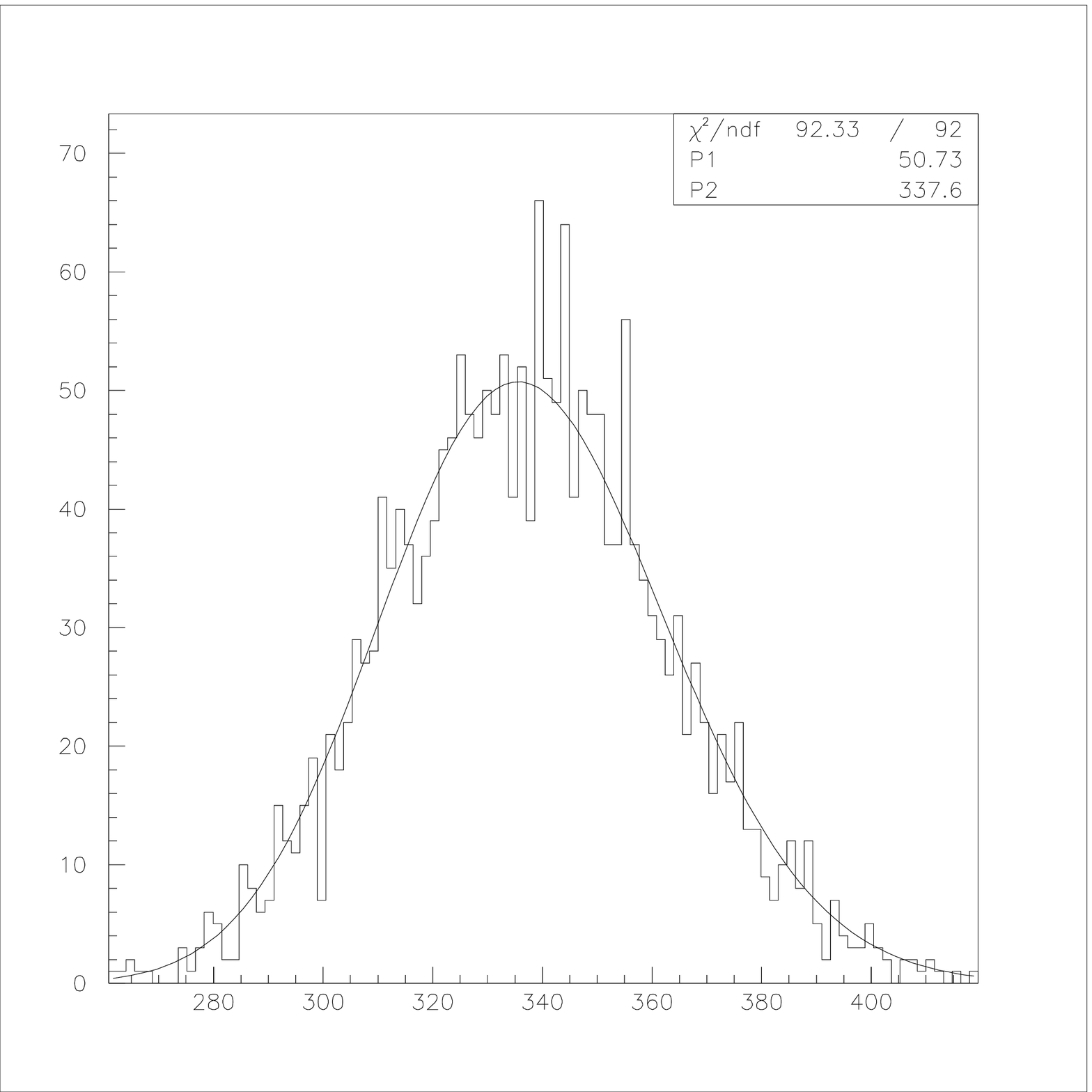}
  }
\caption{Plots showing distributions of $\chi^{2}$ of the difference of
         simulated and real azimuth angle distributions for every half hour
         with and without correction. The expected number of degrees of
         freedom is 337.5. Both histograms have the same number of entries
         (2073).}
\label{fig:chi_squared:azimuth}
\end{figure}

In order to illustrate how well the correction works real and generated
distributions are compared using a $\chi^{2}$ test. The zenith angle
distribution is split into 90 bins, and azimuth one into 360. A word of
precaution is appropriate: the real and generated distributions are not
independent, one is produced from the other using swapping technique. Due
to implementation of the time swapping method (8 hour integration time
window) and because distributions are accumulated every 30 minutes, the
number of degrees of freedom in the $\chi^{2}$ test should be lower than
one would expect from the usual histogram bin considerations by 1 for
every 16 bins or 84.25 and 337.5 for 90 and 360 bin histograms
respectively. This is supported by Monte Carlo simulations presented on
figure (\ref{fig:chi_squared:mc}) where the values of $\chi^{2}$ resulted
from the test are histogramed.

Figure (\ref{fig:chi_squared:zenith}) demonstrates zenith angle
distribution comparisons with and without the correction. From these plots
it is clearly seen how a bias of the standard method is removed. A similar
study was carried out for azimuth angle distribution. Figure
(\ref{fig:chi_squared:azimuth}) demonstrates the results of the study and
shows that azimuth variation, if any, is on a much smaller scale than the
zenith one. Note, that no azimuthal correction was applied, therefore, a
slight change in the azimuth distribution test is due to possible
correlation between zenith and azimuth angles of the detected events.
Thus, the improved stability assumption (equation
\ref{equation:sloshing:zenith_modulation}) is considered to be valid for
the 8 hour duration of timing integration.

%/////////////////////////////////////////////////
\section{Known Anisotropies.\label{chapter:background:anisotropy}}

It was, so far, conceded that no anisotropy on the sky is present. This,
together with the stability assumption had lead to the equation
(\ref{equation:sloshing:stability_assumption}). In fact, if there are
known sources on the sky, then the number of registered events is given
by:

\[
 d N(x,t) = S(x,t) \cdot G(x) \cdot R(t) \; dx dt
\]

where $S(x,t)$ describes the strength of the sources as function of local
coordinates and time. (This is how anisotropy can mimic detector
instability.)

There are two known sources, actually sinks, on the sky: the Sun and the
Moon \cite{sun_moon}. Because the anisotropy function $S(x,t)$ produced by
them is not known, the anisotropy is handled by vetoing the $\pm
5^{\circ}$ degree regions around the objects. This entails treating them
as part of the source bin during Monte Carlo integration and disregarding
generated events in the background estimation $N_{b}$ if they fall within
the veto region. In general, known small scale anisotropies must be vetoed
as described, known large scale ones have to be incorporated into the
stability assumption. These become part of the null hypothesis.

\chapter{Alternative Hypothesis.}

The construction of the alternative hypothesis, or more precisely, of
critical regions is based on the knowledge about the Galactic gamma ray
emission available to date \cite{ong,egret}. Among all experiments
operated at different energies the EGRET telescope is the one of closest
energy range to Milagro and made a convincing detection of emission. The
data made available by the EGRET collaboration
\cite{stan_hunter_private,egret} is the foundation of the alternative
hypothesis. Because of the change of the primary emission mechanism
between MeV and GeV energy regions from bremsstrahlung to $\pi^{0}$ decay
\cite{model_bertsch}, only highest EGRET energy data points are used in
the formulation. Figure \ref{fig:alternative:egret} plots the longitude
dependence of the EGRET emission flux in the $\pm2^{\circ}$ latitude band
along the Galactic equator at energies 10-30 GeV. Figure
\ref{fig:alternative:milagro} shows the Milagro Galactic disk exposure as
a function of Galactic longitude. The expected significance which is
roughly proportional (equation (\ref{equation:significance})) to the
product of the signal flux and the square root of the exposure is
presented on figure \ref{fig:alternative:significance}.

Thus, the region of Galactic disk with $l \in (20^{\circ},100^{\circ})$ is
the region where highest significance is expected provided the emission
model of EGRET is valid at TeV energies. Therefore, the region of $l \in
(20^{\circ},100^{\circ})$ is selected as Milagro inner Galaxy. Due to the
fact that the inverse Compton mechanisms may result in a wider latitude
distribution than that of the gas column density, the region with
$|b|<5^{\circ}$ is also considered. The two regions with $l \in
(140^{\circ},220^{\circ})$, $|b|<2^{\circ}$ and $|b|<5^{\circ}$ are also
selected as Milagro outer Galaxy. These parts of the disk are selected due
to their high exposure. These regions are analyzed separately, however
$|b|<2^{\circ}$ and $|b|<5^{\circ}$ ones are not independent for the same
longitude range. The regions were selected before the analysis was
performed.

% ====== Pictures.....===========
\begin{figure}
\centering
  \subfigure[Total flux at 10-30 GeV in $|b|<2^{\circ}$ disk.]{
    \label{fig:alternative:egret}   
    \includegraphics[width=4.5in]{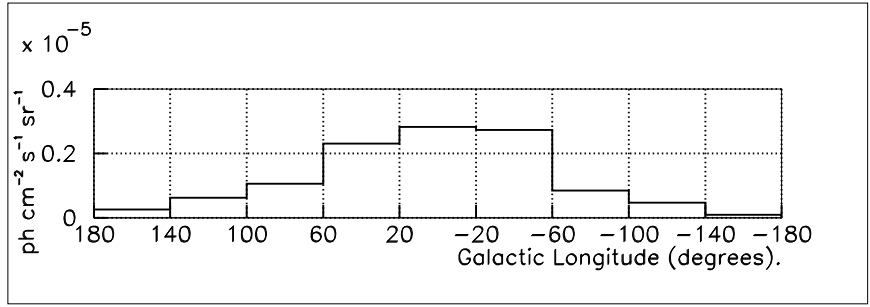}
  }
  \subfigure[Milagro exposure of $|b|<2^{\circ}$ disk.]{
    \label{fig:alternative:milagro}
    \includegraphics[width=4.5in]{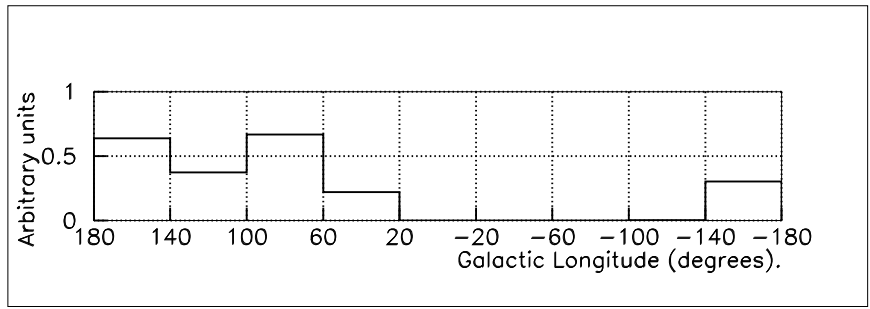}
  }
  \subfigure[The expected relative significance.]{
    \label{fig:alternative:significance}
    \includegraphics[width=4.5in]{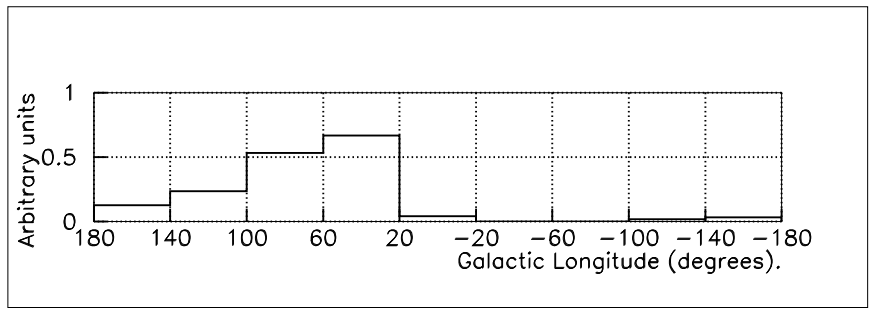}
  }

\caption{Plots showing the construction of the critical region for the
         alternative hypothesis based on the EGRET data.}
\label{fig:alternative}
\end{figure}

\chapter{Results.}

%/////////////////////////////////////////////
\section{Results of the Measurements.}

The results presented below are obtained after the analysis of data
collected between July 19, 2000 and September 10, 2001 by the Milagro
detector. The starting date corresponds to the commissioning of the hadron
rejection algorithm, the ending date is when the experiment was turned off
for a major scheduled maintenance.

To ensure high quality of the data, the periods with abnormal event rate
with and without the hadron rejection cut were eliminated from the
analysis. These may indicate possible detector problems. Parts of the data
were also discarded if the Milagro log-book showed entries of some
hardware/software problems or indications of unstable operation. Periods
of abnormal zenith angle distribution of events were also discarded. Other
cuts applied include $N_{fit}>20$ to ensure good angular resolution and
the zenith angle cut $0^{\circ}<z<50^{\circ}$ due to stability
considerations (appendix \ref{chapter:zenith_correction}).

The null hypothesis tested consists of:

\begin{itemize}
  \item There is no emission from the Galactic plane except for
        background.
  \item All of the detected events are due to background and it is 
        isotropic except for $\pm 5^{\circ}$ region around the Sun and the
        Moon. (Events from these regions are vetoed.)
  \item Detected events satisfy the stability equation
        (\ref{equation:sloshing:zenith_modulation}).
  \item The null hypothesis also includes certain aspects of shower
        development and detector operation important for the detection and
        reconstruction of events. These include reconstructions algorithms
        and detector calibration procedures.
\end{itemize}

Because the emission from the Galactic plane is under investigation,
events arriving from the $\pm 7^{\circ}$ band along the Galactic equator
are excluded from the background generation (see section
\ref{section:background:signal_events}). Four regions of the disk
(selected a priory) were studied. The Milagro inner Galaxy with Galactic
longitude $l \in (20^{\circ},100^{\circ})$, $|b|<2^{\circ}$ and
$|b|<5^{\circ}$; and Milagro outer Galaxy with $l \in
(140^{\circ},220^{\circ})$, $|b|<2^{\circ}$ and $|b|<5^{\circ}$. The
results are given in the table \ref{table:results:results}.

\begin{table}[htbp]
\begin{center}
\begin{tabular}{|c|c|c|} \hline
                   &  \multicolumn{2}{c|}{Inner Galaxy}                    \\ \cline{2-3}
                   &   $|b|<2^{\circ}$               &   $|b|<5^{\circ}$ \\ \hline
$N_{s}$            & $(43446.6 \pm 6.6)\cdot 10^{3}$ & $(10823.8 \pm 1.0)\cdot 10^{4}$ \\ \hline
$N_{b}$            & $(43426.7 \pm 2.7)\cdot 10^{3}$ & $(10819.7 \pm 0.5)\cdot 10^{4}$ \\ \hline
$N_{\gamma}/N_{b}$ & $(+4.57 \pm 1.64)\cdot 10^{-4}$ & $(+3.77 \pm 1.08)\cdot 10^{-4}$ \\ \hline
significance       &     +2.8                        &      +3.5                       \\ \hline \hline
                   &\multicolumn{2}{c|}{Outer Galaxy} \\ \cline{2-3}
                   &   $|b|<2^{\circ}$               &   $|b|<5^{\circ}$ \\ \hline
$N_{s}$            & $(46409.1 \pm 6.8)\cdot 10^{3}$ & $(11588.6 \pm 1.1)\cdot 10^{4}$ \\ \hline
$N_{b}$            & $(46413.0 \pm 3.0)\cdot 10^{3}$ & $(11590.1 \pm 0.6)\cdot 10^{4}$ \\ \hline
$N_{\gamma}/N_{b}$ & $(-0.82 \pm 1.6)\cdot 10^{-4}$  & $(-1.28 \pm 1.06)\cdot 10^{-4}$ \\ \hline
significance       &     -0.5                        &      -1.2                       \\ \hline

\end{tabular}
\end{center}
\caption{Summary of the results for the analysis of the Milagro inner and
         outer Galaxy ($N_{\gamma}\equiv N_{s}-N_{b}$).}
\label{table:results:results}
\end{table}

As seen from the table \ref{table:results:results} the results of the
analysis for the inner Galaxy indicate violation of the null hypothesis.
The null hypothesis is rejected with the significance of 3.5 or the
probability that it is rejected by chance fluctuations is only
$2.3\cdot10^{-4}$ (see table \ref{table:significance_probability}). A
detection of a source is usually claimed if the chance probability is less
than 0.1\%.

In order to be able to interpret the result as a presence of excess of
events in the direction of the Galactic plane, other constituents of the
null hypothesis have to be examined. The aspects of detector operation and
event reconstruction are unlikely to produce localized structure on the
sky, they should rather show up as global changes such as in total event
rate, in zenith angle distribution of detected events. These were
carefully monitored. A violation of the stability assumption may, however,
``produce'' large scale structures on the sky (and vise versa).

Isotropy of the background is another ingredient.  Because of the random
bending of cosmic rays in the galactic magnetic fields which are
responsible for the isotropy of the background, small deviations from
isotropy are expected to be smooth and therefore reveal themselves as
large scale structures too. According to indications in the literature
from underground muon experiments \cite{kamiokande_anisotropy} a slight
anisotropy may be present which is parametrized by the authors as:

\[
  R(\alpha) = 1 + r_{0} \cos(\alpha - \alpha_{0})
\]

where $r_{0} = (5.6 \pm 1.9) \cdot 10^{-4}$, $\alpha_{0} = 8.0^{\circ} \pm
19.1^{\circ}$. To study the systematic error caused by a possible neglect
of the effect, a computer simulation was employed. In this simulation a
data set equivalent to the 18 times the real one was generated. The
anisotropy was simulated at the specified level but standard analysis was
applied. The results of this simulation are presented in the table
\ref{table:results:systematic}.

\begin{table}[htbp]
\begin{center}
\begin{tabular}{|c|c|c|} \hline
             &   $|b|<2^{\circ}$           &   $|b|<5^{\circ}$ \\ \hline
Inner Galaxy & $(-3.7 \pm 3.8)\cdot 10^{-5}$ & $(-4.0 \pm 2.5)\cdot 10^{-5}$ \\ \hline
Outer Galaxy & $(-5.4 \pm 3.7)\cdot 10^{-5}$ & $(-1.7 \pm 2.4)\cdot 10^{-5}$ \\ \hline

\end{tabular}
\end{center}
\caption{Results of the systematic error study for $N_{\gamma}/N_{b}$.}
\label{table:results:systematic}
\end{table}

Despite the computational error bars that are only about 4.3 times smaller
than the ones in the real data (limited by the computer time), this is
enough to show that this anisotropy is also unlikely to be responsible for
the observed result. It should be noted that the discussed muon anisotropy
need not be identical to that of the cosmic ray air showers in the Milagro
energy range. However, the result may be interpreted as a systematic
correction if the anisotropy is established to exist
\cite{systemat_error_facts_fiction}. Within limits of the calculation, the
correction is not inconsistent with being zero.

% ////////////////////////////////////////////////////////////
% ======  Pictures.....===========
\begin{figure}
\centering
\includegraphics[width=5.0in]{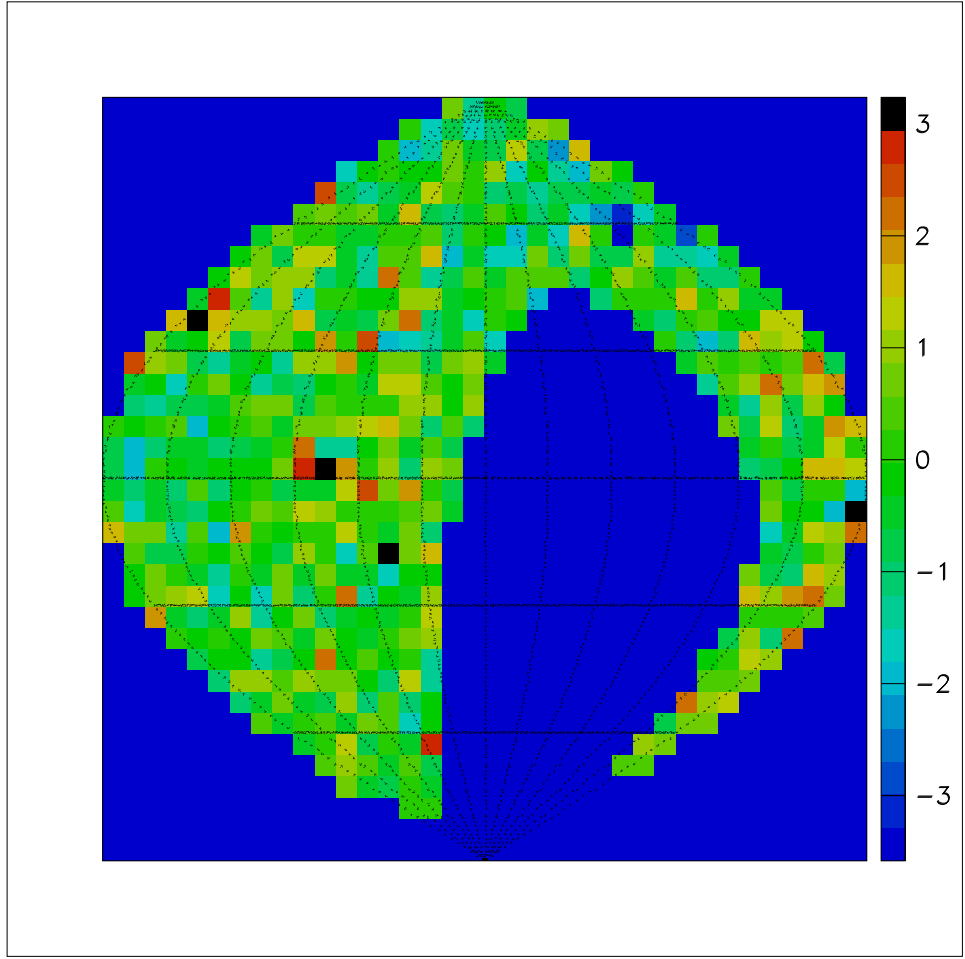}
\caption{Significance map of the sky in Galactic coordinates. Emission
         from the inner Galactic region is seen. Grid lines are plotted
         every $30^{\circ}$ in longitude and latitude. The Galactic
         center, not visible by Milagro, is in the middle of the
         map. Galactic longitude increases to the left.}
\label{fig:results:significance_map}
\end{figure}

% ////////////////////////////////////////////////////////////
% ======  Pictures.....===========
\begin{figure}
\centering
\includegraphics[width=5.0in]{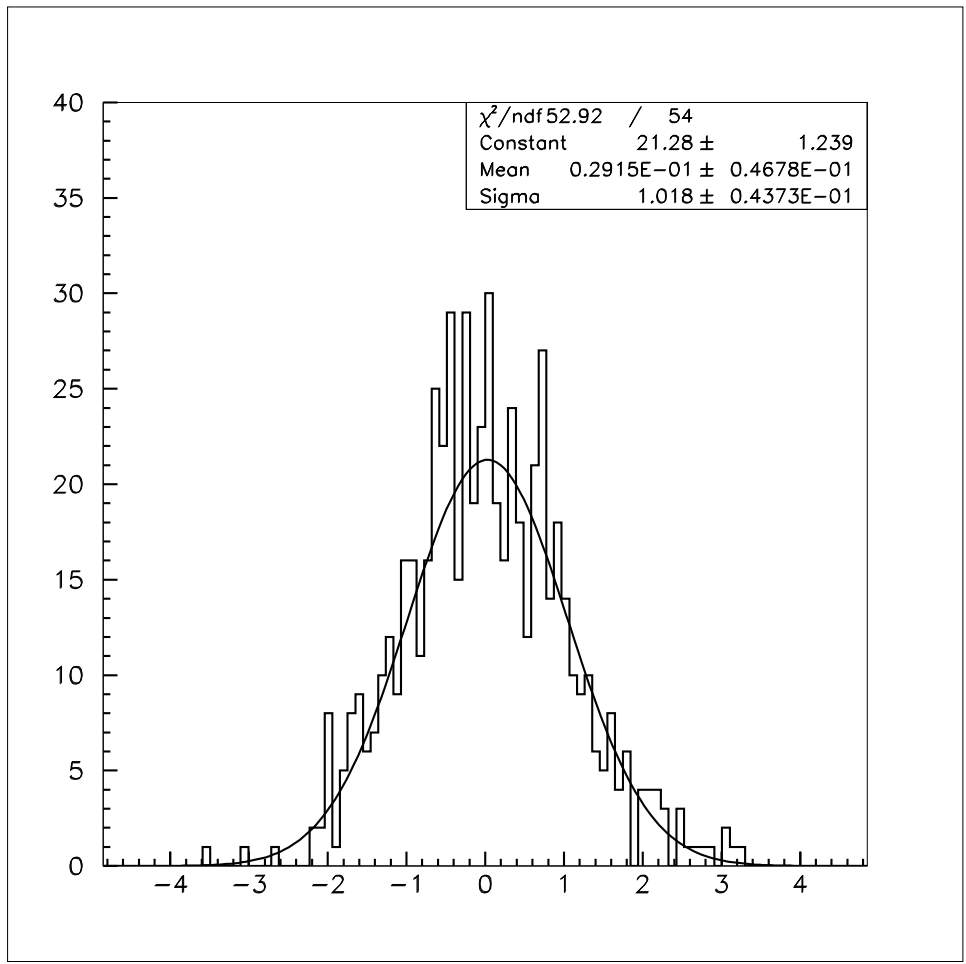}
\caption{The distribution of significances of individual cells of figure
         \ref{fig:results:significance_map}. The best fit of a Gaussian
         distribution is also shown. It agrees well with the standard
         normal distribution (zero mean and unit variance).}
\label{fig:results:significance_distrib}
\end{figure}

% ====== Pictures.....===========
\begin{figure} 
\centering
  \subfigure[Latitude profile of $N_{\gamma}/N_{b}$ for inner Galaxy.
             The value of $\chi^{2}$ with respect to 0 is 72.10 with 83
             bins.]{
    \label{fig:results:profile:ig}
    \includegraphics[width=5.0in]{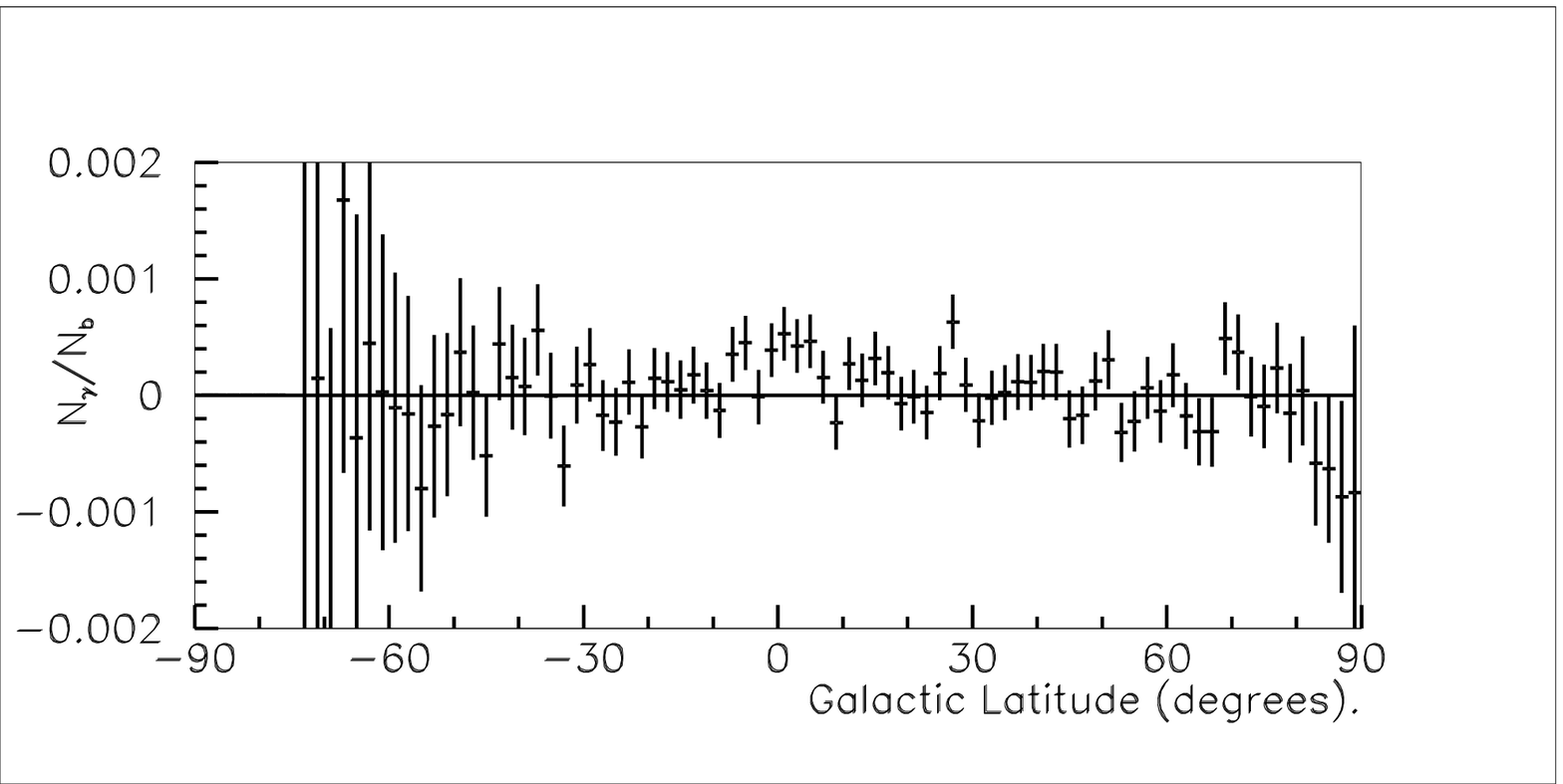}
  }
  \subfigure[Latitude profile of $N_{\gamma}/N_{b}$ for outer Galaxy. The
             value of $\chi^{2}$ with respect to 0 is 116.9 with 84   
             bins.]{
    \label{fig:results:profile:og}
    \includegraphics[width=5.0in]{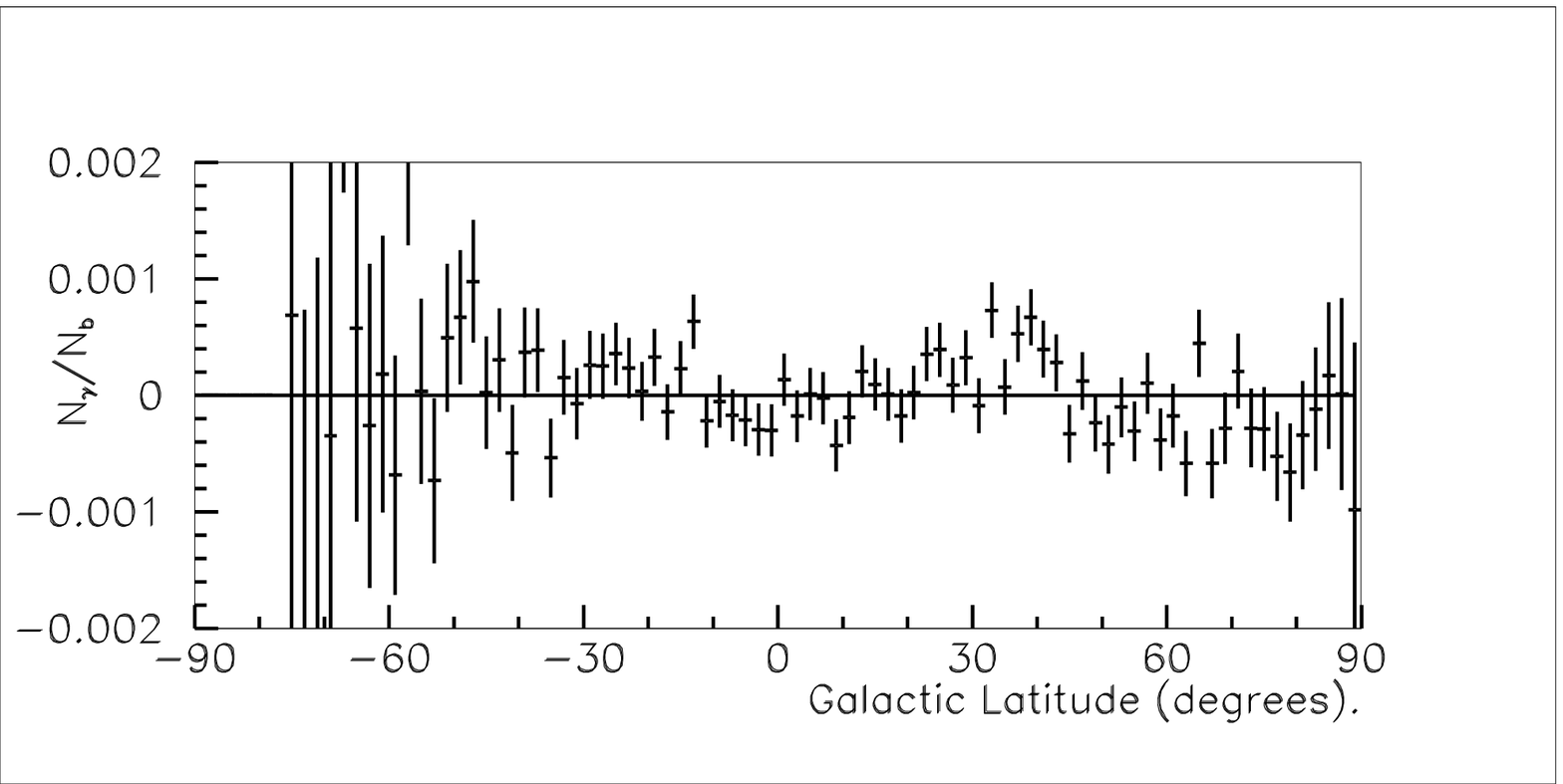}
  }

\caption{Latitude profiles of $N_{\gamma}/N_{b}$ for the inner  and outer
         Galaxy.}
\label{fig:results:profile}
\end{figure}

In the process of background estimation using the time swapping method,
the background was generated and saved for the entire sky, not only for
the regions of interest. This may serve to provide additional information.

Figure \ref{fig:results:significance_map} shows the 2 dimensional map of
the significance in the Galactic coordinates with coarse bins appropriate
for a large scale structure. An enhancement in the region of inner Galaxy
is seen. Another feature visible on the plot is a valley at
$l=-150^{\circ}$, $b=+60^{\circ}$. It may be interpreted as an indication
of a possible small anisotropy of the cosmic rays. Haverah Park
\cite{haverah_park} and Yakutsk \cite{yakutsk} experiments have found a
deficit of flux from the Galactic North in the energy range of
$10^{15}-10^{19}$ eV, however, the existence of such an anisotropy in the
Milagro energy range is unknown. Figure
\ref{fig:results:significance_distrib} presents the distribution of
significances of individual cells of figure
\ref{fig:results:significance_map}. The distribution is expected to be the
standard normal one (see discussion following equation
(\ref{equation:significance})) with which it agrees well. Therefore, the
observed excesses and deficits are statistically allowed.

A signature of the galactic nature of the emission is presence of a ridge
along the galactic equator. Therefore, another check is provided by a
lateral distribution plot of $N_{\gamma}/N_{b}$ (figure
\ref{fig:results:profile}). The plot for the inner Galaxy (figure
\ref{fig:results:profile:ig}) shows the excess in the region around the
equator ($b=0$) as expected and is flat outside suggesting that it is the
disk which produces the excess.  The plot for the outer Galaxy (figure
\ref{fig:results:profile:og}) does not show presence of excess in the
plane region, however, indicates a possible existence of a large scale
anisotropy. The exact interpretation of the values of $\chi^{2}$ given in
the captions to the figures is, however, difficult. Calculation of values
and errors for the points plotted outside of the excluded region suffers
from the dependence problem such as illustrated on figure
\ref{fig:galactic_latitide:mc}. Also, the points in adjacent bins are
positively correlated, therefore, number of degrees of freedom is
effectively lower than what is expected from the number of bins
considerations. However, if it is assumed that anisotropy is present and
is modeled by a straight line in the inner Galactic region and by a
parabola in the outer one, then it is possible to estimate its effect by
fitting a straight line in the region of $5^{\circ}<|b|<60^{\circ}$ for
the case of the inner Galaxy and quadratic polynomial in the region of
$5^{\circ}<|b|<40^{\circ}$ for the case of the outer Galaxy and
interpolating it into the plane $|b|<5^{\circ}$. This results in the
estimated contribution of $(+0.54 \pm 0.34) \cdot 10^{-4}$ (inner Galaxy)
and $(-0.48 \pm 0.65) \cdot 10^{-4}$ (outer Galaxy) to the measured ratios
$N_{\gamma}/N_{cr}$. These values have to be subtracted from the results
for the ratios of table \ref{table:results:results} as systematic
corrections if the anisotropy is known to be of the form discussed
\cite{systemat_error_facts_fiction}. They also are not inconsistent with
being zero.

It is therefore concluded that the outer Galaxy does not show excess and
the measurements may be used to set upper limits for the emission. The
results for the inner Galaxy, if interpreted as gamma ray emission, may
be used to calculate the gamma ray flux.

%/////////////////////////////////////////////
\section{Interpretation of the Results.}

In what follows it is assumed that the emission from the Galactic plane is
known to exist and that the observed excess in the galactic plane is due
to gamma rays. With these assumptions in mind, the results of the
measurements can be used to establish some properties of the emission.

Assuming simple power law emission spectrum independent of Galactic
coordinates for Galactic gamma rays in the Milagro energy range or,
equivalently, assuming gamma ray signal function to be in the form of
equation (\ref{equation:source_function_prototype}) and given power law
spectrum of cosmic-ray flux at the Earth ($F_{cr}(>1TeV) = 1.2 \cdot
10^{-5} (cm^{-2} s^{-1} str^{-1})$ \cite{Grigorov}, $\alpha_{cr} = 2.7$)
it is possible to use results of the table \ref{table:results:results} to
relate integral flux of gamma rays above 1 TeV ($F_{\gamma}(>1TeV)$) with
their assumed spectral index $\alpha_{\gamma}$ (equation
\ref{equation:flux_ratio}). The dimensionless coefficient
$\eta(\alpha_{\gamma}, \alpha_{cr})$ needed for this conversion is plotted
on figure \ref{fig:results:eta} for $\alpha_{cr} = 2.7$. The value of this
coefficient reflects efficiency of the cosmic ray rejection algorithm and
for $\alpha_{\gamma}=\alpha_{cr}$ has a direct interpretation of a ratio
of effective areas after the cut. The integral flux of gamma rays above 1
TeV obtained is plotted on figure \ref{fig:results:spectral_index} as
solid lines.

% ////////////////////////////////////////////////////////////
% ======  Pictures.....===========
\begin{figure}
\centering
\includegraphics[width=5.0in]{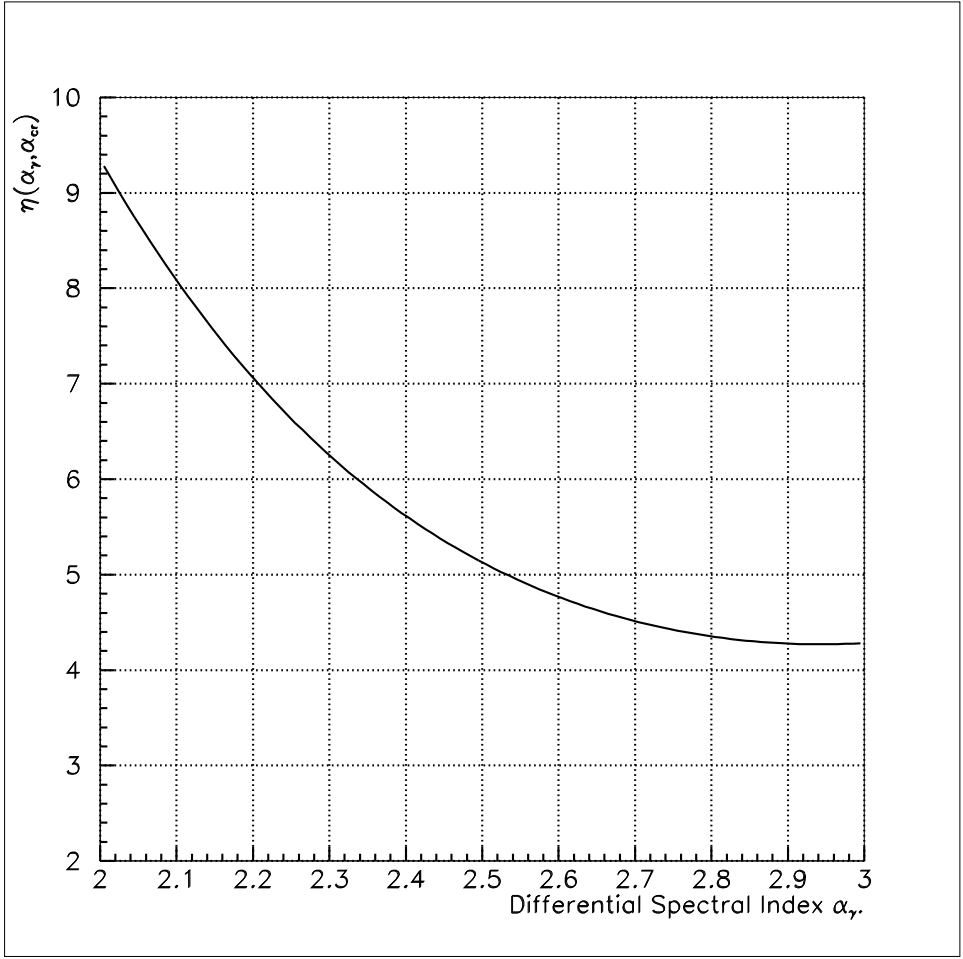}
\caption{Dimensionless coefficient $\eta(\alpha_{\gamma}, \alpha_{cr})$
         plotted as function of $\alpha_{\gamma}$ ($\alpha_{cr}=2.7)$ for
         the inner Galaxy with $|b|<5^{\circ}$. Other regions of the
         Galaxy considered have very similar exposure and therefore the
         presented coefficient is almost identical.}
\label{fig:results:eta}
\end{figure}

\begin{table}[htbp]
\begin{center}
\begin{tabular}{|c|c|c|} \hline
             &   $|b|<2^{\circ}$           &   $|b|<5^{\circ}$ \\ \hline
Inner Galaxy & $(1.82 \pm 0.25)\cdot 10^{-6}$ & $(1.18 \pm 0.14)\cdot 10^{-6}$ \\ \hline
Outer Galaxy & $(1.90 \pm 0.81)\cdot 10^{-7}$ & $(3.45 \pm 0.60)\cdot 10^{-7}$ \\ \hline

\end{tabular}
\end{center}
\caption{Average integral flux of photons ($cm^{-2}sr^{-1}s^{-1}$) in the
         10-30 GeV energy range measured by EGRET in the corresponding
         regions of the Galactic plane. Isotropic diffuse component has
         been subtracted \cite{{egret},{stan_hunter_private}}.}

\label{table:results:egret}
\end{table}

% ====== Pictures.....===========
\begin{figure} 
\centering
  \subfigure[Inner Galaxy, $|b|<2^{\circ}$.]{
    \label{fig:results:spectral_index:ig_2}
    \includegraphics[width=2.6in]{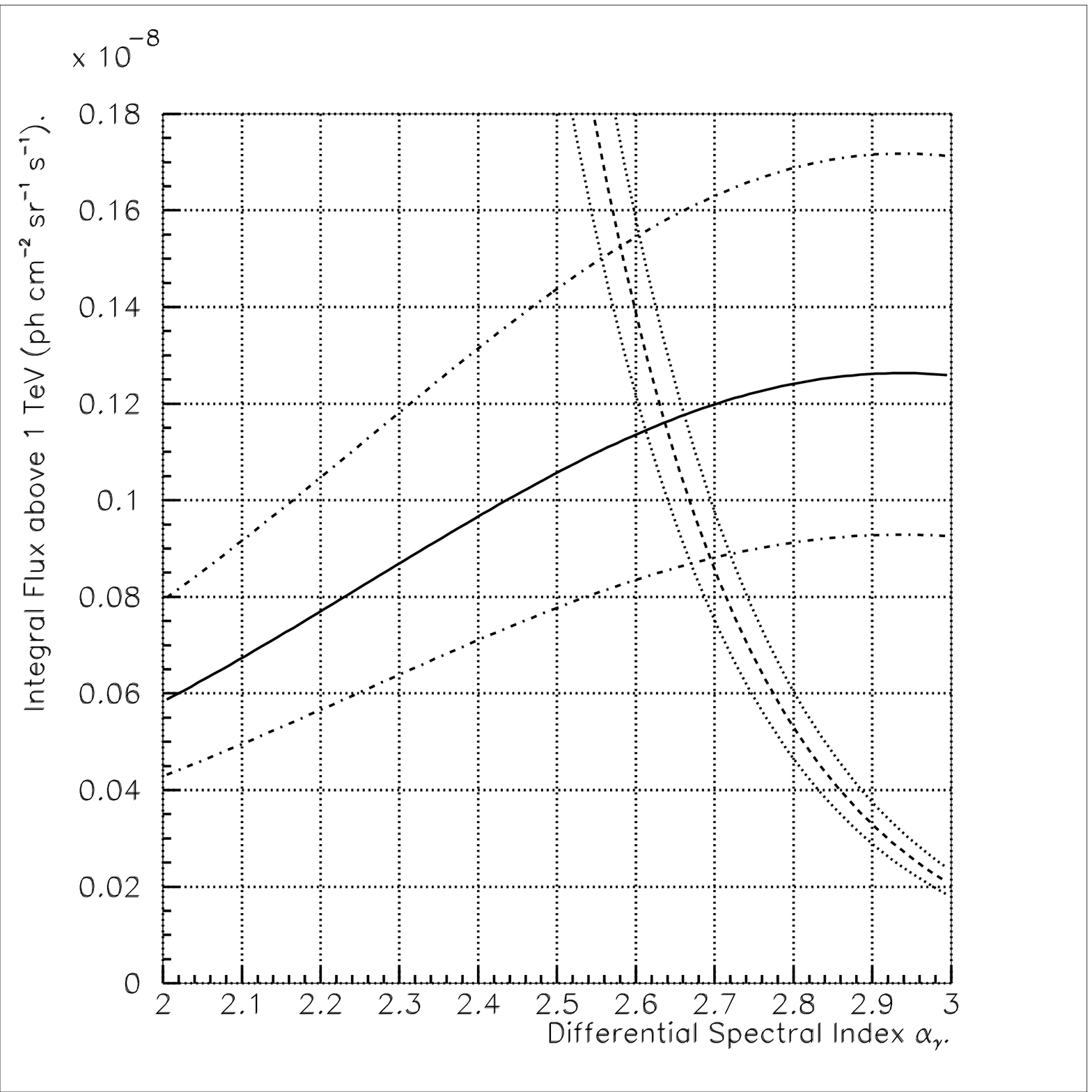}
  }
  \subfigure[Inner Galaxy, $|b|<5^{\circ}$.]{
    \label{fig:results:spectral_index:in_5}
    \includegraphics[width=2.6in]{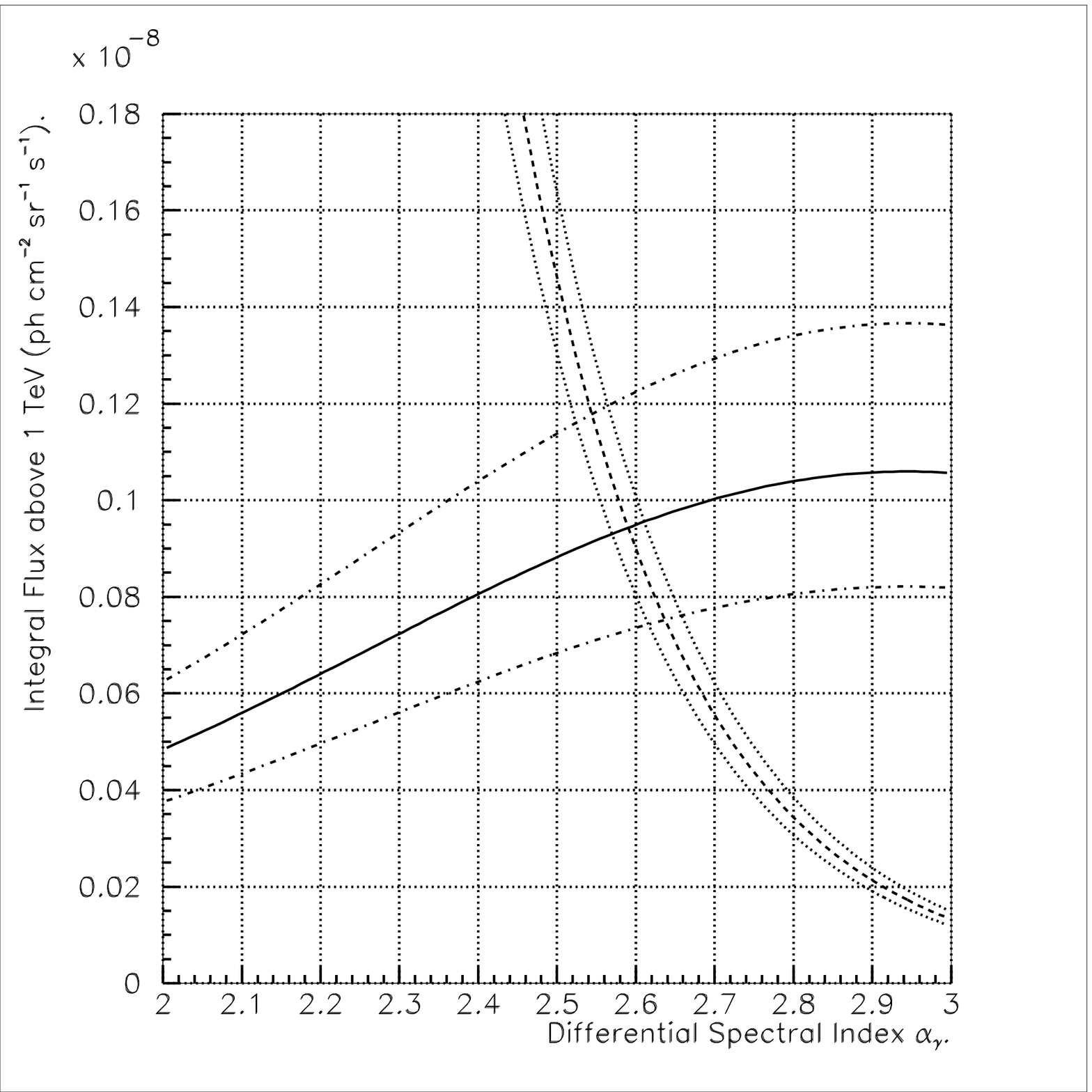}
  }

  \subfigure[Outer Galaxy, $|b|<2^{\circ}$.]{
    \label{fig:results:spectral_index:og_2}
    \includegraphics[width=2.6in]{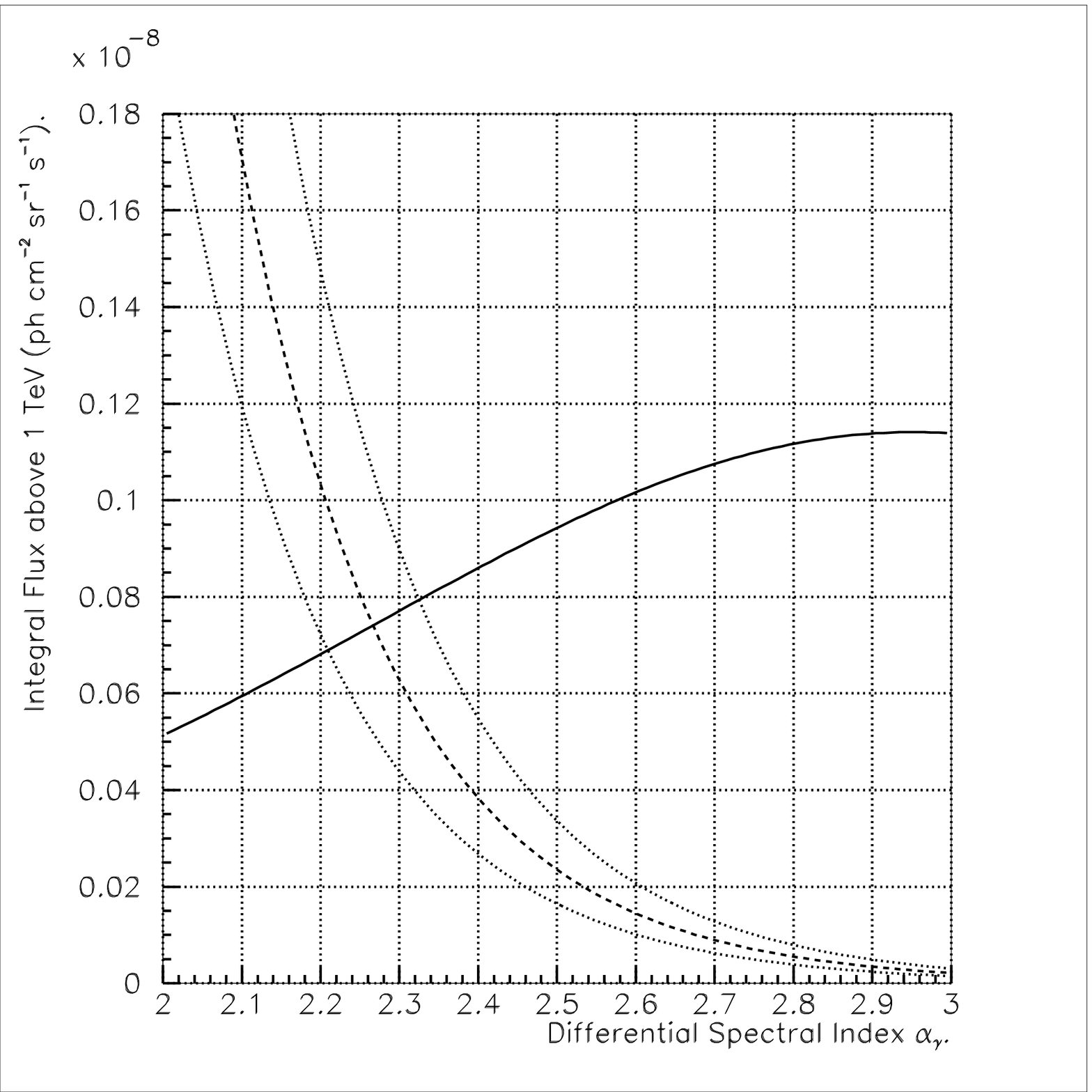}
  }
  \subfigure[Outer Galaxy, $|b|<5^{\circ}$.]{
    \label{fig:results:spectral_index:og_5}
    \includegraphics[width=2.6in]{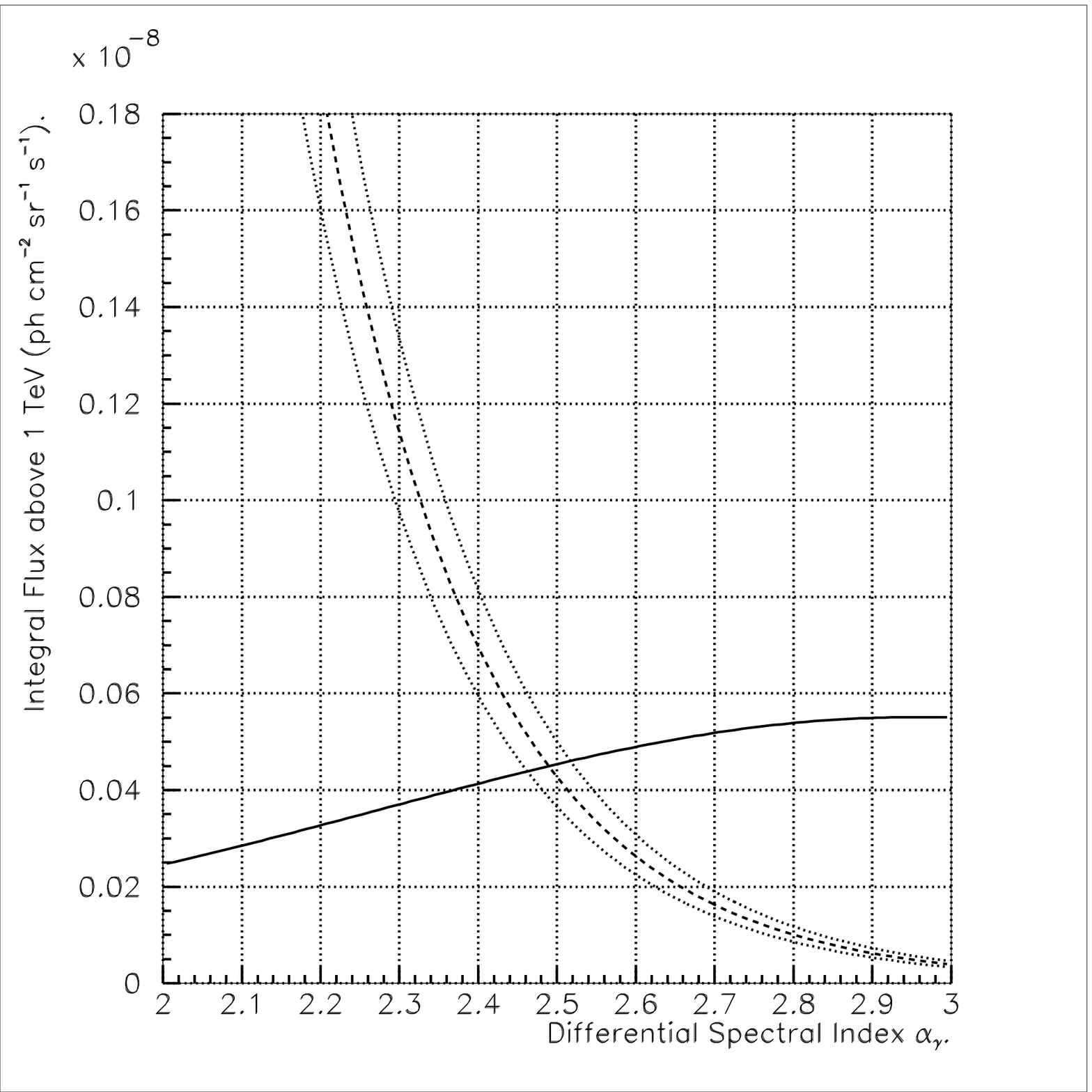}
  }

\caption{Integral flux of photons in different parts of Galactic disk with 
         energies above 1 TeV or corresponding 99.9\% upper limit for
         assumed differential spectral index $\alpha_{\gamma}$ are shown
         as solid lines. The dashed lines represent the extrapolated EGRET
         measurements in the 10-30 GeV energy  range and the same filed of
         Galaxy. The error corridors are shown as lines with dots.}
\label{fig:results:spectral_index}
\end{figure}

On the other hand, given the measurements of EGRET (table
\ref{table:results:egret}) and assuming that power law index
$\alpha_{\gamma}$ is constant above 10 GeV it is possible to extrapolate
the EGRET measurement in the highest energy bin 10-30 GeV to the 1 TeV
region. This is shown on figure \ref{fig:results:spectral_index} as dashed
lines. Both EGRET and Milagro measurements are presented together with the
corresponding error corridors. The intersection of the corridors provides
the spectral index and integral flux that is consistent with the EGRET and
Milagro data. In the case of Milagro upper limit, the limit itself is
presented. This comparison leads to estimation of the spectral index
$\alpha_{\gamma} = 2.59 \pm 0.07$ and flux $F(>1 TeV) = (9.5 \pm 2.0)
\cdot 10^{-10} \; cm^{-1}\; sr^{-1} \; s^{-1}$ in the region of inner
Galaxy and to a lower limit of $\alpha_{\gamma} > 2.49$ and flux $F(>1
TeV) < 4.5 \cdot 10^{-10} \; cm^{-1}\; sr^{-1} \; s^{-1}$ in the outer
Galaxy. (Regions with $|b|<5^{\circ}$ are used for final calculations
because of greater sensitivity of Milagro.) These findings may be used to
exclude models which predict a strong enhancement of the diffuse flux
compared to conventional mechanisms. Model predictions attempting to
explain the excess flux in the GeV region measured by EGRET by assuming an
increased inverse Compton component \cite{pohl_esposito} or by harder
proton spectrum \cite{galactic_flux3} or by considering contribution from
unresolved supernova remnants \cite{galactic_flux2} are given in
literature for different ranges in Galactic latitude and longitude and can
not be directly compared with the presented results. However, it is
possible to find spectral indexes in the energy range above 1 GeV as
measured by EGRET averaged over the Milagro inner and outer Galaxies
(table \ref{table:results:egret_index}) and compare them with the above
results. While the EGRET spectral index does not contradict to the Milagro
upper limit for the region of outer Galaxy, measurements of the index in
the inner Galaxy exclude the possibility of constant spectral index and
require softening of the spectrum between EGRET and Milagro energy
regions. This is also illustrated on figures
\ref{fig:results:integral_spectr} for the regions of inner and outer
Galaxy with $|b|<5^{\circ}$.

\begin{table}[htbp]
\begin{center}
\begin{tabular}{|c|c|c|} \hline
             &   $|b|<2^{\circ}$  &   $|b|<5^{\circ}$ \\ \hline
Inner Galaxy & $2.346 \pm 0.028$  &  $2.365 \pm 0.016$ \\ \hline
Outer Galaxy & $2.560 \pm 0.054$  &  $2.487 \pm 0.059$ \\ \hline

\end{tabular}
\end{center}
\caption{Spectral index $\alpha_{\gamma}$ derived from the EGRET data in the
         1-30 GeV energy range in the corresponding 
         regions of the Galactic plane. (The data was made available by
         \cite{stan_hunter_private}.)}
\label{table:results:egret_index}
\end{table}

% ====== Pictures.....===========
\begin{figure} 
\centering
  \subfigure[Inner Galaxy.]{
    \label{fig:results:integral_spectr:ig}
    \includegraphics[width=2.6in]{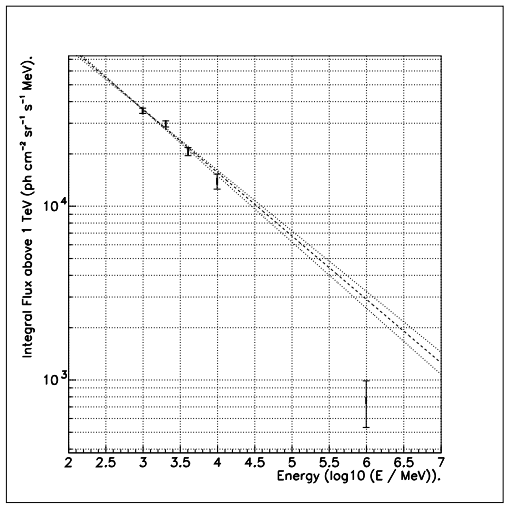}
  }
  \subfigure[Outer Galaxy.]{
    \label{fig:results:integral_spectr:og}
    \includegraphics[width=2.6in]{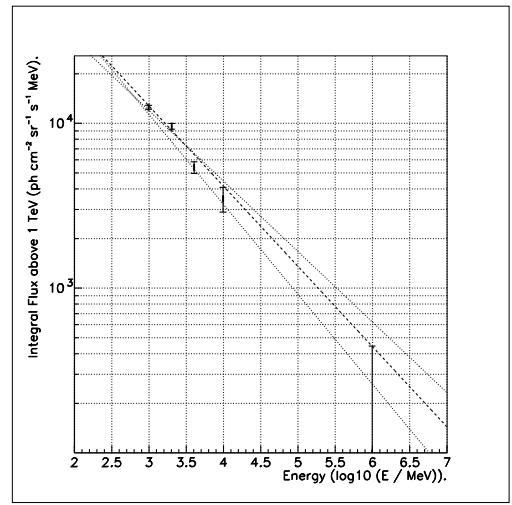}
  }

\caption{Integral flux of photons from $|b|<5^{\circ}$ region multiplied
         by photon energy $E$ is plotted. EGRET data points in the energy
         range of 1-30 GeV are also shown. The dashed line is the spectrum
         derived from the EGRET data above 1 GeV.}
\label{fig:results:integral_spectr}
\end{figure}

\chapter{Conclusions.}

A search for diffuse gamma ray emission from the Galactic plane has been
performed using data collected by the Milagro detector between July 19,
2000 and September 10, 2001. Four regions of the disk were studied. An
excess has been observed from the regions of the Milagro inner Galaxy
defined by $l \in (20^{\circ}, 100^{\circ})$, $|b| < 2^{\circ}$ and $|b| <
5^{\circ}$ with the significance of $2.6 \cdot 10^{-3}$ and $2.3 \cdot
10^{-4}$ respectively. The emission from the region of the Milagro outer
Galaxy defined by $l \in (140^{\circ}, 220^{\circ})$, $|b| < 2^{\circ}$
and $|b| < 5^{\circ}$ is not inconsistent with being that of background
only.

The results are interpreted assuming a simple power law energy spectrum.
Given the EGRET measurements, the integral gamma ray flux with energies
above 1 TeV has been obtained, yielding $F(>1 TeV) = (9.5 \pm 2.0) \cdot
10^{-10} \; cm^{-1}\; sr^{-1} \; s^{-1}$ with spectral index
$\alpha_{\gamma} = 2.59 \pm 0.07$ for the region of inner Galaxy. The
99.9\% upper limit for the diffuse emission in the region of outer Galaxy
is set at $F(>1 TeV) < 4.5 \cdot 10^{-10} \; cm^{-1}\; sr^{-1} \; s^{-1}$
assuming a differential spectral index of $\alpha_{\gamma} = 2.49$. These
observations are compatible with earlier upper limits obtained by the
Whipple observatory at 500 GeV \cite{whipple}. The upper limit for the
outer Galaxy is also consistent with the extrapolation of EGRET
measurements between 1 and 30 GeV. Extrapolation of the EGRET measurements
between 1 and 30 GeV for the region of inner Galaxy using constant power
law spectral index is incompatible with the Milagro data. This indicates
softening of the spectrum at energy between 10 GeV and 1 TeV. These
observations may be used to constrain models of Galactic emission
\cite{galactic_flux3,galactic_flux2,pohl_esposito}.

During preparation of this manuscript, the Milagro detector continued to
operate and collected more data. The analysis of these data will confirm
or refute the observations. Regardless of the outcome, the combined data
set will provide for more sensitive measurement of the diffuse gamma ray
emission at energies near 1 Tev. The Milagro detector is being
continuously upgraded both on software and hardware level. New event
reconstruction algorithms may become available in the near future. These,
however, will have no affect on the data collected so far. In terms of
hardware, Milagro is being supplimented by an outrigger array, which will
enable better hadron rejection as well as energy determination and angular
resolution. As more and better quality data become available, it will be
possible to study the question of cosmic ray anisotropy extensively using
the Milagro data itself. This presents an interest not only in the context
of the search for the Galactic gamma ray emission, but is valuable by
itself as the detection of Compton-Getting effect may become possible
\cite{Compton-getting}.

%%%%%%%%%%%%%%%%%%%%%%%%%%%%%%%%%%%%%%%%%%%%%%%%%%%%%%%%%%%%%%%%%%%%%%%%%
%                     Appendices
%%%%%%%%%%%%%%%%%%%%%%%%%%%%%%%%%%%%%%%%%%%%%%%%%%%%%%%%%%%%%%%%%%%%%%%%%
\appendix                           % appendices
\chapter{Arguments pertaining to equation \ref{equation:significance}. 
         \label{appendix:poisson_arguments}}

The Poisson distribution of an integer random variable $n$ is given by:

\[
   P_{\mu}(n) = \frac{\mu^{n}}{n!} e^{-\mu}
\]
where $\mu$ is a parameter of distribution. One can show that both average
and dispersion of $n$ are equal to $\mu$. It is intended to show that for
large $\mu$, the Poisson distribution approaches Gaussian:

\[
   G(x) = \frac{1}{\sqrt{2\pi}\sigma}
          e^{-\frac{(x-x_{0})^{2}}{2\sigma^{2}}} 
\]
with average $x_{0} = \mu$ and dispersion $\sigma^{2} = \mu$.

For the case of large $\mu$ and $n$ being not too different from it, the
Stirling formula for the factorial can be applied:

\[
  n! \approx \sqrt{2 \pi n} \; n^{n} e^{-n}
\]
with relative error of order of $(e^{\frac{1}{12n}}-1)$, leading to

\[ P_{\mu}(n) \approx \frac{e^{n-\mu}}{\sqrt{2 \pi n}} \;
    \left(\frac{\mu}{n}\right)^{n} = e^{y(n)}
\]

The exponent $y(n)$ can be expanded in Taylor series around $\mu$:

\begin{center}
\begin{tabular}{rcl}
$y(\mu)$        & = &  $-\ln\sqrt{2\pi\mu}$    \\
$y'(\mu)$       & = &  $-\frac{1}{2\mu}$    \\
$y''(\mu)$      & = &  $-\frac{1}{\mu}\left(1+\frac{1}{2\mu}\right)$ \\
$y'''(\mu)$     & = &  $+\frac{1}{\mu^{2}}\left(1-\frac{1}{\mu}\right)$    \\
$y^{(4)}(\mu)$  & = &  $-\frac{2}{\mu^{3}}\left(1-\frac{3}{2\mu}\right)$
\end{tabular}
\end{center}

\[
y(n) = -\ln\sqrt{2\pi\mu} -\frac{1}{2\mu} (n-\mu)
       -\frac{1}{2\mu}\left(1+\frac{1}{2\mu}\right) (n-\mu)^{2}
       +
\]
\[ 
       +\frac{1}{6\mu^{2}}\left(1-\frac{1}{\mu}\right) (n-\mu)^{3}
       +\cdots
       +O\left(\frac{(n-\mu)^{k}}{\mu^{k-1}} \right)
\]

In this Taylor series (recalling that $\mu$ is large) the second term can
be neglected compared to the leading if $|n-\mu| \ll
2\mu\ln\sqrt{2\pi\mu}$, third if $|n-\mu| \ll
\sqrt{2\mu\ln\sqrt{2\pi\mu}}$, fourth if $|n-\mu| \ll
(6\mu^{2}\ln\sqrt{2\pi\mu})^{1/3}$, all higher terms if $|n-\mu| \ll
(\mu^{k-1}\ln\sqrt{2\pi\mu})^{1/k}$. It is seen that the first condition
is least restrictive one, second, however, is the most limiting. Thus,
keeping only two leading terms in $|n-\mu|$ will result:

\[ y(n) \approx -\ln\sqrt{2\pi\mu}
          -\frac{1}{2\mu}\left(1+\frac{1}{2\mu}\right)(n-\mu)^{2} 
\]
with restriction $|n-\mu| \ll (6\mu^{2}\ln\sqrt{2\pi\mu})^{1/3}$. This
means that $n$ has to be not too far off $\mu$ and when $\mu$ is large the
use of the Sterling formula is justified. The Poisson distribution is
approximated by Gaussian in the central region:

\[ 
 P_{\mu}(n) = \frac{\mu^{n}}{n!} e^{-\mu} \approx
      \frac{1}{\sqrt{2\pi\mu}} e^{-\frac{(n-\mu)^{2}}{2\mu}},
      \;\;\;
      |n-\mu| \ll (6\mu^{2}\ln\sqrt{2\pi\mu})^{1/3},
      \;\;\;
      \mu \gg 3/2
\]

Examining the equation \ref{equation:significance} if null hypothesis is
true, it follows from the above, that the numerator is distributed under
the Gauss law with zero mean if both $N_{1}$ and $N_{2}$ are large and are
not too different from $\mu_{1,2} = \lambda t_{1,2}$. By construction, if
null hypothesis is true and $N_{1,2} \sim \mu_{1,2}$, the dispersion of
$S$ is approximately unity, therefore, the statistic $S$ has normal
distribution in $-S_{0}<S<S_{0}$ region around zero. The bound $S_{0}$ is
defined by the smallest of the $N$'s and is on the order of

\begin{equation}
   S_{0} \ll \left(6N^{1/2}\ln\sqrt{2\pi N}\right)^{1/3}
  \label{equation:lima_bound}
\end{equation}

A Monte Carlo simulation helps to verify the statement. Figure
\ref{fig:lima} shows the results of such a simulation.

% ////////////////////////////////////////////////////////////
\begin{figure}
\centering
  \subfigure[Null hypothesis with parameters $\mu_{2}=5000$,
             $\mu_{1}=\alpha\mu_{2}$, $\alpha = 0.1$ was used. According
             to equation (\ref{equation:lima_bound}), $S_{0} \ll 8$.]{
    \label{fig:lima:a}
    \includegraphics[width=2.7in]{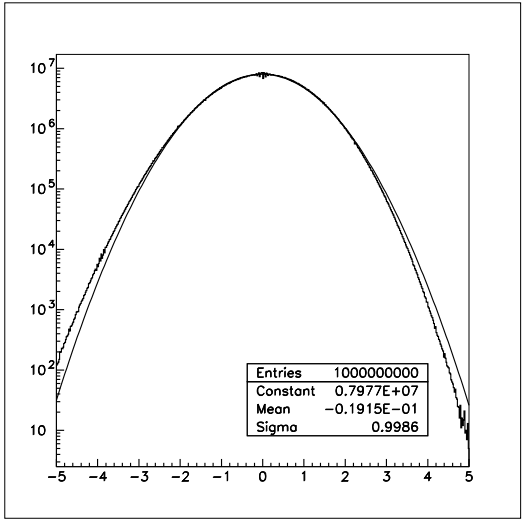}
  }
  \subfigure[Null hypothesis with parameters $\mu_{2}=15\cdot 10 ^{6}$,
             $\mu_{1}=\alpha\mu_{2}$, $\alpha = 0.1$ was used. According 
             to equation (\ref{equation:lima_bound}), $S_{0} \ll 39$.]{
    \label{fig:lima:b}
    \includegraphics[width=2.7in]{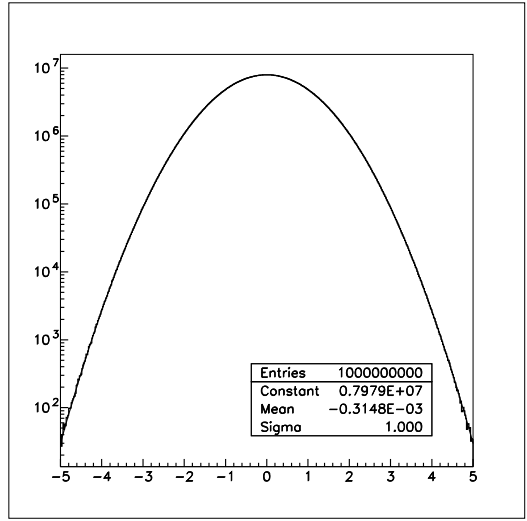}
  }
\caption{Distributions of statistic $S$ (equation
         \ref{equation:significance}) obtained in two runs of Monte Carlo
         simulations. Best fit to Gauss distribution is made, parameters
         of the fit are presented. The two curves should agree in
         the $S_{0}$-neighborhood around zero. Number of entries in each
         run (about $10^{9}$) was chosen to provide reasonable accuracy in
         the region plotted.}
\label{fig:lima}
\end{figure}

\chapter{Solution of Background Equations.\label{chapter:background_equations}}

The problem of exclusion of the source region from the background
estimation had led to the problem of solution of the integral equations:

\[
  \left\{
  \begin{array}{lcl}
   N_{out}(x) & = & G(x) \int \phi(x,t') \cdot R(t') \; dt'  \\
   R_{out}(t) & = & R(t) \int \phi(x',t) \cdot G(x') \; dx'
  \end{array}
  \right|
\]

with respect to $G(x)$ and $R(t)$. $N_{out}(x)$ and $R_{out}(t)$ represent
the input data collected from the off-source region, $\phi(x,t)$ is the
function which defines its bounds: it is one if $(x,t)$ points into the
off-source region and zero otherwise. It is first noted that both $G(x)$
and $R(t)$ enter into equations only as a product $G(x) \cdot R(t)$,
therefore, normalization of either of them does not make any difference as
long as the product is preserved. Also, if there is a point $x_{0}$ in the
local coordinates which is never exposed into the off-source region, that
is $\phi(x_{0},t) = 0, \; \forall t$, then $N_{out}(x_{0}) = 0$ and the
first equation becomes:

\[
    0 = G(x_{0}) \cdot 0 
\]

leading to $G(x_{0})$ being undefined. Such regions, if any, are marked
with a negative value, on-source events with local coordinates from these
regions are discarded as having no corresponding background estimate. In
such cases, measurement is impossible within the environs of assumptions.

% ////////////////////////////////////////////////////////////
% ======  Pictures.....===========
\begin{figure}
\centering
\includegraphics[width=5in]{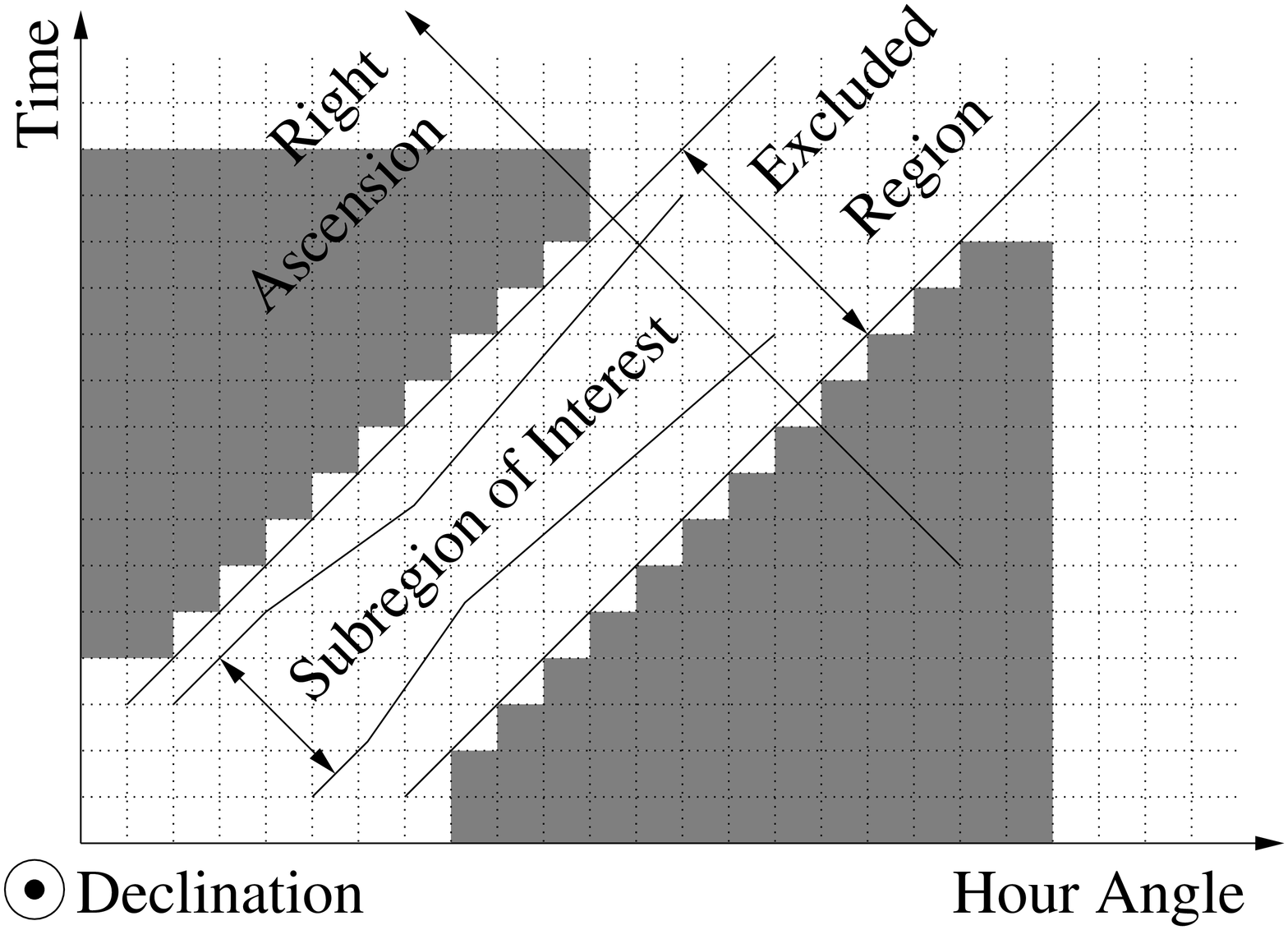}
\caption{To the formulation of backgroud equations.}
\label{fig:background:solution}
\end{figure}

% ////////////////////////////////////////////////////////////
% ======  Pictures.....===========
\begin{figure}
\centering
\includegraphics[width=5in]{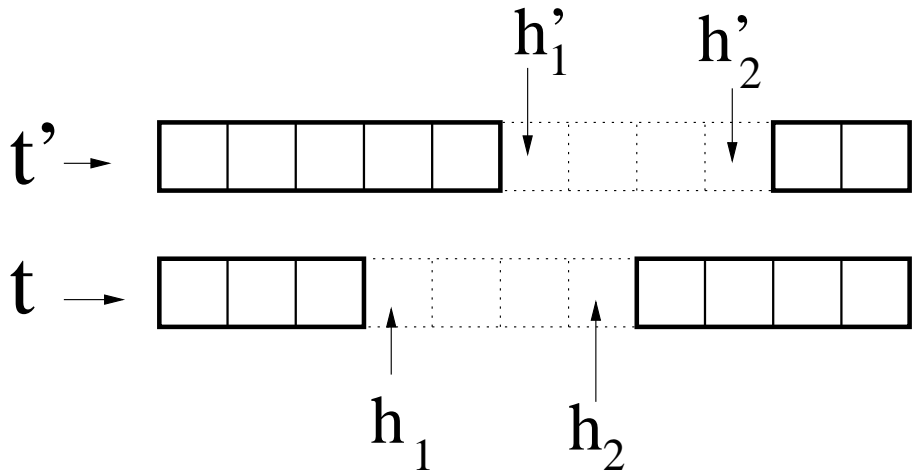}
\caption{To the derivation of recursion relations.}
\label{fig:background:recursion}
\end{figure}

The system of equations is solved numerically. The equations are
discretized on a fine grid corresponding to $0.1^{\circ}$ in $x$ and $t$,
the integrals become sums, $x$ and $t$ become indexes. Since most often
one is interested in celestial objects, the off-source region is defined
in declination and right ascension $(\delta, \alpha)$ and due to
discretization presents the bound indexes $\alpha_{1}$ and $\alpha_{2}$,
$\alpha_{1} < \alpha_{2}$ for every declination index $\delta$. Therefore,
for this problem, it is natural to consider local coordinates $x$ as
declination and hour angle $(\delta, h)$ rather than zenith and azimuth.
The situation is illustrated on figure \ref{fig:background:solution}. The
shaded area is the off-source region bounded by $\phi(x,t)$ in its
discrete form. The region of interest, the on-source region, is defined by
some other conditions which are irrelevant for the background equations as
long as both regions do not overlap.

Thus, the excluded region, defined by interval $(\alpha_{1}, \alpha_{2})$
is translated to local coordinates $h$ and $t$ as region inside intervals
$(h_{1},h_{2})$ and $(t_{1},t_{2})$, where

\[
  \begin{array}{l}
    \left\{
    \begin{array}{lcl}
     h_{1} & = & t - \alpha_{2} \; \bmod 360^{\circ}  \\
     h_{2} & = & h_{1} + \alpha_{2} - \alpha_{1}
    \end{array}
    \right.
    \\
    \\
    \left\{
    \begin{array}{lcl}
     t_{1} & = & h + \alpha_{2} \; \bmod 24 \; hours  \\
     t_{2} & = & t_{1} + \alpha_{2} - \alpha_{1}
    \end{array}
    \right.
  \end{array}
\]

Because of these relations it is convenient to quantize all quantities in
the same fashion. This, however, is not required. Here, it is adopted that
$\alpha_{1}$ is the index of the first excluded bin, $\alpha_{2}$ is the
index of the last excluded bin. So will be $h_{1},t_{1}$ and
$h_{2},t_{2}$. The integration is performed using recursion formulas which
are illustrated on the $\rho(t) = \int \phi(h',t) G(h') dh'$ example.
After integral is replaced by sum, it becomes:

\[
  \rho(t) = \sum_{h=0}^{h_{1}-1} G(h) + \sum_{h=h_{2}+1}^{N-1} G(h)
\]

where $N$ is the number of cells from $0$ to $N-1$ created by the
discretization of the hour angle between $0^{\circ}$ and
$360^{\circ}$. The limits of the summation are included, that is

\[
  \sum_{h=a}^{b} G(h) = G(a) + G(a+1) + \cdots + G(b-1)+G(b)
\]

Then, it is seen (figure \ref{fig:background:recursion}) that the integral
at some other time $t'$ is given by

\[
  \rho(t') = \sum_{h=0}^{h_{1}'-1} G(h) + \sum_{h=h_{2}'+1}^{N-1} G(h)
           = \rho(t) + \Delta_{1} - \Delta_{2}
\]

where

\[
  \begin{array}{l}
    \left\{
    \begin{array}{lr}
     \Delta_{1}  =  +\sum_{h_{1}}^{h_{1}'-1} G(h),    &  h_{1}' > h_{1} \\
     \Delta_{1}  =  -\sum_{h_{1}'}^{h_{1}-1} G(h),    &  h_{1}' < h_{1} \\
     \Delta_{1}  = 0,                                 &  h_{1}' = h_{1}
    \end{array}
    \right.
    \\
    \\
    \left\{
    \begin{array}{lr}
     \Delta_{2}  =  +\sum_{h_{2}+1}^{h_{2}'} G(h),    &  h_{2}' > h_{2} \\
     \Delta_{2}  =  -\sum_{h_{2}'+1}^{h_{2}} G(h),    &  h_{2}' < h_{2} \\
     \Delta_{2}  = 0,                                 &  h_{2}' = h_{2}
    \end{array}
    \right.
  \end{array}
\]

This scheme is applied for every declination. Integration over time is
performed in the similar fashion. It is important to note, that case of
$\alpha_{1} = \alpha_{2}$ corresponds to exclusion of one bin. Absence
of excluded region for a declination is encoded as $\alpha_{2} =
\alpha_{1} -1$, which is the only case with $\alpha_{2} < \alpha_{1}$
allowed. The system is solved by iterations where $R_{out}(t)$ is used as
first approximation to $R(t)$. Then, the first equation is solved with
respect to $G(x)$ which is used in the second equation to update
$R(t)$. The process continues until desired accuracy is reached. In
practice, 13-digit precision is achieved after about 15 iterations.

\chapter{Zenith Correction.\label{chapter:zenith_correction}}

\begin{table}[htbp]
\[
\begin{array}{|c|c|c|c|c|c|c|} \hline
 Period      &    a      &    b      &      c    &     d     &     e     & A \\
  (MJD)      & (10^{-6}) & (10^{-3}) & (10^{-4}) & (10^{-1}) & (10^{-2}) & (10^{0}) \\ \hline
 1745   &&&&&& \\
\updownarrow & +203.0670 & -174.6125 & +10721.45 & -16.27103 & -5.631589 & -122.02 \\ 
 1813   &&&&&& \\
\updownarrow & -27.00866 & -19.20310 & +291.3698 & +4.232470 & -129.1226 & -698.0 \\
 1876   &&&&&& \\
\updownarrow & +56.50721 & -32.53573 & +1355.857 & +1.621818 & -108.8645 & -551.0 \\
 1937   &&&&&& \\
\updownarrow & +124.5009 & -12.71605 & +171.8406 & +3.488231 & -116.1735 & -1300.0 \\
 2000   &&&&&& \\
\updownarrow & -4.604483 & -10.42552 & +4.944296 & +4.006497 & -121.4843 & -1260.0 \\
 2069   &&&&&& \\
\updownarrow & -32.60869 & -8.961481 & -297.4525 & +5.175355 & -133.4420 & -1100.0 \\
 2117   &&&&&& \\
\updownarrow & -27.23581 & -11.71815 & +27.33091 & +4.266234 & -126.5341 & -1110.0 \\
 2163   &&&&&& \\ \hline
\end{array}
\]
\caption{Coefficients of the zenith correction function
$\Delta(z)$ derived for data which passes the hadron rejection cut
$X_{2}>2.5$. MJD, the Modified Julian Day, identifies the validity time
period. For instance, MJD 1745 corresponds to July 19, 2000; MJD 2163
corresponds to September 10, 2001 \cite{julian_calculator}.}
\label{table:zenith:coefficients}
\end{table}

% ////////////////////////////////////////////////////////////
% ======  Pictures.....===========
\begin{figure}
\centering   
\includegraphics[width=5in]{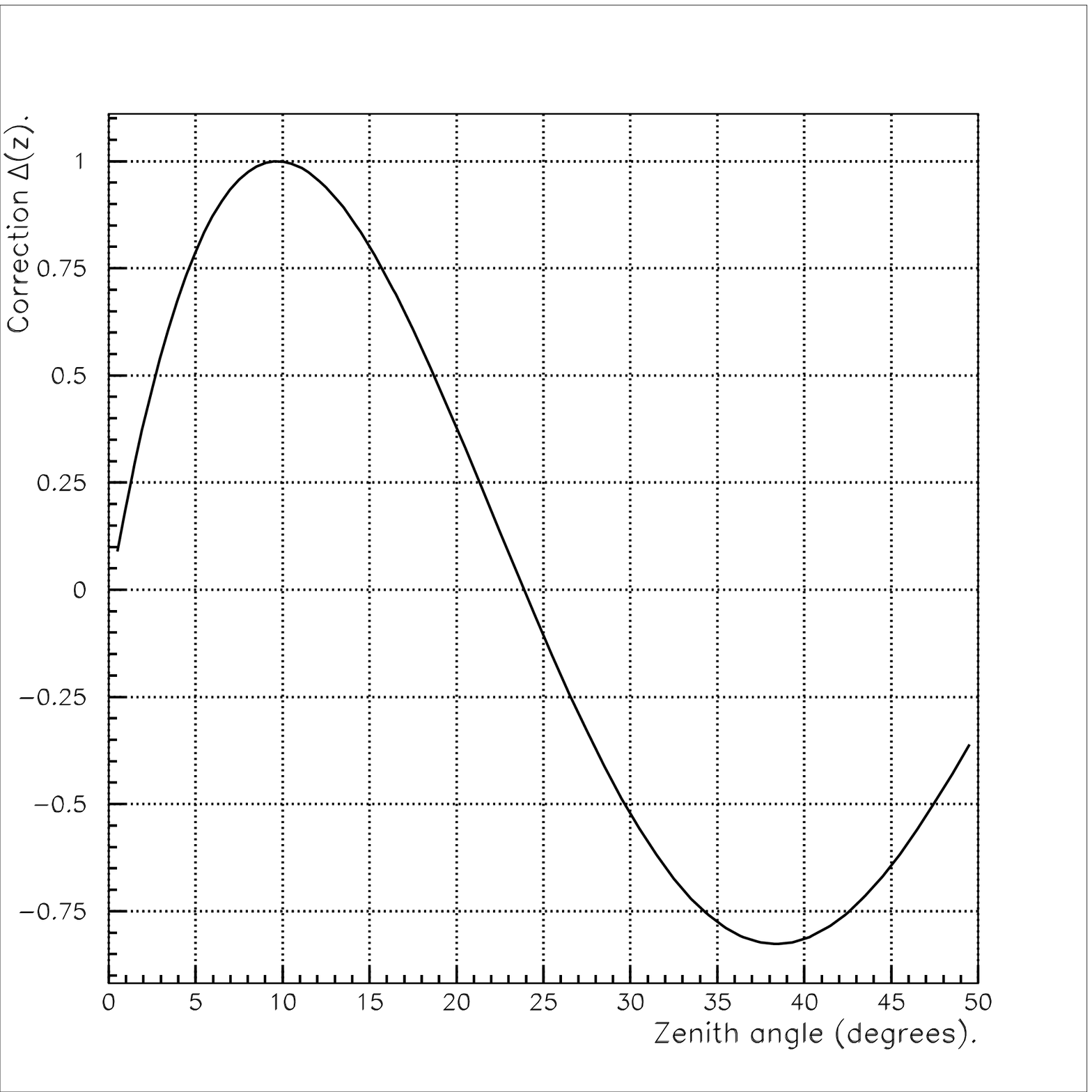}
\caption{Correction function $\Delta(z)$. Coefficients corresponding to
period MJD 1745 - MJD 1813 were used.}
\label{fig:zenith_variation}
\end{figure}

The zenith correction function $\Delta(z)$ describes deviation of the
current shape of the zenith distribution from the average one over the
same UT day. It is obtained in the iteration process in the form of
polynomial of 5-th degree starting with some zero-th approximation. The
differences between the two normalized histograms are accumulated
according to the sign of amplitude $\theta$ of the fit of a difference
into $\theta\Delta(z)$. When all differences according to their
corresponding sing of amplitudes are accrued, coefficients of the
$\Delta(z)$ are adjusted to provide the best fit. This is the updated
function $\Delta(z)$. The procedure is repeated until coefficients of the
polynomial stop changing. Then, $\Delta(z)$ is rescaled so that its
maximum is equal to one. This is the final form of $\Delta(z)$. The $K(z)$
function is obtained from  the $\Delta(z)$ by

\[
  K(z) = \frac{\Delta(z)}{G(z)}
\]

where $G(z)$ is the average zenith distribution. Coefficients of the
correction function $\Delta(z)$ are given in the table
\ref{table:zenith:coefficients} with its shape defined by:

\[
 \Delta(z) = A(y^{5} + e y^{4} + d y^{3} + c y^{2} + b y + a)
\]

where $y=z/90$, $0^{\circ}< z < 50^{\circ}$.

The example of the $\Delta(z)$ curve is presented on figure
\ref{fig:zenith_variation} The field of view of the detector was
artificially made limited to $0^{\circ}< z < 50^{\circ}$ because zenith
angle distribution variations outside of this range could not be
characterized by such a simple dependence.

%%%%%%%%%%%%%%%%%%%%%%%%%%%%%%%%%%%%%%%%%%%%%%%%%%%%%%%%%%%%%%%%%%%%%%%%%
%                       bibliography (MUST BE LAST)
%%%%%%%%%%%%%%%%%%%%%%%%%%%%%%%%%%%%%%%%%%%%%%%%%%%%%%%%%%%%%%%%%%%%%%%%%

\end{thesisbody}
%\today           % to see when it was compiled remove in final version

\begin{thebibliography}{999}

% ///////////////////////////  introduction

\bibitem{galactic_flux3}
F. A. Aharonian and A. M.  Atoyan, astro-ph/0009009, 2000.

\bibitem{galactic_flux2}
E. G. Berezhko and H. J. V\"{o}lk, Astrophysical Journal, 540, 923-929,
2000.

\bibitem{casa-mia}
A. Borione et al., Astrophysical Journal, 493, 175, 1998.

\bibitem{galactic_flux}
P. Chardonnet et al., Astrophysical Journal, 454, 774, 1995.

\bibitem{cos-b}
F. Lebrun et al., Astrophysical Journal, 274, 231, 1983.

\bibitem{sas2}
R.C. Hartman et al., Astrophysical Journal, 230, 597-606, 1979.

\bibitem{egret}
S. D. Hunter et al., Astrophysical Journal, 481, 205, 1997.

\bibitem{whipple}
P. T. Reynolds et al., Astrophysical Journal, 404, 206, 1993.

\bibitem{ong}
R. Ong, Physics Reports, 305, 93, 1998.

\bibitem{egretweb}
http://cossc.gsfc.nasa.gov/cossc/EGRET.html.


% /////////////////////////// Diffuse emission
\bibitem{neutralino_urban}
M. Urban et al., Phys. Lett. B, 293, 149 1992.

\bibitem{bloemen}
H. Bloemen, Annual Review Astronomy and Astrophysics, 27, 469-516, 1989.

\bibitem{pohl_esposito}
M. Pohl and J. A. Esposito, Astrophysical Journal, 507, 327, 1998.

\bibitem{model_bertsch}
D. L. Bertsch et al., Astrophysical Journal, 416, 587, 1993.

% /////////////////////////// Shower development

\bibitem{Longair}
M.S. Longair, High Energy Astrophysics, vol. 1, Cambridge Unversity Press,
1992.

\bibitem{Rossi_Greisen}
B. Rossi and K. Greisen, Rev. Mod. Phys., 13, 240 (1941). \\
B. Rossi, High Energy Particles, Prentice-Hall, Englewood Cliffs, N.J.,
1952.
% (from Hayakawa page 141.)

\bibitem{Sciascio}
B. D'Ettorre Piazzoli and G. Di Sciascio, Astropart. Phys. 2
(1994) 1999; Erratum, 2 (1994) 327. \\
G. Di Sciascio, B. D'Ettorre Piazzoli and M, Iacovacci, Astropart. Phys. 6
(1997) 313.

\bibitem{particle_book}
see for example Review of Particle Properties, Physical Review D50, 1173,
1994.

\bibitem{Becker}
R. Becker-Szendy et al. Nuclear Instruments and Methods, A, 324, 363
(1993).

\bibitem{gaisser}
T. K. Gaisser, Cosmic Rays and Particle Physics, Cambridge University
Press, 1990.

% /////////////////////////// Detector

\bibitem{milagrito:nim}
R.. Atkins et al., Nuclear Instruments and Methods, A449, 478,
2000.

\bibitem{milagro_water}
D. Coyne and M. Schneider, First Results on Milagro Water Attenuation
Using Measurements from the Upgraded TUBE, Internal Milagro Memo, August
14, 2002.

\bibitem{corsika_web}
http://ik1au1.fzk.de/$\sim$heck/corsika/

\bibitem{geant_web}
http://wwwinfo.cern.ch/asd/geant/index.html



% /////////////////////////// Calibration


\bibitem{IMB_calibration}
R. Becker-Szendy et al. Nuclear Instruments and Methods, A, 352, 629
(1995).

\bibitem{tped_memo}
Lazar and Roman Fleysher and Peter Nemethy, with Isabel Leonor, Timing
Pedestals, Internal Milagro Memo, July 7, 1997.

\bibitem{tped_icrc}
L. Fleysher for the Milagro Collaboration, Proc. 26-th ICRC, Salt Lake
City, USA, OG4.4.03, 1999.


\bibitem{isabel_memo}
Isabel Leonor, TOT-PE Conversion using Occupancy Method, Internal
Milagro Memo, April 22, 1998.

\bibitem{Milagro_calibration_manual}
Lazar Fleysher and Roman Fleysher, Milagro Calibration System, Internal
Milagro Manual, June 28, 2000.

\bibitem{random_invitation}
Lazar Fleysher and Roman Fleysher, An Introduction to Random Functions,
Internal Milagro Memo, May 7, 1999.

\bibitem{random_pugachev}
V. S. Pugachev, Theory of Random Function and its Applications to Control
Problems. Oxford; New York; Pergamon, Mass. Addison-Wesley, 1965.

\bibitem{slewing_extrapolation}
Lazar Fleysher, Roman Fleysher, Allen Mincer, Peter Nemethy, Slewing
Extrapolation, Internal Milagro Memo, April 20, 2000.

\bibitem{stability_study}
Peter Nemethy, Lazar Fleysher, Roman Fleysher, Time Calibration: stability
studies, Internal Milagro Memo, June 5, 2000.


% /////////////////////////// Event reconstruction
\bibitem{Gaurang:energy_memo}
Gaurang B. Yodh, Event Energy Determination. Internal Milagro Memo, March
14, 2000.

\bibitem{Joe_Bussy}
Joe F. McCullough and Javier Bussonos-Gordo, Tuning the Angle Fitter,
Sampling and Curvature Corrections for Milagro. Internal Milagro Memo,
April 3, 2000.

\bibitem{andy:likelihood}
Andy Smith, My Likelihood Fitter. Internal Milagro Memo, July 1997.


\bibitem{Greg:corefitter}
Greg Sullivan and Andy Smith, An off-pond Core Finder. Internal Milagro
Memo, July 7, 2000.

\bibitem{Gaurang:gamma_proton}
Gaurang B. Yodh and Robert Atkins, A Simple Algorithm for Hadron Rejection
in Milagro. Internal Milagro Memo, February 24, 1999.

\bibitem{Gus:gamma_proton}
Gus Sinnis, Hadron Rejection in Milagro. Internal Milagro Memo, July 17,
2000.

% /////////////////////////// Performance of the detector

\bibitem{cygnus:moon}
D. E. Alexandreas et al., Phys. Rev. D 43, 1743, 1991.

\bibitem{cygnus:deleo} 
D. E. Alexandreas et al., Nuclear Instruments and Methods, A311, 350,
1992.

\bibitem{milagro_crab}
Milagro collaboration, "Observation of the Crab Nebula in the Milagro
Gamma Ray Observatory using the Particle Imaging Technique", in
preraration.


% /////////////////////////// Coordinates, technique
\bibitem{sky_coordinate_definiton}
W. M. Smart, Textbook on spherical astronomy, Cambridge University Press,
1977.

\bibitem{geodesy}
P.S. Zakatov, Course on Higher Geodesy. (Russian) Nedra Publisher, Moscow,
1976.

\bibitem{galactic_coefficients}
http://www.seds.org/$\sim$spider/spider/ScholarX/coords.html


\bibitem{handbook_space_astro}
M. V. Zombeck, Handbook of Space Astronomy and Astrophysics, Cambridge
University Press, 1990.

\bibitem{neyman_pearson}
J. Neyman and K. Pearson, Philosophical Transactions of the Royal Society
of London, Series A. vol. 231, pp. 289-337 (1933).

\bibitem{korn_korn}
G. A. Korn and T. M. Korn, Mathematical Handbook for Scientists and
Engineers, McGraw-Hill Book Company, 1968.

% \bibitem{li_ma}
% T. P. Li and Y. Q. Ma, Astrophysical Journal, 272, 317, 1983.



% /////////////////////////// Time sloshing
\bibitem{cygnus_methods}
D. E. Alexandreas et al., Nuclear Instruments and Methods, A328, 570,
1993.

\bibitem{flyeye_methods}
G. L. Cassiday et al., Phys. Rev. Lett., 62, 383 (1989).

\bibitem{zoshka}
developed with Lazar Fleysher, Milagro collaboration.

\bibitem{numerical_recipes}
see for example W. H. Press, S. A. Teukolsky, W. T. Vetterling and
B. P. Flannery, Numerical Recipes, Cambridge
University Press, 1994.

\bibitem{todd:systematics}
Todd Haines, Systematic Errors in ``Standard'' Background Estimates of
Milagro Data, Internal Milagro Memo, December 14, 1999.

\bibitem{fisher}
Sir Ronald A. Fisher, Statistical Methods for Research Workers,
Edinburgh, Oliver and Boyd, 1970.

\bibitem{diurnal_memo}
Roman Fleysher and Peter Nemethy, Diurnal Effects and Systematic Errors in
Background Determination, Internal Milagro Memo, May 22, 2001.

\bibitem{Eadie} 
see for example: W.T. Eadie, D. Drijard, F.E. James, M. Ross, B. Sadoulet.  
Statistical Methods in Experimental Physics. North-Holland Publishing
Company, 1982.

\bibitem{sun_moon}
F. W. Samuelson for the Milagro Collaboration, Proc. 27-th ICRC, Hamburg,
Germany, HE130, 2001.

%///////////////////////// appendix. Julian days
\bibitem{julian_calculator}
http://www.nr.com/julian.html


% /////////////////////////// Alternative hypothesis
\bibitem{stan_hunter_private}
private communication with Stanley D. Hunter, EGRET collaboration.


% /////////////////////////// Results
\bibitem{kamiokande_anisotropy}
K. Munakata et al., Physical Review D, 56, 23, 1997.

\bibitem{systemat_error_facts_fiction}
R. Barlow, preprint hep-ex/0207026

\bibitem{haverah_park}
S. M. Astley et al., Proc. 17-th ICRC, Paris, France, 2 156, 1981.

\bibitem{yakutsk}
N. N. Efimov et al., Proc. 18-th ICRC, Bangalore, 2 149, 1983.

\bibitem{Grigorov}
N. Grigorov et al., Proc. 12-th ICRC, Hobart, Australia, 5 1746, 1971.

\bibitem{Compton-getting}
A. H. Compton and I. A. Getting, Physical Review, 47, 817, 1935.


\end{thebibliography}
\end{document}